\documentclass{article}
\usepackage{jheppub,esint,shuffle,psfrag}
 \usepackage[utf8]{inputenc}

\usepackage{varioref}
\usepackage{amsmath,amsfonts,amsthm,mathrsfs}
\usepackage[normalem]{ulem}
\usepackage{multirow,color,graphics}
\usepackage{amsmath}
\usepackage{amsfonts}
\usepackage{amssymb}
\usepackage{jheppub}
\usepackage{enumerate}
\usepackage{fancyvrb}
\usepackage{verbatim}
\usepackage{wrapfig}
\usepackage{appendix}
\usepackage{amstext}
\usepackage{amssymb}
\usepackage{graphicx}
\usepackage{color}
\usepackage{varioref}
\usepackage{multirow,graphics}
\usepackage{epstopdf}
\usepackage{bbm}
\usepackage{slashed}
\usepackage{relsize}

\newcommand{\nn}{\nonumber}

\numberwithin{equation}{section}

\usepackage{tikz}
\usetikzlibrary{plotmarks,calc,decorations, decorations.pathmorphing, patterns}
\usetikzlibrary{arrows}
\tikzstyle dynkin node=[very thick,shape=circle,draw,inner sep=0pt,minimum size=5mm]
\tikzstyle dynkin line=[very thick]
\tikzstyle inverse line=[gray,line width=1.46pt,line cap=round, dash pattern=on 0pt off 2\pgflinewidth]
\tikzstyle red phase=[red,decoration={snake,amplitude=0.1mm,segment length=1.6mm},decorate]
\tikzstyle blue phase=[blue,decoration={snake,amplitude=0.1mm,segment length=0.9mm},decorate]
\tikzstyle green phase=[green,decoration={snake,amplitude=0.1mm,segment length=0.9mm},decorate]
\tikzstyle brown phase=[brown,decoration={snake,amplitude=0.1mm,segment length=0.9mm},decorate]

\usetikzlibrary{decorations.pathmorphing}
\tikzstyle arrow=[thick,rounded corners=18pt,-latex]
\tikzstyle box=[draw,rounded corners,outer sep=4pt]
\tikzstyle B node=[outer sep=0pt]
\tikzstyle Q node=[inner sep=1pt,outer sep=0pt]
\definecolor{purple_nice}{rgb}{0.4,0.2,0.7}
\definecolor{fuel_blue}{RGB}{42,162,185}

\def\<{\langle}
\def\>{\rangle}

\newcommand{\p}{\partial}

\newcommand{\tr}{{\text{tr}}}

\newcommand{\la}[1]{\label{#1}}
\newcommand{\eq}[1]{(\ref{#1})}

\usepackage{bbold}

\def\ga{{{\color{brown}\mathbb{0}}}}
\def\gb{{{\color{brown}\mathbb{1}}}}
\def\gc{{{\color{brown}\mathbb{2}}}}

\def\Ga{G_\ga}
\def\Gb{G_\gb}
\def\Gc{G_\gc}

\makeatletter
\@ifundefined{usebibtex}{} {}
\makeatother

\newcommand{\beq}{\begin{equation}}
\newcommand{\eeq}{\end{equation}}
\newcommand{\beqq}{\begin{equation*}}
\newcommand{\eeqq}{\end{equation*}}
\newcommand\beqa{\begin{eqnarray}}
\newcommand\eeqa{\end{eqnarray}}
\newcommand\beqaa{\begin{eqnarray*}}
\newcommand\eeqaa{\end{eqnarray*}}
\newcommand\bea{\begin{array}}
\newcommand\eea{\end{array}}
\newcommand\beaa{\begin{array}}
\newcommand\eeaa{\end{array}}

\def\e{\epsilon}

\def\Tr{\text{Tr}}

\title{
\Large  Exact Correlation Functions in Conformal Fishnet Theory    }

\author[a,b]{Nikolay Gromov}
\author[c,d]{~Vladimir Kazakov}
\author[e]{~Gregory Korchemsky}

\affiliation[a]{Mathematics Department, King's College London,
The Strand, London WC2R 2LS, UK.}
\affiliation[b]{St.Petersburg INP, Gatchina, 188 300, St.Petersburg,
Russia}
\affiliation[c]{Laboratoire de Physique Th\'eorique, D\'epartement de Physique de l'ENS, \'Ecole Normale Sup\'erieure, rue Lhomond 75005
Paris, France}
\affiliation[d]{Universit\'e Paris-VI, PSL Research University, Sorbonne Universit\'es, UPMC Univ. Paris 06, CNRS, 75005 Paris, France}
\affiliation[e]{Institut de Physique Th\'eorique\footnote{Unit\'e Mixte de Recherche 3681 du CNRS}, Universit\'e Paris Saclay, CNRS, CEA, F-91191 Gif-sur-Yvette }

\abstract{   We compute exactly various $4-$point correlation functions  of shortest  scalar operators in bi-scalar planar four-dimensional ``fishnet" CFT.  We apply the OPE to extract from these functions the exact expressions for the scaling dimensions and the structure constants of all exchanged operators with an arbitrary Lorentz spin. In particular, we determine the conformal data of the simplest unprotected two-magnon operator analogous to the Konishi operator, as well as of the one-magnon operator. We show that at weak coupling $4-$point correlation functions can be systematically expanded in terms of harmonic polylogarithm functions and verify our results by explicit calculation of Feynman graphs at a few orders in the coupling.  At strong coupling we obtain that the correlation functions exhibit the scaling behaviour typical for semiclassical description hinting at the existence of the holographic dual.
}

\usepackage{esint} 
\usepackage{breqn}
\def \Tr {\mathop{\rm Tr}\nolimits}
\def \tr {\mathop{\rm tr}\nolimits}
\def \Im {\mathop{\rm Im}\nolimits}
\def \Re {\mathop{\rm Re}\nolimits}

\def \e  {\mathop{\rm e}\nolimits}

\newcommand\lr[1]{{\left({#1}\right)}}

\newcommand \vev [1] {\langle{#1}\rangle}

\newcommand\re[1]{(\ref{#1})}

\def \qqqquad {\qquad\qquad}

\newcommand{\ft}[2]{{\textstyle\frac{#1}{#2}}}

\def\numberbysection{\@addtoreset{equation}{section}
                     \def\theequation{\thesection.\arabic{equation}}}

\begin{document}

\hskip9cm\preprint{ IPhT-T18/...,~LPTENS-04/...}
\maketitle

\flushbottom

\newpage

\section{Introduction}

\begin{figure}
    \centering
    $\bea{ccc}
    \includegraphics[scale=0.4]{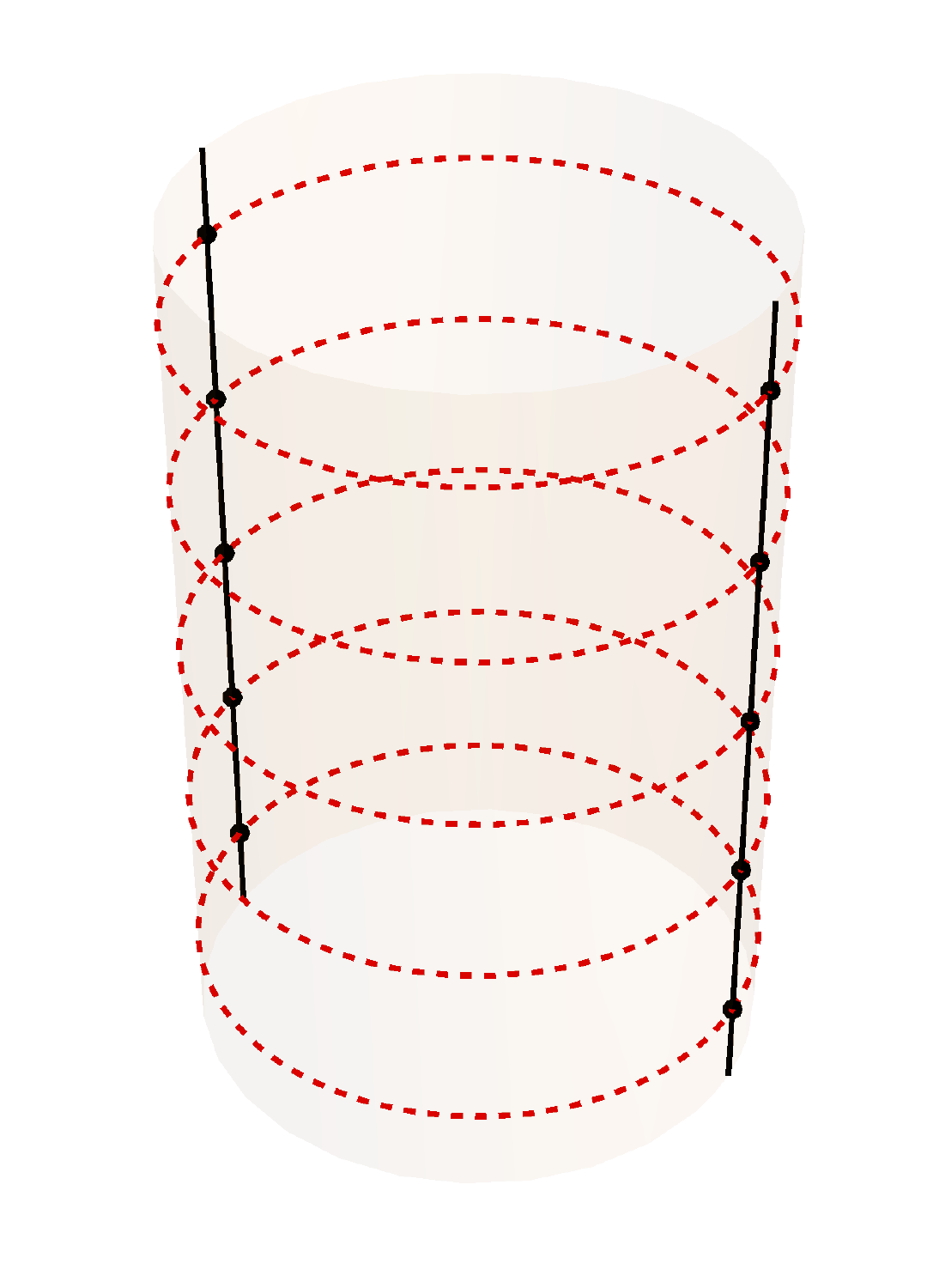}&
    \includegraphics[scale=0.4]{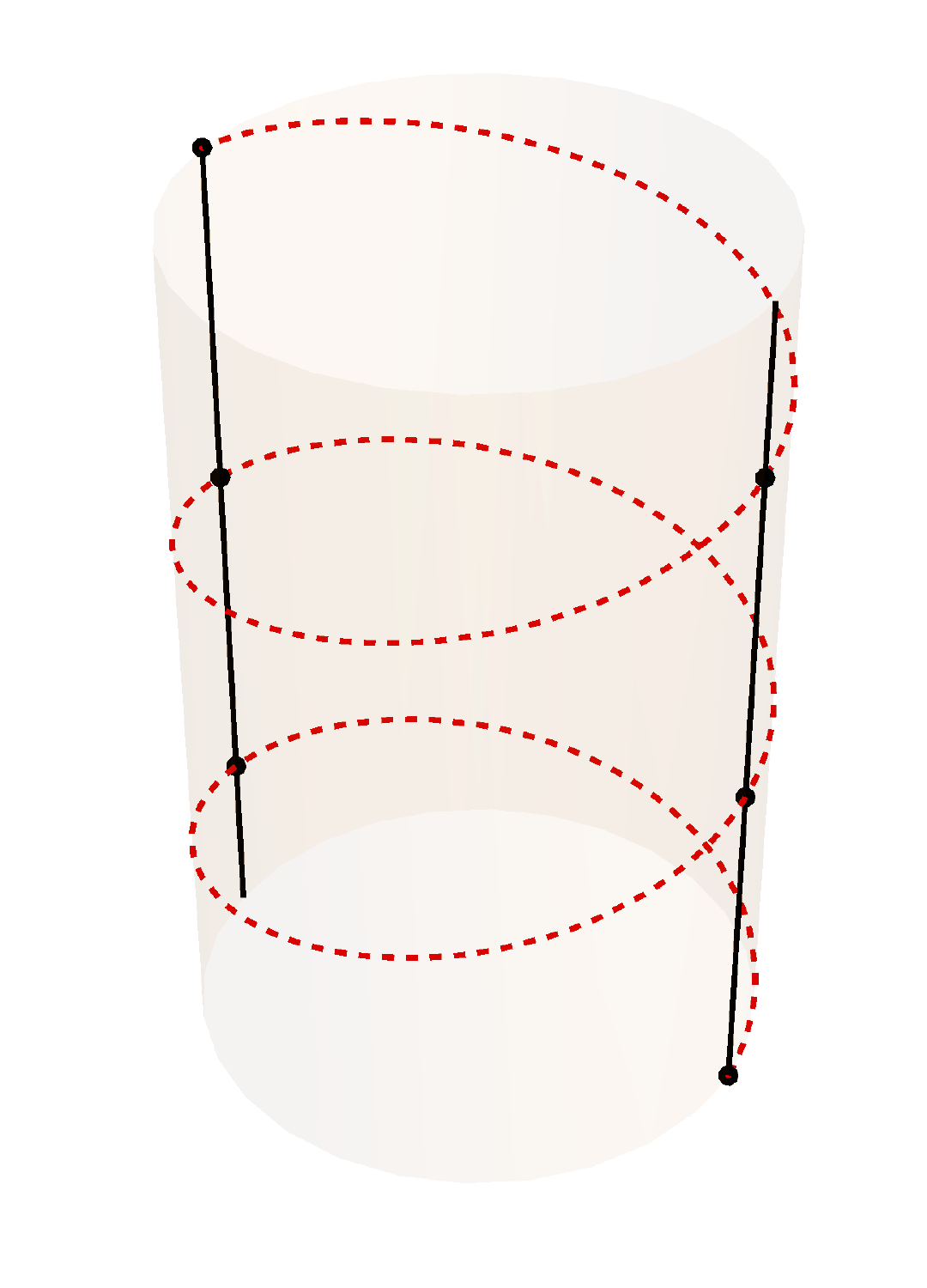}&
    \includegraphics[scale=0.4]{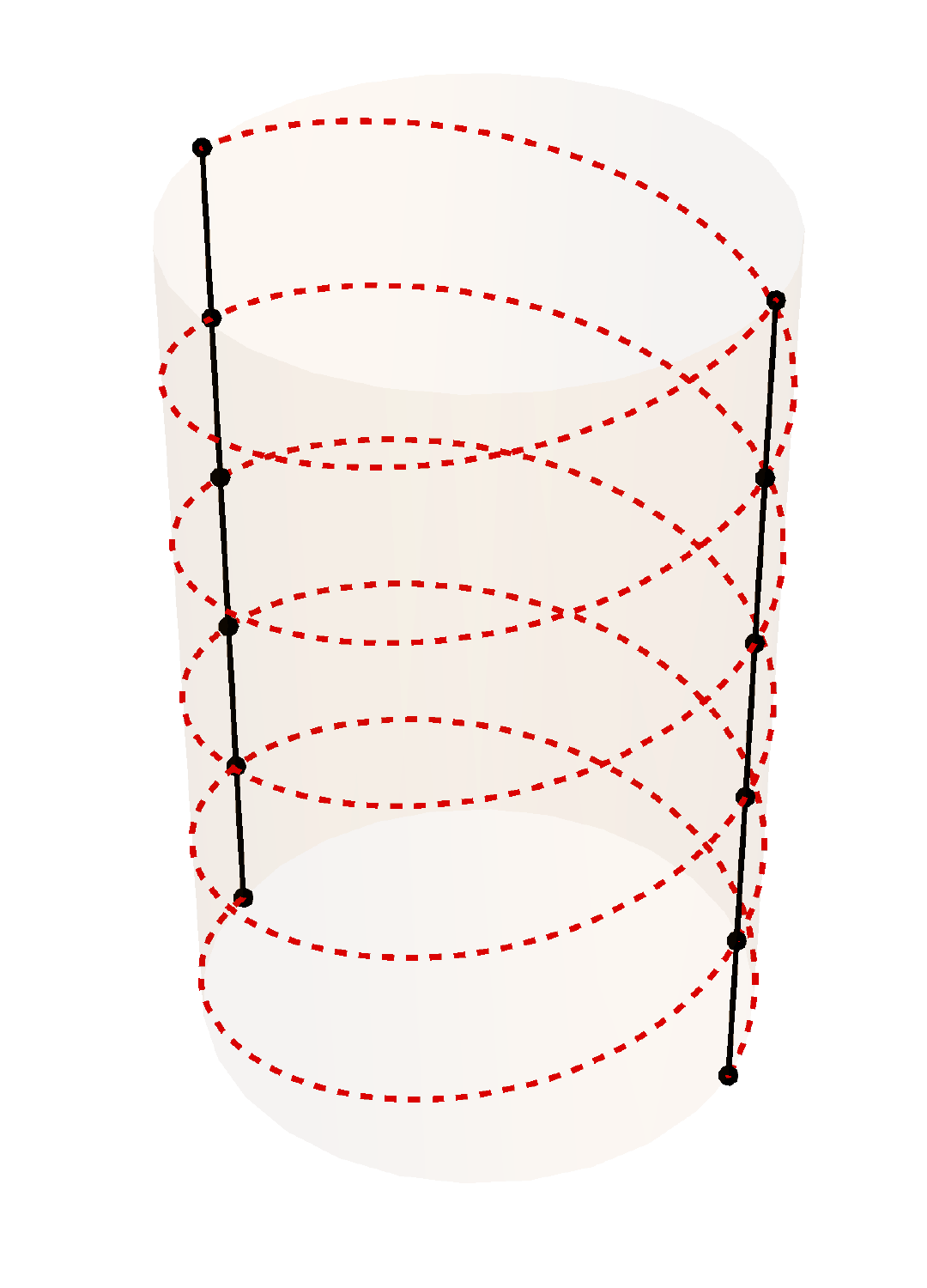}\\
    \Ga&\Gb&\Gc
    \eea$
    \caption{Three types of the 4-points functions topologies.
    These cases differ by the number of ``particles" (red dashed lines) transfered from the bottom to the top. These particles can be associated in terms of the ABA with magnons in the intermediate states on the OPE.
    We refer to them as zero, one and two particle cases correspondingly. The black lines correspond to the $X$-particles.
      }
    \label{fig:types2}
\end{figure}

Conformal  quantum field theories (CFT) have demonstrated their  importance for very diverse fundamental problems in physics, its applications ranging from  the physics of phase transitions (see \cite{Rychkov:2016iqz} and citations therein) to various problems of fundamental interactions and cosmological scenarios~(see \cite{Shaposhnikov:2018xkv} and citations therein) and QCD~\cite{Braun:2003rp}.   Whereas in $d=2$ dimensions the CFTs are well studied and classified~\cite{DiFrancesco1997},  for $d>2$  both the classification and the tools for study of CFTs are notoriously incomplete.  The supersymmetric QFTs, and in particular the supersymmetric Yang-Mills theories include  a rather large class of CFTs which are relatively well classified and, at least qualitatively, 
understood~\cite{Maldacena:1997re,Leigh:1995ep,Lunin:2005jy,Kazakov:2015efa,Cordova:2015,Cordova:2016,Cordova:2016xhm}, especially due to the discovery of the AdS/CFT correspondence. In rare cases, such as 4-dimensional \({\cal\ N}=4\) SYM theory or 3-dimensional  ABJM theory, the integrability allows us to study in-depth, at least in the 't~Hooft limit, the basic quantities of operator product expansion: 
all-loop anomalous dimensions \cite{Beisert:2006ez,Gromov:2009tv,Beisert:2010jr,Cavaglia:2014exa}, where the   comprehensive and efficient solution is given by the quantum spectral curve (QSC) approach~\cite{Gromov2014a,Gromov:2014caa} (see also recent reviews \cite{Gromov:2017blm,Kazakov:2018ugh}), 
OPE coefficients (structure constants) can be studied in various limits \cite{Escobedo:2010xs,Basso:2015zoa,Cavaglia:2018lxi,Giombi:2018qox}  and even obtain some non-perturbative information on multi-point correlators of local operators \cite{Fleury:2016ykk,Eden:2016xvg}, cusped Wilson loops~\cite{Correa:2012hh,Gromov:2015dfa}  and 
\(1/N\)-corrections~\cite{Bargheer:2017nne}.  

Much less is known about non-supersymmetric four-dimensional  CFTs.  Mostly, they are known to be  IR or UV fixed points of various renormalization group flows. Usually these CFTs are strongly coupled at these fixed points and, apart from a few rather exotic cases, such as the Banks-Zaks CFT~\cite{Banks:1981nn},  their Lagrangian description is unknown.  Such theories have been recently quite efficiently studied by conformal bootstrap methods~\cite{ElShowk:2012ht,Rattazzi:2008pe}   (using basic assumptions for CFTs, such as unitarity, crossing, symmetry etc.) which involve  heavy numerical methods and can give very accurate predictions of OPE data. However, as for any numerical approaches, the physics behind these computations often remains obscure. 

In this respect, various integrable deformations of \({\cal\ N}=4\) SYM, breaking partially or even entirely the 
supersymmetry~\cite{Leigh:1995ep,Lunin:2005jy,Frolov:2005dj,Beisert:2005if}, open a unique window into the dynamics of four-dimensional CFTs. In particular,  the \(\gamma \)-deformed \({\cal\ N}=4\) SYM, where breaking of the \(R\)-symmetry leads to the complete loss of supersymmetry, 
is a  CFT with the well-defined classical action. In order for the theory to be consistent at the quantum level, one has to add to the action a finite number of particular scalar double-trace terms, for which the couplings have to be fine tuned to special values corresponding to the
fixed points of the underlying beta-functions, or rather   functions of the Yang-Mills coupling~\cite{Sieg:2016vap,Grabner:2017pgm}. The only drawback of such a CFT is the loss of unitarity, since the double-trace couplings induced by the renormalization~\cite{Tseytlin:1999ii,Jin:2013baa,Fokken:2013aea,Fokken:2014soa} take complex values at these fixed points.  This poses an interesting challenge of construction of nontrivial unitary non-supersymmetric CFTs. On the positive side, the \(\gamma \)-deformed   \({\cal\ N}=4\) SYM at the fixed point  seems to be a genuine CFT, well defined by its explicit action, including the double-trace terms ~\cite{Grabner:2017pgm}. Last but not the least, quantum integrability property of planar \(\gamma\)-deformed \({\cal N}=4\) SYM~\cite{Beisert:2005if} described in terms of \(\gamma\)-deformed quantum spectral curve formalism in~\cite{Kazakov:2015efa}, occurs precisely at these fixed points, as was conjectured and argued in~\cite{Grabner:2017pgm}. The powerful machinery of quantum integrability allows us to study in great detail its complicated non-perturbative dynamics~\cite{Gromov:2010dy,Ahn:2011xq}.                          

Moreover,  in  a specific double scaling  limit proposed in    \cite{Gurdogan:2015csr}, combining weak coupling with strong imaginary \(\gamma\)-twists,  
the \(\gamma\)-deformed \({\cal N}=4\)~SYM drastically simplifies and gives rise to a family of chiral non-unitary CFTs  with 3 effective couplings describing the scalar and Yukawa   interactions of three complex scalars and three fermions.  
Its  spectrum of anomalous dimensions, scalar and fermion amplitudes have been studied in a series of papers~\cite{Gurdogan:2015csr,Caetano:2016ydc,Grabner:2017pgm,Gromov:2017cja,Chicherin:2017cns,Chicherin:2017frs}. In the particular, single coupling version of these models, the bi-scalar ``fishnet" CFT, studied in this paper, the four-point correlation function of certain protected operators was computed  in~\cite{Grabner:2017pgm}, providing a rich non-perturbative OPE data for the exchange operators with an arbitrary spin. These results have been generalized to any dimension \(d\) in~\cite{Kazakov:2018qbr}, where the \(d\)-dimensional version of the bi-scalar model was proposed. 
In this paper we extend these results to more general correlators. In addition to the wheel graphs we also consider single and double spiral graphs as shown on Fig.\ref{fig:types2}. We also analyse the results at weak and strong coupling. 

An important feature of such models is the drastic simplification of their weak coupling expansion, 
where in many particular cases (when we turn on a single coupling) it is dominated by various kinds of ``fishnet" Feynman graphs~\cite{Caetano:2016ydc}.  
These graphs represent integrable two-dimensional statistical-mechanical systems by themselves~\cite{Zamolodchikov:1980mb} and can be efficiently studied by the quantum spin chain methods and the double-scaled version QSC~\cite{Gromov:2016rrp,Gromov:2017cja}. 
In particular, the individual, so called ``wheel" multi-loop Feynman graphs can be computed exactly in terms of  multiple zeta values (MZV)~\cite{Gromov:2017cja}.

Importantly, the explicit graph-by-graph integrability property in such models sheds some light on the origins of the planar integrability of their ``mother"-theory -- the  \({\cal N}=4\)~SYM, 
where the perturbation theory is much more complicated and the reasons for integrability are still obscure.  In particular, in the bi-scalar CFT discussed in this paper the integrability is manifest due to the explicit integrability of fishnet graphs dominating the perturbation theory in this model  \cite{Grabner:2017pgm}.

\subsection{The conformal ``fishnet" theory}

\begin{figure}
        \centering
        \includegraphics[width = \textwidth]{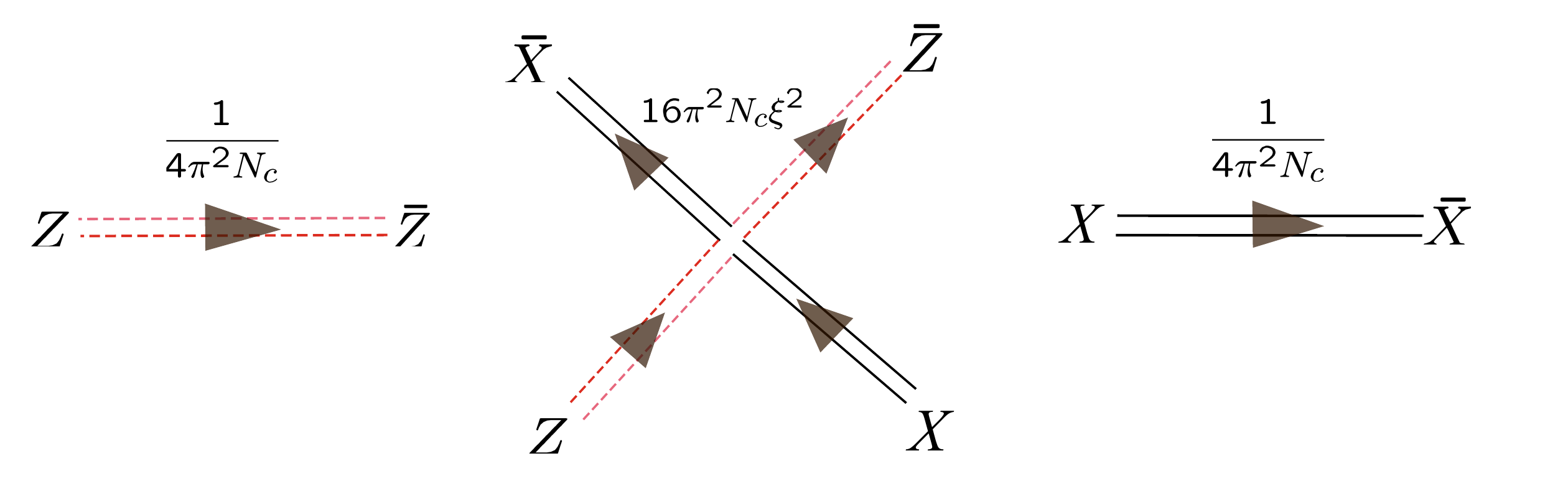}
        \caption{Feynman rules}
        \label{fig:frules}
\end{figure}

We will focus in this paper on a particular example of strongly \(\gamma\)-deformed \({\cal N}=4\)~SYM --   the bi-scalar theory  \cite{Gurdogan:2015csr}. At the classical level, the Lagrangian of the bi-scalar theory is 
given by\footnote{In the literature one also uses notations $\phi_1=X$ and $\phi_2=Z$. We use `bar' for the Hermitian conjugation.}
\begin{align}
    \label{bi-scalarL}
\mathcal L = N_c\tr\Big( \partial^\mu  \bar X \partial_\mu X+\partial^\mu \bar Z \partial_\mu Z 
+ (4\pi)^2 \xi^2  \bar X  \bar Z X Z\Big)\,,
  \end{align}
  where \(X,Z\) are complex \(N_c\times N_c\) matrix fields and \(\bar X\equiv X^\dagger,\bar Z\equiv Z^\dagger\) are their hermitian conjugates.
The model retains the $SU(N_c)$ global symmetry which is a remnant of the gauge symmetry of the original \({\cal N}=4\)~SYM theory. The effective coupling constant $\xi^2=g_{\rm YM}^2 N_c\e^{-i\gamma_3}/(4\pi)^2$ 
 is given by the product
of the Yang-Mills coupling and the complex deformation parameter. The 
general $\gamma-$deformed $\mathcal N=4$~SYM theory depends on three deformation parameters $\gamma_1,\gamma_2,\gamma_3$. The Lagrangian \re{bi-scalarL} arises
in the double scaling limit, $g^2\to 0$ and $\Im \gamma_3\to \infty$ with $\xi^2$ and $\gamma_{1,2}$ fixed. In this limit, all fields except two scalars get decoupled leading to \re{bi-scalarL}.

\begin{figure}
        \centering
        \includegraphics[width = \textwidth]{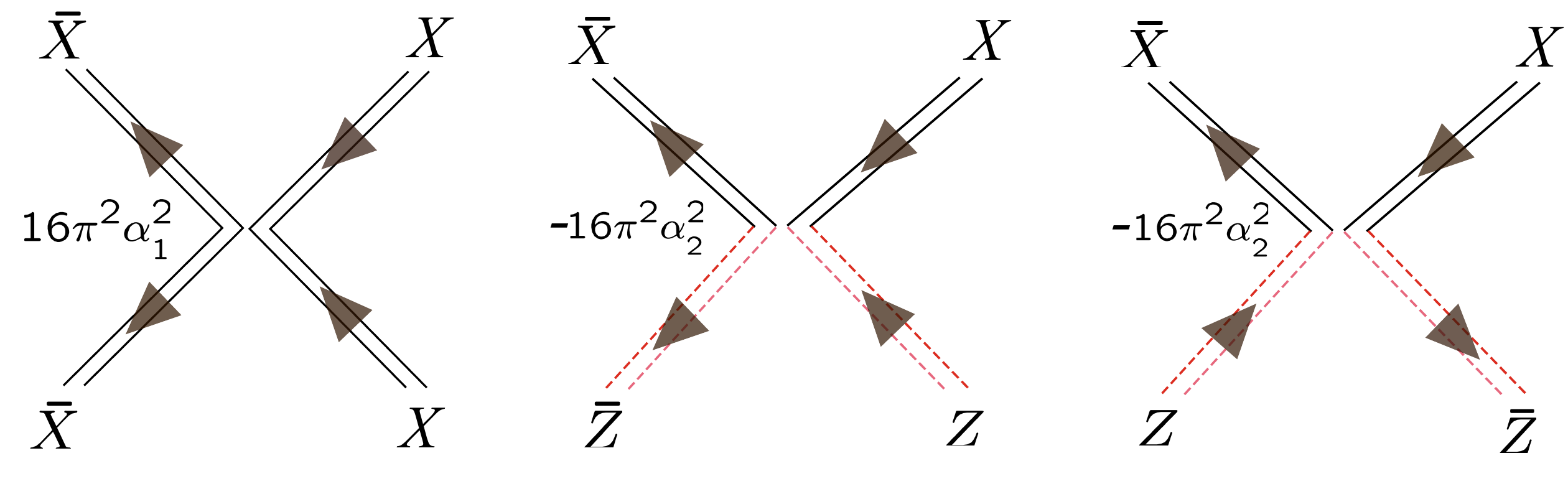}
        \caption{Feynman rules for all three types of double-trace vertices. 
        \label{fig_2traceFrules} }
\end{figure}

On the quantum level, to make the theory conformal we have to add various double-trace terms with well-tuned couplings~\cite{Sieg:2016vap,Fokken:2013aea,Fokken:2014soa,Grabner:2017pgm}:
\begin{align}\notag\label{L-dt}
\mathcal L_{\rm dt} {}& =(4\pi)^2 \alpha_1^2 
\left[\tr(X^2) \tr(\bar X^{2}) + \tr(Z^2) \tr(\bar Z^{ 2}) \right]
\\[2mm]
{}& -(4\pi)^2\alpha_2^2\left[\tr(XZ) 
\tr(\bar X \bar Z)+\tr(X\bar Z) \tr(\bar X Z)\right],
\end{align}
where $\alpha_1^2$ and $\alpha_2^2$ are new induced coupling constants and the factor of $(4\pi)^2$ is introduced for the
convenience. The corresponding Feynman rules for all types of double-trace vertices are presented on Fig.~\ref{fig_2traceFrules}.

The relative coefficients between the operators in both lines of \re{L-dt} follow from the invariance of \eqref{bi-scalarL} under the transformations of fields 
\begin{align}\label{2-sym1}(Z\to \bar Z,\ X\to \bar X)\,,\qquad 
(X\to \bar X^{{\rm t}}\,, \ Z\to  Z^{\rm t})
\,,\qquad (X\to  Z^{\rm t}\,, \ Z\to  X^{\rm t})\,,
\end{align}
 with the conjugate fields \(\bar X,\bar Z\) transforming accordingly. As we show below, these transformations can be used to establish relations between different correlation functions.

The theory with the Lagrangian $\mathcal L + \mathcal L_{\rm dt}$ is renormalizable.  The coupling constants
depend on the renormalization scale and the corresponding beta functions have been computed perturbatively in \cite{Grabner:2017pgm}  in the planar
limit, for  $N_c\to\infty$ and $\xi^2, \alpha_1^2,\, \alpha_2^2 ={\rm fixed}$. Examining zeros of the beta functions, we find
that the theory has two fixed points
\begin{align}\label{fixed}
(\alpha_1^2=\alpha_+^2\,,\ \alpha_2^2=\xi^2) \qquad \text{and} \qquad (\alpha_1^2=\alpha_-^2\,,\ \alpha_2^2=\xi^2)\,,
\end{align}
where $\alpha_\pm^2$ is given at weak coupling by the following expression
\begin{align}\label{apm}
\alpha_{\pm}^2=\pm {i\xi^2\over 2} -{\xi^4\over 2} \mp {3i\xi^6\over 4} + \xi^8 \pm {65i \xi^{10}\over 48} -{19\xi^{12}\over 10} + 
O(\xi^{14})\,.
\end{align}
Notice that the expansion of $\alpha_{\pm}^2$  runs in powers of $i\xi^2$ with real coefficients.

The planar  
bi-scalar theory \eqref{bi-scalarL} possesses a conformal symmetry at the fixed points \re{fixed}~ \cite{Sieg:2016vap,Grabner:2017pgm} and is integrable~\cite{Gurdogan:2015csr,Gromov:2017cja}. Viewed as a function
of $\xi^2$, the relation (\ref{fixed}) defines two lines of the fixed points.
It has been argued in \cite{Grabner:2017pgm} that the ``mother" theory of the bi-scalar model -- the  $\gamma-$deformed $\mathcal N=4$ SYM -- is a nonunitary  CFT on a line of (complex) fixed points of double couplings as functions of the 't~Hooft coupling, also integrable in planar limit.  It is also natural to expect the existence of such a complex conformal trajectory  even at finite \(N\).

 The bi-scalar "fishnet" CFT \eqref{bi-scalarL} is  the most studied case of the abovementioned chiral CFTs proposed in~[41]. The spectrum of long local operators of the type \(\tr(Z^mX^n)\)+permutations, can be efficiently investigated by the asymptotic Bethe ansatz (ABA) equation~\cite{Caetano:2016ydc} -- the double scaled version of the Beisert-Staudacher ABA equations for   \({\cal N}=4\)~SYM~\cite{Beisert:2006ez}. 
  The short operators of this and other types (also with insertions of  derivatives and \(\bar Z,\bar X \) fields) can be studied by QSC methods~~\cite{Gromov:2017cja} and by the quantum non-compact spin chain methods~\cite{Gurdogan:2015csr,Gromov:2017cja}, when the spins take values on conformal group \(SU(2,2)\). The spin-chain approach to this theory is very promising since it would allow us to study there  non-perturbative physics starting from the first principals, without any assumptions.  Unfortunately, efficient methods of study of non-compact, 
 Heisenberg spin chains are not very well developed, especially for principal series representation in physical space and for higher ranks symmetries, such as \(SU(2,2)\), though an important progress has been made in the study of spectral problem for \(SL(2,C)\) spin chain, in relation with high-energy (Regge) limit of QCD \cite{Derkachov:2001yn,DeVega:2001pu}.
Another remarkable observation in bi-scalar theory concerns the planar scalar amplitudes: they are dominated by a single multiloop fishnet graph with open boundary and  obey a Yangian symmetry, potentially allowing for their computation~\cite{Chicherin:2017cns,Chicherin:2017frs}.
    A particular, single-trace four-point correlation function given by such fishnet graph was explicitly computed in \cite{Basso:2017jwq}\footnote{Their result can be interpreted as a correlator of the form $\tr[X^n(x_1) Z^m(x_2) \bar X^n(x_3) \bar Z^m(x_4)]$. Alternatively one can interpret it as a leading weak coupling contribution to the $4$ point correlator of $4$ single trance operators.}. 

We refer to the single trace operators ${\rm tr}(X^n Z^m)+{\rm perm.}$ for $n\geq m$ as $m$-magnon states, in accordance with the ABA description, where the asymptotic anomalous dimension is described by the Bethe state with $m$ Bethe roots and conformal spin chain of length $n$.
The related Feynman graphs have been described in ~\cite{Caetano:2016ydc}. They have a shape of multi-spiral when \(m\) radial lines of the field \(Z\) coming out of the point where the operator is placed, are  
``braided" by \(m\)  parallel spirals, as shown on the Fig.\ref{fig:types2}(in the middle for a single spiral, and on the right for the double spiral).  
 
Correspondingly, the simplest set of non-trivial single trace operators has length $n=2$ and numbers of magnons $m=0,1,2$.
In addition, one can also introduce Lorentz spin $S$ by inserting light-cone derivatives in the following way: ${\rm tr}(X^n (\partial_-)^S Z^m)+{\rm perm}$. 
The most efficient way of studying these operators is to extract their conformal data from the OPE of $4$-point functions. Accordingly, we will analyze the $4$-point functions of $3$ different topologies, corresponding to the number of the magnons (see Fig.\ref{fig:types}). The simplest, 
zero-magnon four-point correlation function was computed to all orders of weak-coupling expansion in~\cite{Grabner:2017pgm}. It is dominated by the wheel graphs containing only two ``spikes".
It was shown in~\cite{Grabner:2017pgm} that this quantity has correct conformal properties in the perturbation theory only if one takes into account the double-trace interactions.   

  In the current paper, we will review the findings of \cite{Grabner:2017pgm} and continue to study the properties of the four-point correlation functions. In addition, we compute a few related four point-functions of short local operators, corresponding to one- and two-magnon cases. These three types of four-point functions are distinguished by the relative simplicity: for  their computation one does not need to appeal to the integrability -- the conformal symmetry is enough for this purpose.   We then use the obtained results for the four point correlation functions to extract explicit expressions for the anomalous dimensions and structure constants. 

All these  results represent a unique  opportunity of  study   properties of bi-scalar CFT, in the hope to better understand the non-perturbative  structure of  non-supersymmetric  CFTs in \(d>2\) dimensions.  
They provide  rich data for the future integrability based calculations of the correlation functions.

\subsection{Correlation functions and their perturbative structure}

In this paper, we exploit conformal symmetry to find exact expressions for correlation functions 
of local protected dimension--two operators  
\begin{align}\notag\label{O-dim2}
& O_{XZ}(x) = \tr(XZ)(x)\,,&& O_{X\bar Z}(x) = \tr(X\bar Z)(x)\,,  
\\[2mm]
& O_{\bar X Z}(x) = \tr(\bar X Z)(x)
\,,&& O_{\bar X \bar Z}(x) = \tr(\bar X \bar Z)(x)\,,
\end{align}
as well as of bi-local operators of a ``one-magnon''  type 
\begin{align}
O_{XZX}(x_1,x_2) = \tr(X(x_1)Z(x_1) X(x_2)).
\label{bi-localO}\end{align}
The reason for the  choice  \re{O-dim2}   is that, in the planar limit, the two-point correlation functions of  operators \re{O-dim2} are
protected at the fixed points \re{fixed}
\begin{align}\label{2pt-prot}
\vev{O_{XZ}(x)O_{\bar X \bar Z}(0) } = \vev{O_{X\bar Z}(x)O_{\bar X  Z}(0) } = {c^2\over (x^2)^2}+O(1/N_c^2)\,,
\end{align}
where the normalization factor $c=1/(4\pi^2)$ comes from  free scalar 
propagator.

The pair correlation  function of bi-local operators of the type \eqref{bi-localO} defined below as type \(C\) will represent another type of four-point functions, having  one-magnon exchange states in OPE, dominated by single-spiral graphs of the type shown in Fig.~\ref{fig:types2}. 
\begin{figure}[t!] 
\centerline{\parbox[c]{\textwidth}{ \includegraphics[width = \textwidth]{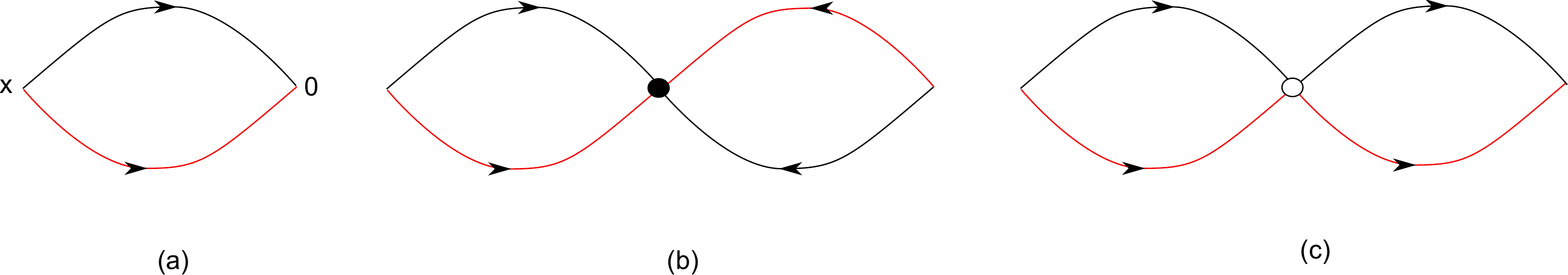}}}
\caption{Feynman diagrams contributing to the correlation function 
$\vev{O_{XZ}(x)O_{\bar X  \bar Z}(0) }$ 
 in the Born approximation (a) and at one-loop (b,c).  Filled and open circles in diagrams (b) and (c) denote single- and double-trace vertices, respectively. Black and red lines represent the propagators of   scalar fields \(X\) and \(Z\), respectively, with arrows pointing from \(X\) to \(\bar X\) and from from \(Z\) to \( \bar Z\).   The sum of diagrams (b) and (c) vanishes at the fixed point \re{fixed}.}
\label{fig:2pt}
\end{figure}

In what follows we consider the simplest unprotected four-point correlation functions of the local operators \re{O-dim2}, of the following
two types:
\begin{itemize}
\item Type A\\[-10mm]
\begin{align} \label{4pt-fun-A}
G_A=\vev{O_{XZ}(x_1) O_{X\bar Z}(x_2)O_{\bar X Z}(x_3)O_{ \bar X \bar Z}(x_4)}\,.
\end{align}
\item Type B\\[-10mm]
\begin{align}\notag\label{4pt-fun}
{}&G_{B}=\vev{O_{XZ}(x_1) O_{XZ}(x_2)O_{\bar X \bar Z}(x_3)O_{\bar X \bar Z}(x_4)}\,,
\\ {}&G_{B'}=\vev{O_{X\bar Z}(x_1) O_{X\bar Z}(x_2)O_{\bar X Z}(x_3)O_{\bar X Z}(x_4)}\,.
\end{align}

The remaining four-point correlation functions of the operators \re{O-dim2} vanish due to nonzero total $U(1)$ charge.
Notice that type \(B\) correlation function expansion in small $x_{12}^2\equiv (x_1-x_2)^2$ limit is saturated by the two-magnon operators.  
Two such spin-zero operators, \(\tr(XZXZ)\) and \(\tr(XXZZ)   \), are not protected and mix with each other, in such a way that their dimensions are related by the change \(\xi^2\to -\xi^2\). The analogous operator in \({\cal\ N}=4\) SYM theory is Konishi operator \(\tr[X,Z]^2\), where as the second operator with the same \(R\)-charge, \(\tr( 2XXZZ+XZXZ)\), is protected\footnote{This combination is obtained by acting on
the $1/2$-BPS operator $\tr(XXXX)$ with $su(2)$ lowering operator.}.    
\item\ Type C 

We will also define the type-C four-point functions containing one-magnon exchange states:  
\begin{align} \label{4pt-fun-C}\notag
&G_{C'}= \vev{\tr(X(x_1)Z(x_1) X(x_2))\tr(\bar X (x_3)\bar X(x_4)\bar Z(x_4))}
\\
&G_{C''}= \vev{\tr(X(x_1)Z(x_1) X(x_2))\tr(\bar Z (x_3)\bar X(x_3)\bar X(x_4))}\,.
\end{align}
We can also define a similar pair of correlation functions 
\begin{align} \label{4pt-fun-Ct}
& \vev{\tr(Z(x_1)\bar X(x_1) \bar X(x_2))\tr(X(x_3)\bar Z(x_4)X(x_4))}\,\notag\\ 
&\vev{\tr(Z(x_1)\bar X(x_1) \bar X(x_2))\tr(X(x_3)\bar Z(x_3)X(x_4))}\,,
\end{align}   which, due to   the    relations \re{2-sym1}, coincide with $G_{C'}$ and $G_{C''}$, respectively.

\item\ Type D

In addition to \re{4pt-fun-A} and \re{4pt-fun}, we also consider a four-point correlation function of scalar fields computed in~\cite{Grabner:2017pgm} 
\begin{align}\label{4pt-aux}
G_{D} = \vev{\tr(X(x_1) X(x_2))\tr(\bar X(x_3)\bar X(x_4))}\,.
\end{align}
As will become clear in a moment, its calculation is closely related to that of \(G_A \) \, given by  \re{4pt-fun-A} (see \re{relG} below). The function $G_D$
is obtained from $G_A$ by Wick contracting two pairs of $Z$ and \(\bar Z\) fields.

\end{itemize}

At the fixed point \re{fixed}, the correlation functions \re{4pt-fun-A}, \re{4pt-fun} and \re{4pt-aux} are finite functions of the coupling $\xi^2$.
The  correlation functions \re{4pt-fun} are related to each other through the first of transformations
\eqref{2-sym1} and, therefore, they coincide \begin{align}
G_B=G_{B'}\,.
\end{align}
For the correlation function \re{4pt-fun-A}, the relations \re{2-sym1}
imply that  $G_A$ should  be invariant under the exchange of
points
\begin{align}\label{GA-sym}
G_A \Big|_{x_1\leftrightarrow x_3,\,  x_2\leftrightarrow x_4} =G_A \Big|_{x_2\leftrightarrow x_3} =   G_A\,.
\end{align}

A remarkable feature of all considered correlation functions is their iterative structure:  the (non-zero)  contribution at each successive order of perturbation theory can be obtained from the previous one by action on it by some graph generating integral operators.
Thus the relevant graphs have
a chain structure and they can be studied 
using
the Bethe-Salpeter equation. 
In addition, the emerging graph generating operators 
commute with the generators of the conformal group and their eigenspectrum can be easily found with a help of the conformal symmetry.  
 Thanks to these features the above mentioned correlation functions of the bi-scalar model are explicitly computable in a relatively simple way.

\section{Relation to skeleton scalar graphs}

\begin{figure}
    \centering
    $\bea{ccc}
    \includegraphics[scale=0.35]{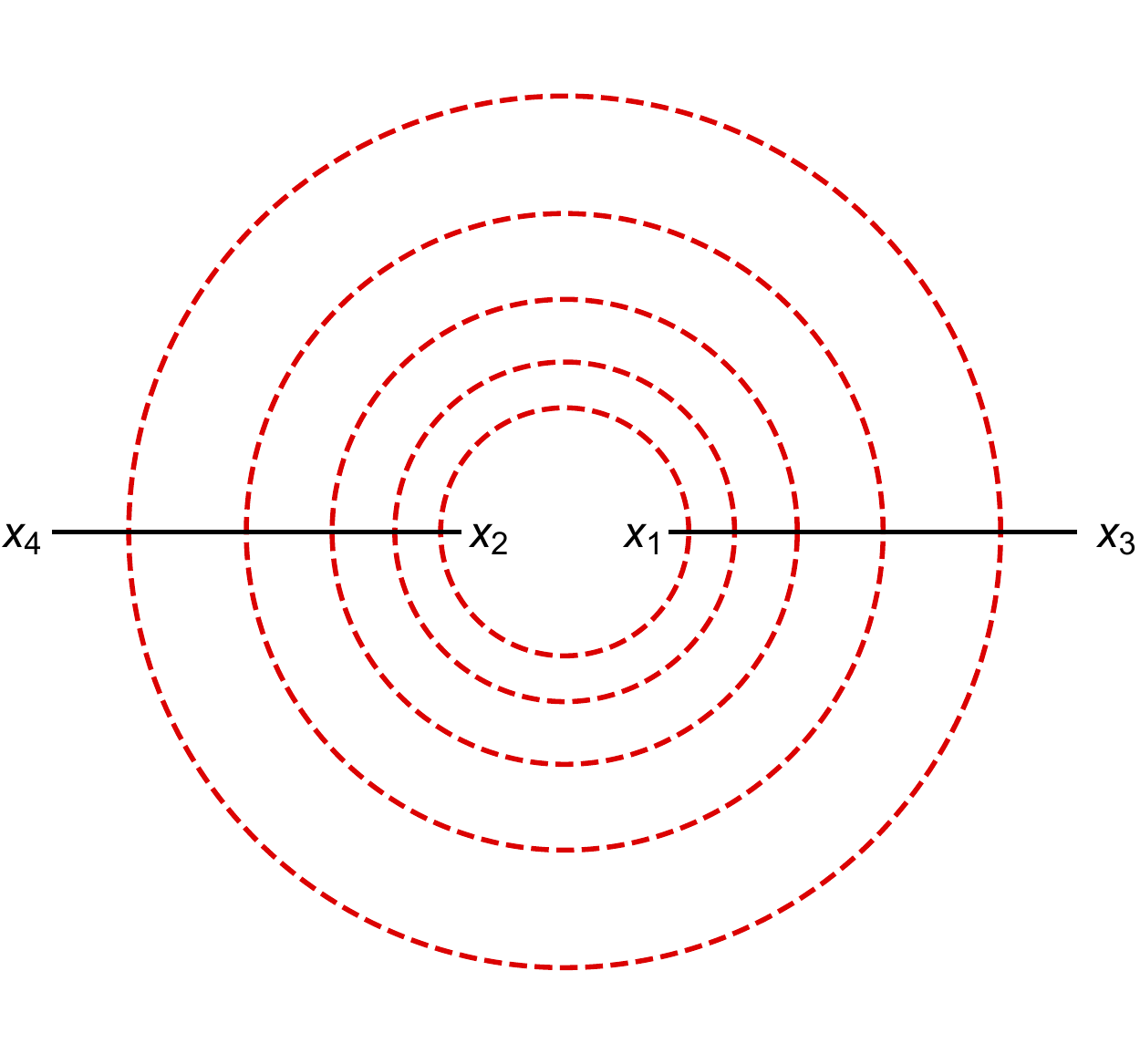}&
    \includegraphics[scale=0.35]{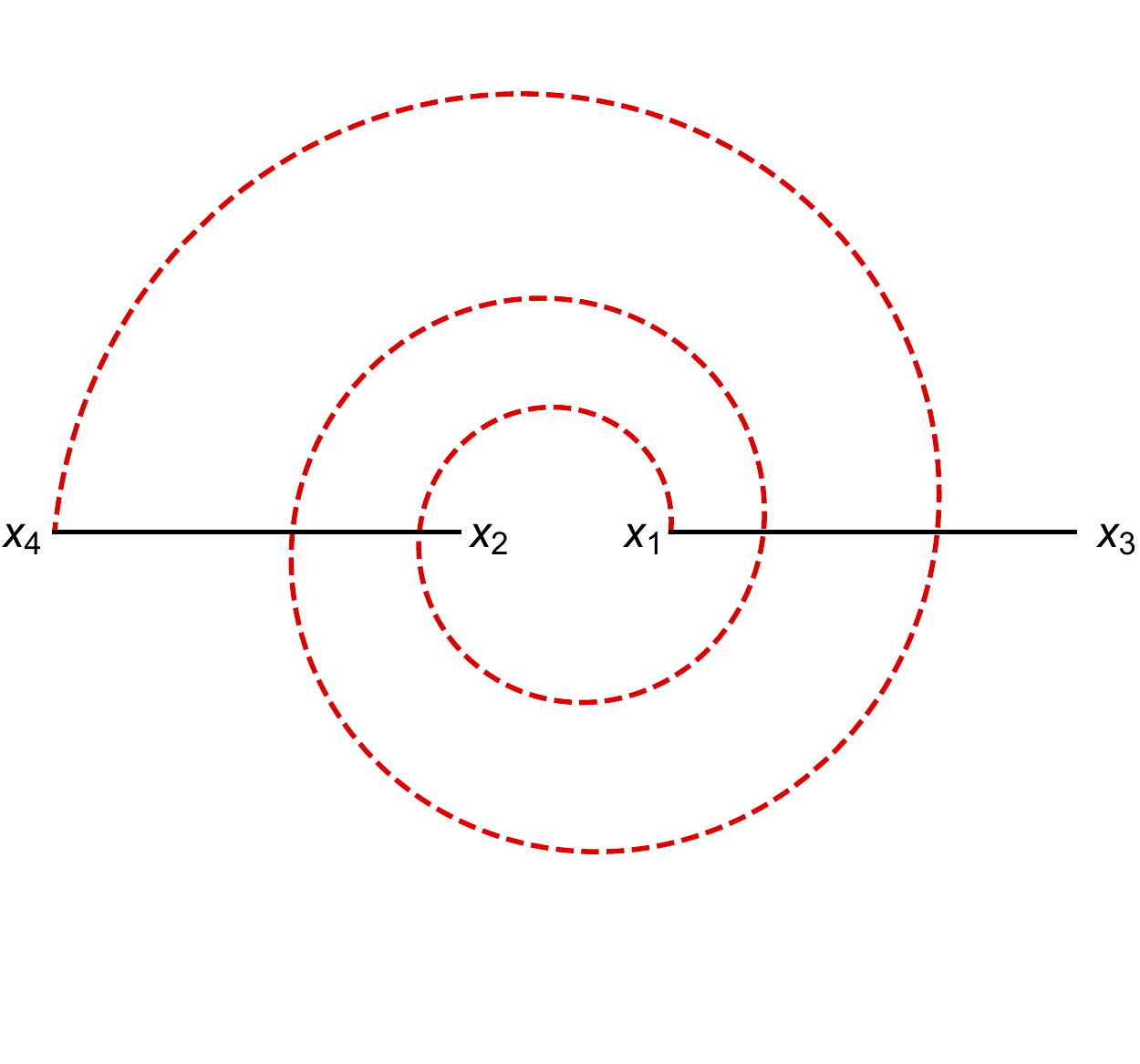}&
    \includegraphics[scale=0.35]{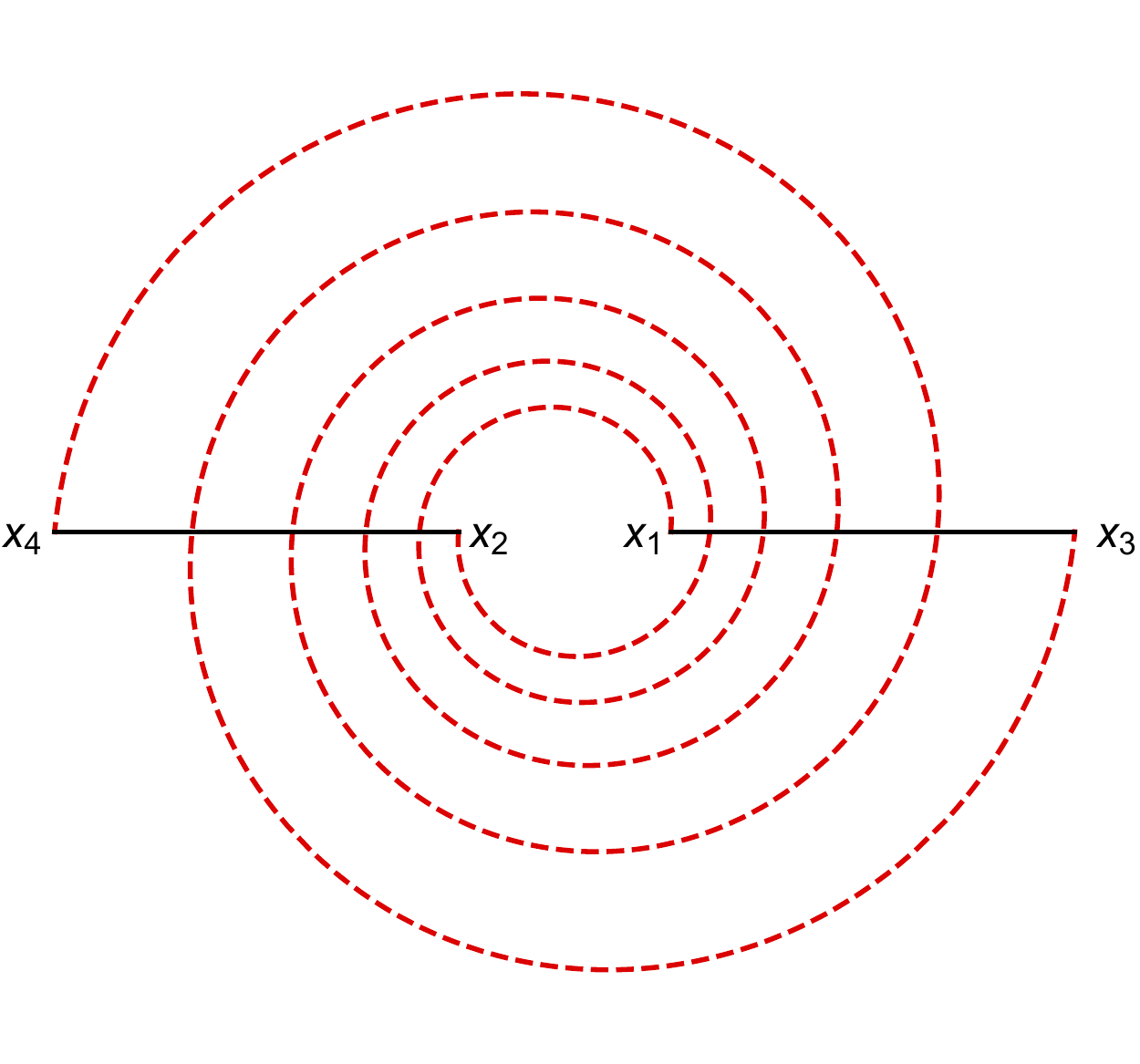}\\
    \Ga&\Gb&\Gc
    \eea$
    \caption{Three basic types of the skeleton graphs corresponding to zero, one and two magnons. They have the structure of wheels, single and double spirals. These basics graphs are the building block of all correlators we discuss in this paper.
    Another way to represent the same graphs given on Fig.\ref{fig:types2}.
     }
    \label{fig:types}
\end{figure}

In this section, we will describe the structure of  Feynman graphs for all    types of the studied 4-point correlation functions.
We express these correlators in terms of the basic generating functions of the wheel graphs ($\Ga$), single ($\Gb$) and double spiral graphs ($\Gc$) as on Fig.\ref{fig:types}.
In the section~\ref{sec:GBO} we evaluate these $3$ types of the graphs by the Bethe-Salpeter method,  by diagonalizing their graph generating kernels.
Consequently,  we compute all the related  structure constants defining the full explicit OPE representation of each of these 4-point functions \eqref{4pt-fun-C}--\eqref{4pt-aux}.  

We will first discuss the connected and disconnected parts of   all these 4-point functions and then study the connected planar part of each of them. For the sake of explicitness, we will restrict   our discussion to    \(d=4\) dimensions, but the final formulas will be readily generalized in section~\ref{sec:anyd} to any  \(d\). 

\subsection{Connected part of the correlation functions}

The four-point correlation functions of the types \(A\) and \(B\) \re{4pt-fun-A} -- \eqref{4pt-fun} receive both connected $G^{(c)}$ and disconnected   (in the sense of factorisation of the coordinate dependence)  contributions $G^{(d)} $. The former are  suppressed with respect to the latter by a power of $1/N_c^2$ 
\begin{align}
G= G^{\rm (d)} + {1\over N_c^2} G^{\rm (c)}\,,
\end{align}
but all the most interesting physics resides of course in the connected part.
The disconnected contribution is given by the product of two-point correlation functions \re{2pt-prot} leading to  
\begin{align} 
G_A^{\rm (d)} {}&  = {c^4\over (x_{14}^2x_{23}^2)^2}\,,
\qquad\qquad
 G_{B}^{\rm (d)} 
= {c^4 \over (x_{14}^2x_{23}^2)^2}+ (x_1\leftrightarrow x_2)\,.
\end{align}
For the correlation function of type \(C''\) and $D$
the disconnected contribution is \begin{align}\label{GCdisc} 
G_{C''}^{\rm (d)} {}&  = {c^3\over (x_{13}^2)^2\,x_{24}^2}\,,\qquad\qquad
G_{D}^{\rm (d)}  = {c^2\over x_{13}^2\,x_{24}^2}\,+(x_3\leftrightarrow x_4)\,.
\end{align}
Due to the color structure of bi-local operators in the definition of $G_{C''}$ and $G_{D}$, 
the disconnected part in both cases is of the same order in \(1/N_c\) as the connected part.

Finally, the disconnected part of the correlation function of type \(C'\) is suppressed by the factor of $1/N_c^2$ as compared with its connected part.

\subsection{Relation of 4-point functions of all types to basic fishnet graphs}

The computation of all above-mentioned types of 4-point functions can be reduced to the evaluating Feynman graphs having 3 basic structures of fishnet graphs. They are depicted  on the Fig.\ref{fig:types2} and they are distinguished by the the magnon numbers - \(0,1,2\) - the number of propagating ``particles" (dotted spirals lines on the Fig.\ref{fig:types2}) in the exchange channel.
We will  denote the related generating functions as \(\Ga\), \(\Gb\) and \(\Gc\)    and call them the \(n\)-magnon functions.
 As we will see shortly, the Feynman graphs for these  functions are computable by the Bethe-Salpeter approach, due to their periodic fishnet structure and conformal properties.  

To be more precise, we will define \(n\)-magnon functions as perturbative expansion w.r.t. the coupling \(\xi^2\) as follows. 

For zero-magnon case (the ``wheels") we have \begin{align}\label{G0-def}
\Ga(x_1,x_2|x_3,x_4) =  \sum_{n\ge 0} (16\pi^2\xi^2)^{2n} \Ga^{(n)}(x_1,x_2|x_3,x_4)\,,
\end{align} 
where  $\Ga^{(n)}(x_1,x_2|x_3,x_4)$ denotes the Feynman  graph depicted on the left of Figs.\ref{fig:types2} and \ref{fig:types} and having \(2n\) interaction vertices (black dots).

 For one-magnon case (``spiral", or ``spiderweb" graphs) the structure function looks as 
\begin{align}\label{G1-def}
\Gb(x_1,x_2|x_3,x_4) =  \sum_{n\ge 0} (16\pi^2\xi^2)^n\Gb^{(n)}(x_1,x_2|x_3,x_4)\,,
\end{align} where  $\Gb^{(n)}(x_1,x_2|x_3,x_4)$ denotes the Feynman  graph depicted in the middle of Figs.\ref{fig:types2} and \ref{fig:types} and having \(n\) interaction vertices (black dots).
Each graph takes a shape of a spiral  consisting of propagators of type \(X\) wounding around two lines of propagators of type \(Z\).
Note that we have two distinguished structures depending  on  the parity of  expansion term w.r.t. \(\xi^2\): for even powers of \(\xi^2\) the spiral on Fig.\ref{fig:types2} starts and ends on the same (black) line of \(Z\)-propagators, whether as for odd
powers of \(\xi^2\) it starts and ends on different \(Z\)-lines.  \\

For two-magnon case (``double spiral") we have the structure function 
\begin{align}\label{G2-def}
\Gc(x_1,x_2|x_3,x_4) =  \sum_{n\ge 0} (16\pi^2\xi^2)^{2n} \Gc^{(n)}(x_1,x_2|x_3,x_4)\,,
\end{align} where  \(\Gc^{(n)}(x_1,x_2|x_3,x_4)\) denotes the Feynman  graph depicted on the right of Figs.\ref{fig:types2} and \ref{fig:types} and having \(2n\) interaction vertices (black dots).   
 
Let us relate the correlation functions \(G_A\), \(G_B\), \(G_C\) and  \(G_D\) to \(\Ga\), \(\Gb\) and \(\Gc\).

\subsubsection{Relations between correlation functions and $n$-magnon functions}

First, we can immediately see from Figs.~\ref{fig:types2}(left) and \ref{fig:types}(left) and from the definitions \eqref{4pt-aux} and \eqref{G0-def} that 
\begin{align}\label{GD-def}
G_D(x_1,x_2|x_3,x_4) =\Ga(x_1,x_2|x_3,x_4)\,+\Ga(x_1,x_2|x_4,x_3)\,.
\end{align}
Further on, the correlation function $G_A$ is given by the following linear combination of the functions \(\Ga\)
\begin{align}\notag\label{GA-G}
G_A {}&= \frac{c^2}{ x_{12}^2  x_{34}^2}\Ga(x_1,x_2|x_3,x_4)  +\frac{c^2}{ x_{12}^2  x_{34}^2}\Ga(x_1,x_2|x_4,x_3)
\\[2mm]
{}&  +\frac{c^2}{  x_{13}^2 x_{24}^2 }\Ga(x_1,x_3|x_2,x_4) +\frac{c^2}{  x_{13}^2 x_{24}^2 } \Ga(x_1,x_3|x_4,x_2) -
\frac{c^4}{ x_{12}^2 x_{13}^2 x_{24}^2 x_{34}^2}\,,
\end{align}
where the last term takes care of double counting of the tree-level diagram in the first $4$ terms.
It is straightforward to verify that the linear combination on the right-hand side of \re{GA-G} satisfies the symmetry relation \re{GA-sym}.  

Comparing \re{GA-G} and \re{GD-def} we notice that the two correlation functions are related to each other as
\begin{align}\label{relG}
G_A = \lr{{G_D \over x_{12}^2 x_{34}^2 } + (x_2\leftrightarrow x_3)} -
\frac{c^4}{ x_{12}^2 x_{13}^2 x_{24}^2 x_{34}^2}\,.
\end{align}
Having determined  $\Ga(x_1,x_2|x_3,x_4)$, we can apply \re{GD-def} and \re{GA-G} to find the correlation functions $G_A$ and $G_D$.

For the one-magnon correlation functions we get the following expressions through the even and odd in \(\xi^2\) parts of the magnon function \(\Gb\): 
\begin{align}\label{G1p-def}
&G_{C'}   =\frac{1}{2}\Gb(x_1,x_2|x_3,x_4)-(\xi^2\to -\xi^2)\,,\notag\\
&G_{C''}   =\frac{1}{2}\Gb(x_1,x_2|x_3,x_4)+(\xi^2\to -\xi^2) .
\end{align}

Finally, the two-magnon correlation function
\eqref{4pt-fun}
coincides with the two-magnon function  \eqref{G2-def}:
\begin{align}\label{GB-def}
G_B(x_1,x_2|x_3,x_4) =\Gc(x_1,x_2|x_3,x_4)\,.
\end{align}  

In the next section we review the general method for computing the magnon functions \(\Ga\), \(\Gb\) and \(\Gc\) based on the Bethe-Salpeter equation and conformal symmetry. The explicit expressions for these functions are derived in section~\ref{sec:GBO}.

\section{Conformal symmetry and  Bethe-Salpeter equations: generalities}\label{sec:BSgen}

As we saw above and  discuss in detail in the next section,
the correlation functions 
introduced in previous sections are given by very specific type of fishnet graphs: in each graph the  periodically repeating configurations of propagators  are connected by pairs of coordinates of  the related     vertices (see Fig.~\ref{fig:types}).   Each graph can be cut into two disconnected parts by splitting only two vertices. The three cases we are going to consider differ, in particular, by the values of dimensions of four
external (protected) operators.
For all correlation functions under considerations we have $\Delta_1=\Delta_4$ and $\Delta_2=\Delta_3$.

This section is based on the observation that three topologically distinct configurations $\Ga$, $\Gb$ and $\Gc$ can be written, each, in terms of a suitable ``graph-building" operator $\hat H$.
In each case we find at the level of the operators
\beq\la{Ggeneral}
\hat G=\left(\frac{c}{x_{12}^{2}}\right)^{-d+\Delta_1+\Delta_2}
\sum_{\ell=0}^\infty \chi^\ell\hat H^{\ell+n}=
\left(\frac{c}{x_{12}^{2}}\right)^{-d+\Delta_1+\Delta_2}
\frac{\hat H^n}{1-\chi\hat H}
\eeq
where $n$ is a nonnegative integer,
the constant $\chi$ is proportional to a fixed power of the coupling constant $\xi^2$ (specified below for each case) and $d$ is the dimension of the space-time. In most of the paper we set $d=4$ although, as we will see in the section~\ref{sec:anyd}, most of the equations discussed here have a natural generalization to general $d$, where the bi-scalar theory can be also formulated~\cite{Kazakov:2018qbr}. 
The operators $\hat H$ and $\hat G$ are represented by the corresponding integration kernels, e.g.
\beq
\langle x_1,x_2| \hat G|x_3,x_4\rangle = G(x_1,x_2|x_3,x_4)\,,
\eeq
in such a way that
\begin{align}\label{H-conv}
\langle x_0,y_0|\hat H^m|x_m,y_m\rangle=
\int \prod_{j=1}^{m-1} d^4 x_j d^4 y_j { H}(x_0,y_0|x_1,y_1) { H}(x_1,y_1|x_2,x_2)\dots { H}(x_{m-1},y_{m-1}|x_m,y_m)\;.
\end{align}
The problem of finding $G(x_1,x_2|x_3,x_4)$ can be split into two main steps. First, we have to solve the eigenvalue problem for $\hat H$ and, then, decompose $\hat G$ over the complete basis of the eigenfunctions of $\hat H$.

Fortunately, the first step is simple in our case. The eigenfunctions \(
\Phi_{\nu,x_0}^{\mu_1,\dots,\mu_S}(x_1,x_2)
\) of $\hat H$ which are defined by 
\begin{align}\la{waveGeneral}
\int d^4 x_1 d^4 x_2\,\Phi^{\mu_1,\dots,\mu_S}_{\nu,x_0}(x_1,x_2){ H}(x_2,x_1|x_4,x_3)={E_{\Delta,S}}
\Phi^{\mu_1,\dots,\mu_S}_{\nu,x_0}(x_3,x_4)\;
\end{align}
where $x_0,\nu$ and $S$ parameterise the eigenstates,   are totally symmetric traceless tensors in $4$-dimensional indices $\mu_1,\dots,\mu_S$.
The form of \(
\Phi_{\nu,x_0}^{\mu_1,\dots,\mu_S}(x_1,x_2)
\)  is completely fixed 
by the conformal symmetry -- if the operator $\hat H$ commutes with the generators of the conformal group, its eigenstates should 
transform covariantly under conformal transformations acting on $x_i$. As such, \(
\Phi_{\nu,x_0}^{\mu_1,\dots,\mu_S}(x_1,x_2)
\) can be represented as a conformal ``triangle" -- three-point correlation function of two scalar operators
at the points \(x_1,x_2\) and some reference operator with dimension \(\Delta=2+2i\nu\) and spin \(S\) at the point \(x_0\).
Explicitly
\begin{align}\notag\label{waveFGeneral}
\Phi_{\nu,S,x_0}(x_1,x_2)
{}& \equiv \Phi_{\nu,x_0}^{\mu_1,\dots,\mu_S}(x_1,x_2)n_{\mu_1}\dots n_{\mu_S}
\\
{}& ={1\over x_{12}^{\Delta_1+\Delta_2-t}x_{10}^{\Delta_1+t-\Delta_2}x_{20}^{\Delta_2+t-\Delta_1}} 
\left(
\frac{2 (nx_{02})}{x_{02}^2}
- \frac{2 (nx_{01})}{x_{01}^2}
\right)^S   
\end{align}
where  \footnote{Following the standard conventions, we shall refer to $t$ at zero value of the coupling constant as twist.} $t=\Delta-S$ and $x_{ab}^\Delta \equiv  (x_{ab}^2)^{\Delta/2}$. In order to simplify tensor structure, 
we projected all Lorentz indices onto a light-like vector $n^\mu$.
In the next section we verify 
explicitly that the functions \re{waveFGeneral} 
diagonalize the graph-building Hamiltonians and find the corresponding eigenvalues $E_{\Delta,S}$.

The scaling dimension $\Delta$ in \re{waveFGeneral} is given by \cite{Tod:1977harm}
\begin{align}
    \Delta=2+2i\nu
\end{align}
where $\nu$ is real nonnegative. For such values of $\Delta$ the functions \(
\Phi_{\nu,x_0}^{\mu_1,\dots,\mu_S}(x_1,x_2)
\) define the complete orthonormal basis of states on the Hilbert space on which the graph building kernel $\hat H$ acts. Viewed as a function of $x_0$,  \(
\Phi_{\nu,x_0}^{\mu_1,\dots,\mu_S}\) belongs to the irreducible principle series representation of the conformal group labelled by real $\nu$ and nonnegative integer Lorentz spin $S$.
Together with \re{waveGeneral} this leads to
\begin{align}\label{H-dec}
H(x_1,x_2|x_3,x_4) = \sum_{S=0}^\infty\frac{(-1)^S}{
{(x^2_{12})^{\Delta_1+\Delta_2-4}}
}\int_0^\infty\frac{d\nu}{c_1(\nu,S)} E_{\Delta,S}\int d^4 x_0\,
{{\Phi^{\mu_1\dots\mu_S}_{-\nu,x_0}}(x_1,x_2)
{\Phi^{\mu_1\dots\mu_S}_{\nu,x_0}}(x_4,x_3)
}{}\;
\end{align}
where $c_1$ is the normalization factor defined in \re{c1general} and we used that $\bar \Phi^{\mu_1\dots\mu_S}_{\nu,x_0}=\Phi^{\mu_1\dots\mu_S}_{-\nu,x_0}$. The values of $\nu$ in \re{H-dec} can be restricted to be $\nu\ge 0$ since the states $\Phi_{\nu,x_0}^{\mu_1,\dots,\mu_S}$ and 
$\Phi_{-\nu,x_0}^{\mu_1,\dots,\mu_S}$ belong to the same representation and  are related to each other through intertwining relation.

Inserting \re{H-dec} into  \eq{Ggeneral} we get the following expression for the correlation function $G$
\beq\label{G-aux}
G(x_1,x_2|x_3,x_4)=
\sum_{S=0}^\infty\frac{(-1)^S}{c^{4-\Delta_1-\Delta_2}}\int_0^\infty\frac{d\nu}{c_1(\nu,S)}\frac{E_{\Delta,S}^n}{1-\chi E_{\Delta,S}}\int d^d x_0
{{\Phi^{\mu_1\dots\mu_S}_{-\nu,S,x_0}}(x_1,x_2)
{\Phi^{\mu_1\dots\mu_S}_{\nu,S,x_0}}(x_4,x_3)
}\;.
\eeq
The integral over $x_0$ can be evaluated explicitly in terms of four-dimensional conformal blocks\;\cite{Dolan2011,Dolan:2000ut,Tod:1977harm} 
\beqa
\label{Phi-blocks}
  &&\int d^4 x_0 \Phi^{\mu_1\dots \mu_S}_{-\nu,S,x_0}(x_1,x_2)\Phi^{\mu_1\dots \mu_S}_{\nu,S,x_0}(x_4,x_3) \\
 \nn&&= \left(\frac{1}{x_{12}x_{34}}\right)^{\Delta _1+\Delta _2} \left(\frac{x_{13}x_{24}}{x_{14}^2}\right)^{\Delta
        _1-\Delta _2} 
\notag\left[
c(\nu,S)g_{\Delta,S}(u,v)+c(-\nu,S)g_{4-\Delta,S}(u,v)
\right]\,,
\eeqa
the expression in the brackets depends only on the cross ratios $u$ and $v$ defined as
\begin{align}\label{zbarz}
u=z\bar z={x_{12}^2 x_{34}^2\over x_{13}^2 x_{24}^2}\,,\qqqquad 
v=(1-z)(1-\bar z)={x_{23}^2 x_{14}^2\over x_{13}^2 x_{24}^2}\,,
\end{align}
where $z$ and $\bar z$ are auxiliary complex variables. 
The conformal block \(g_{\Delta,S}(u,v)\)  depends on $\Delta_1$ and $\Delta_2$ and it is given explicitly in terms of the hypergeometric functions in \eq{defg}.
The coefficient $c(\nu,S)$ is a ratio of the normalisation coefficients $c(\nu,S)=c_1(\nu,S)/c_2(\nu,S)$
defined in \eq{c1general} and \eq{c2general}. 

Combining together \re{G-aux} and \re{Phi-blocks} we obtain\footnote{The constant $c=\frac{1}{4\pi}$ should not be confused with the function $c(\nu,S)$.} 
\beqa\la{GGgen}
{G(x_1,x_2|x_3,x_4)}\equiv
{\left(\frac{c}{x_{12}x_{34}}\right)^{\Delta _1+\Delta _2} \left(\frac{x_{13}x_{24}}{x_{14}^2}\right)^{\Delta
                _1-\Delta _2}}
{\cal G}(u,v)
\;,
\eeqa
where ${\cal G}(u,v)$ admits the following representation
\beqa\label{Ggen}
{\cal G}(u,v)=
c^{-4}\sum_{S\ge 0}(-1)^S\int_{-\infty}^\infty\frac{d\nu}{c_2(\nu,S)}\frac{E_{\Delta,S}^n}{1-\chi E_{\Delta,S}}
g_{\Delta,S}(u,v)\,,
\eeqa
where $\Delta=2+2i\nu$.
Here we combined the two terms on the right-hand side of \re{Phi-blocks} to extend the integration over $\nu$ to the whole real axis and used the identity $c(\nu,S)=c_1(\nu,S)/c_2(\nu,S)$. Note that doing so we required that 
\beq\la{Esym}
E_{4-\Delta,S}=E_{\Delta,S}\;.
\eeq
This property follows from the fact that the states with the same Lorentz spin $S$ and the scaling dimensions $\Delta$ and $4-\Delta$ belong to the same representation of the conformal group.
We will check \re{Esym} explicitly in each case. 

To bring the integral \re{Ggen} to the standard
OPE form we examine the short distance limit $x_{12}\to 0$, or equivalently $u\to 0$ and $v\to 1$. In this limit,
the conformal block scales as
$g_{2+2i\nu,S}(u,v)\sim u^{1+i\nu -S/2} (1-v)^S$ and decays exponentially fast for $\Re (i\nu)\to \infty$.
This allows us to close the integration contour in \re{Ggen} to the lower half-plane and compute the integral on the right-hand
side of \re{Ggen} by 
residues. In terms of the OPE, the condition $\Re(i\nu)>0$ is equivalent to the restriction $\Re \Delta >2$ on the scaling dimension of 
exchanged operators.

The integrand in \re{Ggen} has `physical' poles coming from zeros of the denominator
\beq\la{forspec}
\frac{1}{E_{\Delta,S}}=\chi\,, 
\eeq
and two series of 
`spurious' poles generated by the kinematical factor $c_2(\nu,S)$ and the conformal block $g_{2+2i\nu,S}(u,v)$. We show in Appendix~\ref{App:cancel} that the spurious poles cancel against each other provided that 
$E_{\Delta,S}$ satisfies the following relation
\beq\la{conditionE0}
r_k (E_{3+S+k,S}-E_{3+S,S+k})=0\,,\qquad (k=0,1,2,\dots)\;.
\eeq 
with  $r_k$  defined in \eq{rn}. Then, the correlation function \re{Ggen} is
given by the sum of residues at the physical poles
\re{forspec}. 
Finally, we obtain the following conformal partial wave expansion of the correlation function \re{Ggen}
\beqa\la{Gres}
{\cal G}(u,v)=
\sum_{S,\Delta}C_{\Delta,S}\;
g_{\Delta,S}(u,v)\,,
\eeqa
where the OPE coefficients are given by the residues at the physical poles
\beqa\la{Cstruct}
C_{\Delta,S}=\frac{(-1)^S}{c^4} \; 4\pi\;{\rm res}_{\Delta}\left(\frac{1}{c_2(\nu,S)}\frac{E_{\Delta,S}^n}{1-\chi E_{\Delta,S}}\right)
\eeqa 
 and the sum in \eq{Gres} runs over solutions of \eq{forspec} with ${\rm Re}\;\Delta>2$.
 
 In the next section, we apply the relations \eq{Cstruct} and
 \eq{Gres} to  compute the four-point correlations introduced in the previous section. 
 In each case we shall verify the relations \eq{Esym} and
 \re{conditionE0} which we assumed  in the above derivation.

\section{Four-point correlators and the conformal OPE data}\la{sec:GBO}

In this section, we describe Feynman graphs for the $3$ types of 4-point correlation functions depicted on Fig.\ref{fig:types}, and establish the corresponding graph-building operators ${\hat H}$.
In the previous section, it was shown that the eigenvalue $E_{\Delta,S}$ of these operators is the only input needed in order to write a representation of OPE type for the correlation function. We diagonalize the operators ${\hat H}$ and present explicit expressions for the conformal data (scaling dimensions and the OPE coefficients) for 
each of these 4-point functions. 

\subsection{Zero-magnon case and the wheel-graphs ($\Ga$)}

The zero-magnon correlation function
was studied in detail in~\cite{Grabner:2017pgm}. Here we re-derive the results of~\cite{Grabner:2017pgm} and show how they fit into the general scheme described in the previous section.

As we show below, the zero-magnon case corresponds to the situation when the correlation function \re{Gres} receives contribution from magnon-free operators 
of the type ${\rm tr}(X(n\partial)^S X)$ and ${\rm tr}(X\Box(n\partial)^S X)$\footnote{Interestingly, in our case only single box can appear. In space-time dimension $d\neq 4$ this is not the case.}. 
We will see that only these two types of operators with arbitrary spin $S$ contribute. 

To find the zero-magnon correlation function we have to summed up diagrams shown on Figure~\ref{fig:types}(left). These diagrams contain an arbitrary number of wheels attached to the rest of the diagram at two (single-trace) vertices. It is easy to see that the integral over the position of these vertices develops a ultraviolet (UV) divergence at short distances. This seems to be in contradiction with expected UV finiteness of four-point correlation function of protected operators.
We recall however that quantum corrections induce new double-trace interaction terms \re{L-dt}. In particular, the 
double-trace coupling $\tr(X^2)\tr(\bar X^2)$ affects the 
zero-magnon correlation function. It produces a UV divergent contribution which cancels against that of the wheel graphs in such a way that the four-point correlation function remains UV finite at any order in the weak coupling expansion. Due to form of the double-trace interaction term $\tr(X^2)\tr(\bar X^2)$, it can only affect the contribution of 
partial waves to \re{Gres} with zero Lorentz spin $S=0$. 
We therefore expect that the contribution of the wheel graphs to \re{Gres} is well-defined for $S\neq 0$ whereas for $S=0$ the additional contribution due to double traces should be taken into account. We discuss this issue in more detail in Sec.\ref{sec:doubletrace}.

\begin{figure}[t]
	\centering
	$\bea{ccc}
	\includegraphics[scale=0.4]{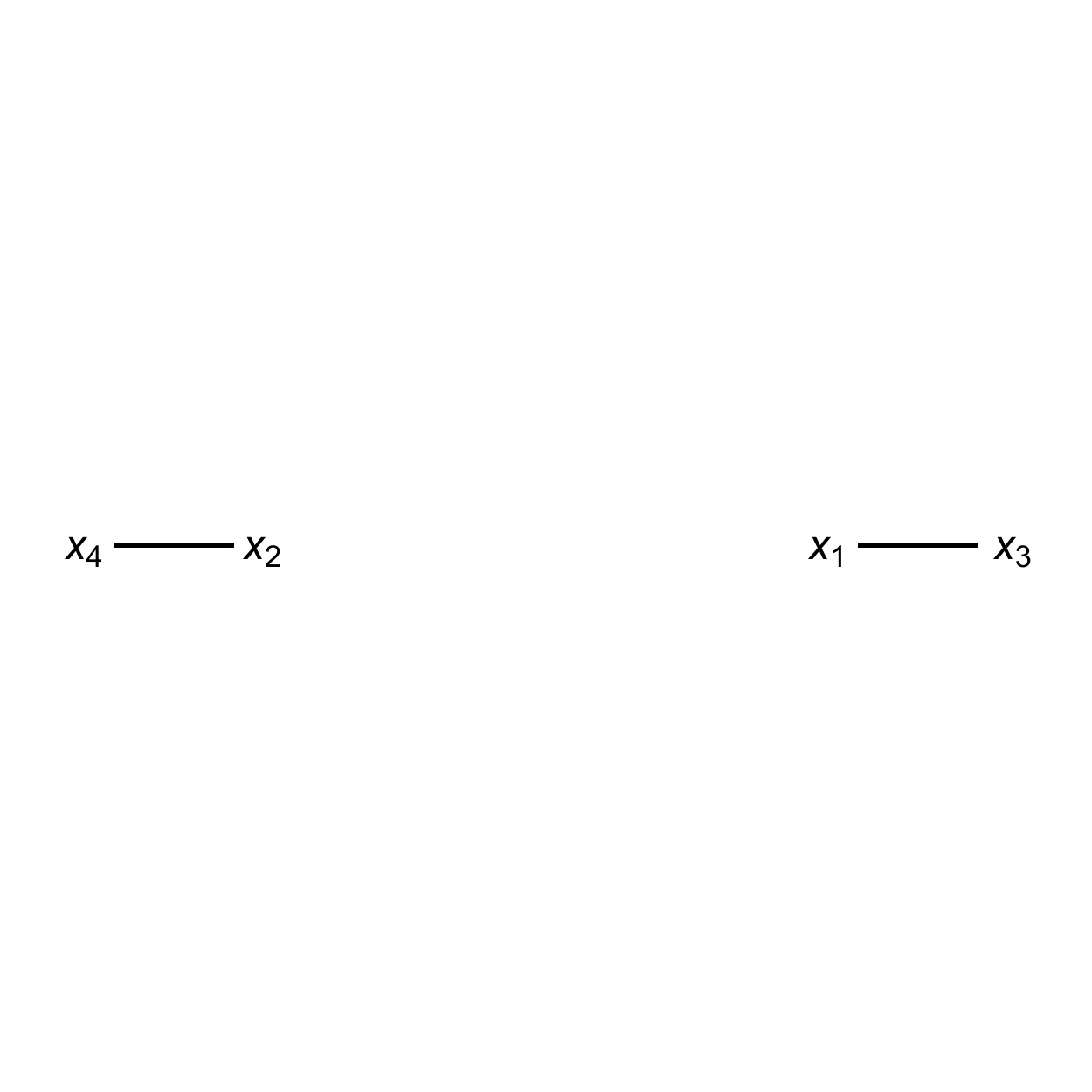}&
	\includegraphics[scale=0.4]{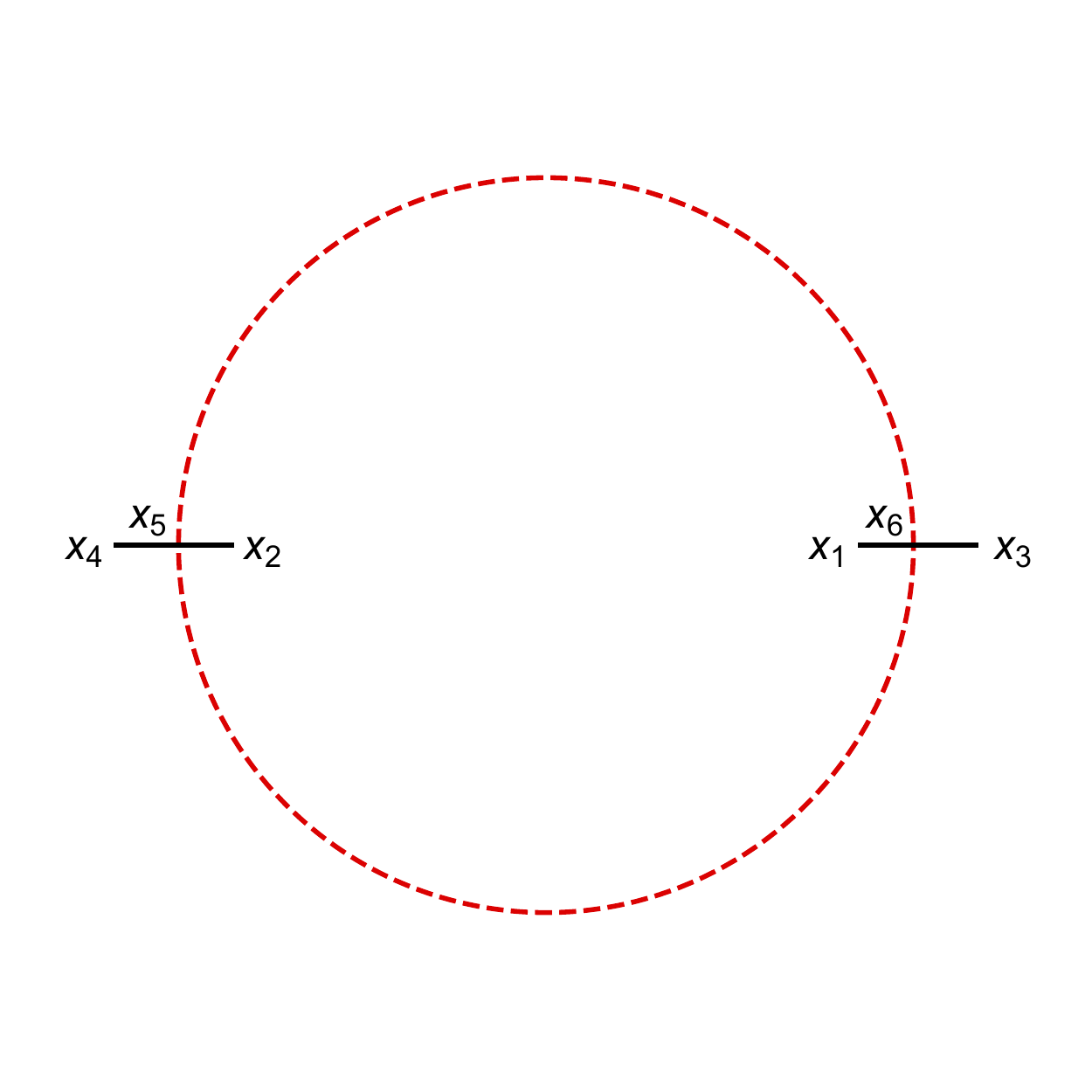}&
	\includegraphics[scale=0.4]{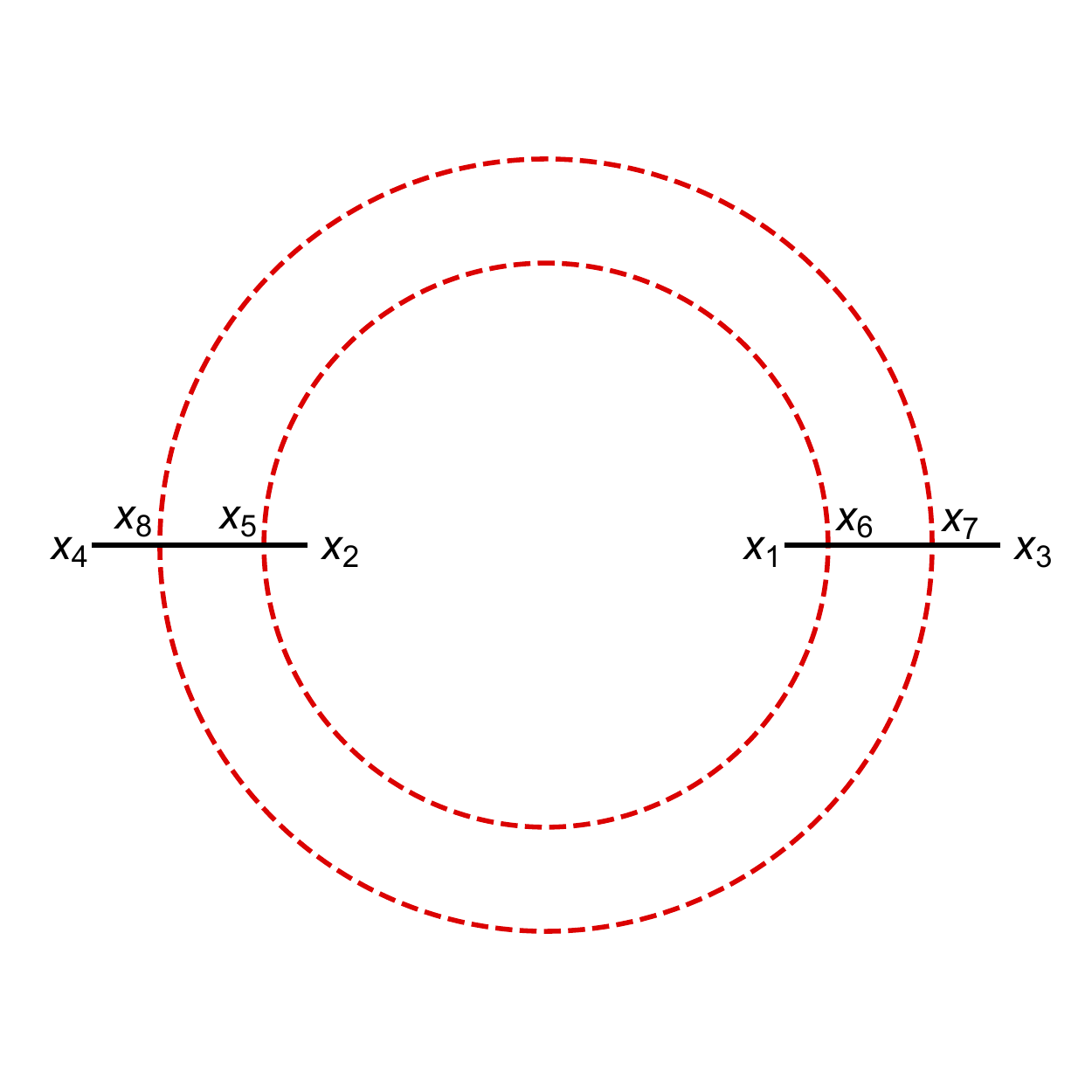}\\
	\xi^0&
	\xi^4&
	\xi^8
	\eea$
	\caption{First $3$ orders contributing to the $\Ga$ correlator. }
	\label{fig:G0loops}
\end{figure}

In this section we proceed without taking the double trace interaction into account and identify the contrubution of wheels graphs to \re{Gres}. We will see that the double trace contributions will be automatically taken into account by correct treatment of the singularity at \(\xi\to 0\) in the forthcoming 
formulas. We start with constructing the graph building operator and identifying the parameters $\Delta_i$, $n$ and $\chi$ introduced in \eqref{Ggeneral}.
We recall that the parameters $\Delta_i$ define the scaling dimension of the external operators. Since the wheel graphs have only one propagator attached to each external leg we have $\Delta_1=\Delta_4=\Delta_2=\Delta_3=1$.

To first two orders of the weak coupling expansion the zero magnon function $\Ga$ is given by the sum of diagrams shown in  Fig.\ref{fig:G0loops}. The expressions corresponding to the first two diagrams are 
\beqa\la{G0firsttwo}
\Ga^{(0)}&=&\frac{c^2}{x_{24}^2 x_{13}^2}\,,\qquad\qquad
\Ga^{(1)}= c^6\int  \frac{d^4 x_5 d^4 x_6}{x_{25}^2 x_{45}^2 x_{16}^2 x_{36}^2 x_{56}^4}
\;,
\eeqa 
where each scalar propagator brings in the factor of $c/x_{ij}^2$.
These expressions can be represented as $1$st and $2$nd powers of the following graph building operator $\hat H_{\ga}$
\begin{align}\label{Hc_kernel0}
{}& H_{\ga}(x_1,x_2|x_3,x_4)=\frac{c^4}{x_{12}^4 x_{13}^2 x_{24}^2}\;\;.
\end{align}
Indeed, we verify that
\begin{align}
\Ga^{(0)} &=\frac{x_{12}^4}{c^2} H_{\ga}
\,,\qqqquad
\Ga^{(1)} =
\frac{x_{12}^4}{c^2}
\int d^4 y_5 d^4 y_6 H_{\ga}(x_1,x_2|y_5,y_6)H_{\ga}(y_5,y_6|x_3,x_4)\,
\end{align}
from where it is clear that for a general graph with $(n+1)$ wheels we get $\hat \Ga^{(n)}=(4\pi^2)^2x_{12}^4 \hat H_{\ga}^{n}$. It is straightforward to verify that the function \re{Hc_kernel0} transforms covariantly under the conformal transformations acting on $x_i$.\footnote{The simplest way to check this is to employ inversions $x_i^\mu \to x_i^\mu/x_i^2$ and take into account that $x_{ij}^2\to x_{ij}^2/(x_i^2x_j^2)$.}
As a consequence, the corresponding integral operator $\hat H_\ga$ commutes with the generators of the conformal group.

Thus the zero-magnon correlation function $\Ga$ can be written as
\beq\label{Ga-div}
\hat \Ga=\sum_{\ell=0}^\infty (16\pi^2\xi^2)^{2\ell}\hat \Ga^{(\ell)}=(4\pi^2)^2x_{12}^4\frac{\hat H_{\ga}}{1-(16\pi^2\xi^2)^2
	\hat H_{\ga}}\,.
\eeq
Comparing with the general expression
\eq{Ggeneral} we deduce that $\chi=(16\pi^2\xi^2)^2$ and $n=1$ in the zero-magnon case.

\subsubsection{Eigenvalue of the zero-magnon graph-building operator}

In order to use the general expression for the correlation function \eqref{Gres}, we have to determine the eigenspectrum of the graph building operator \re{Hc_kernel0}. In virtue of conformal symmetry,
its eigenstates are given in \eq{waveFGeneral} with $\Delta_1=1$ and $\Delta_2=1$. Substitution of \re{Hc_kernel0} into \eq{waveGeneral} leads to an integral,
which can be evaluated using the star-triangle identity as explained in appendix~\ref{app:E0}. Going through the calculation we obtain the following simple result~\cite{Grabner:2017pgm}
\beq\la{E0}
{E}_{\ga}=\frac{16 \pi ^4 c^4}{(-\Delta +S+2) (-\Delta +S+4) (\Delta +S-2) (\Delta
	+S)}\;.
\eeq
It is easy to see that ${E}_{\ga}$ is invariant under $\Delta\to 4-\Delta$, in agreement with \re{Esym}. We can also check that \re{E0} verifies the relation \eq{conditionE0} that ensures the cancellation of spurious poles. In the present case, for $\Delta_1=\Delta_2=1$, it follows from \eq{rn} that  $r_{2n+1}=0$ for integer $n$ and thus \eq{conditionE0} reduces to
\beq
\left.E_{\ga}\right|_{\Delta=3+s+2n,S=s}=\left.E_{\ga}\right|_{\Delta=3+s,S=s+2n}\;\;,\;\;n,s=0,1,2,\dots\,,
\eeq
which is indeed satisfied for \eq{E0}.

\subsubsection{Spectrum  for zero-magnon exchange operators}
\begin{figure}[t!]
	\centering
	\includegraphics[scale=0.61]{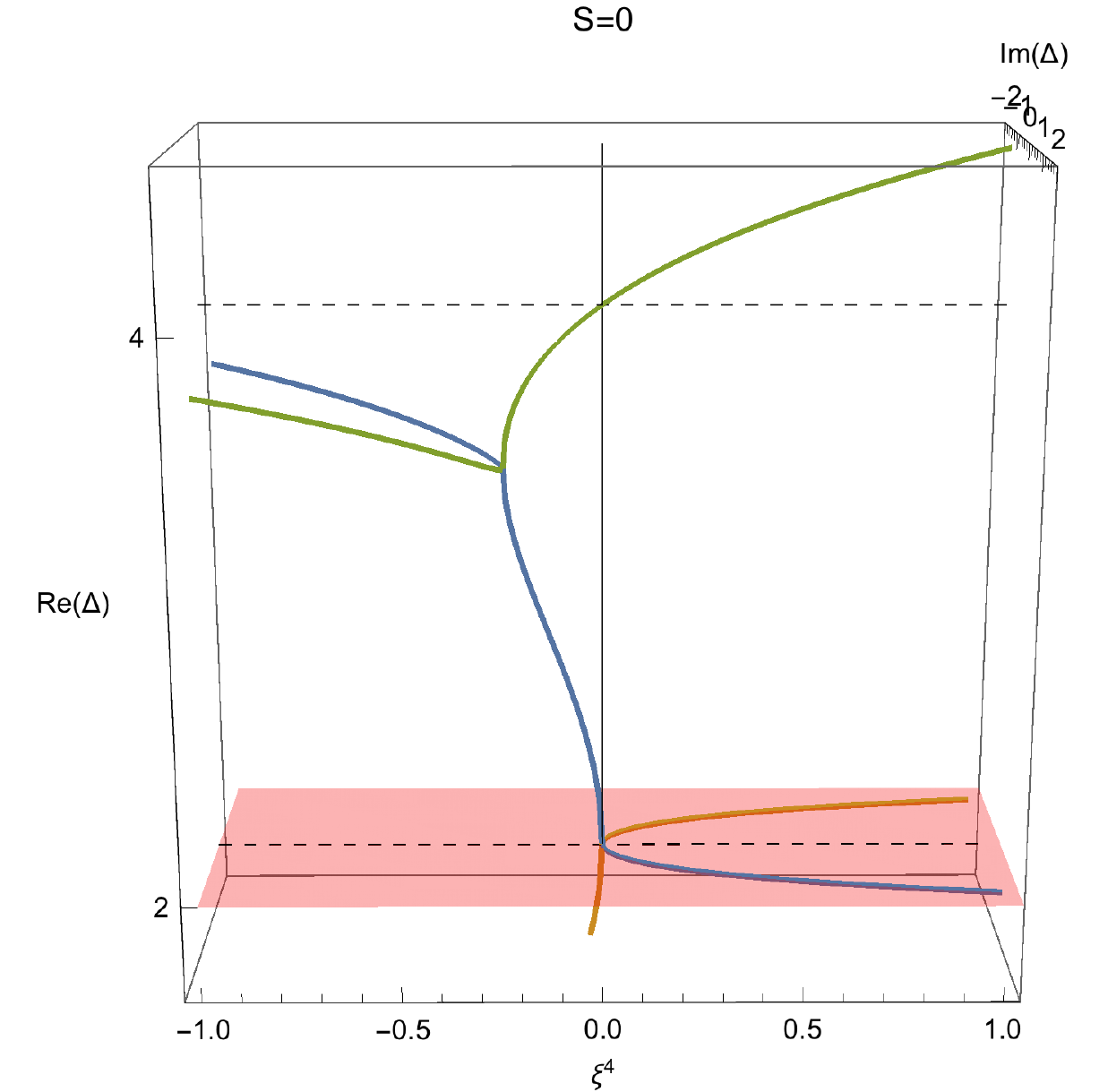}
	\includegraphics[scale=0.61]{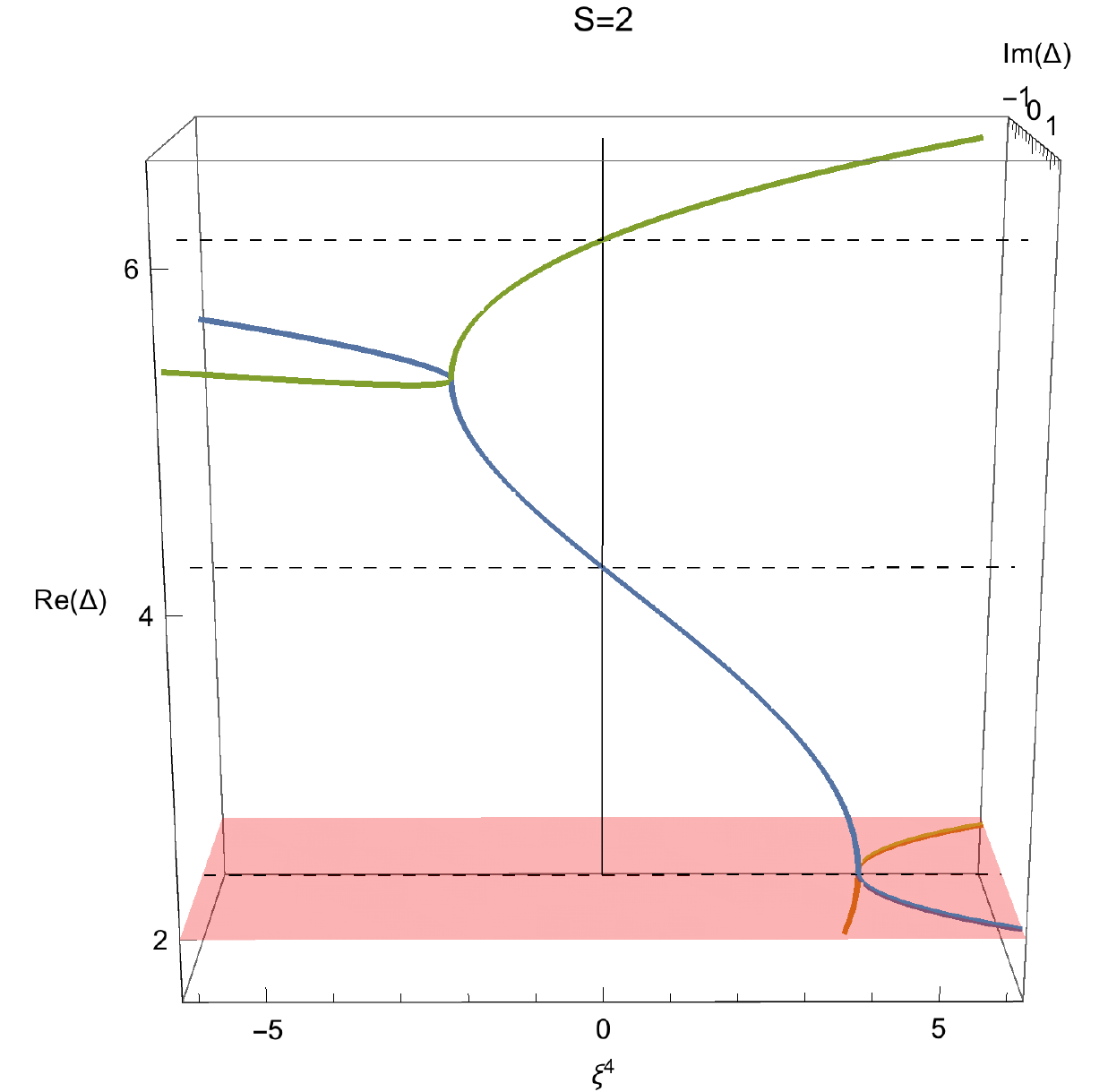}
	\caption{Structure of the spectrum $\Delta_{\ga}(\xi)$ for $S=0$ (left) and $S=2$ (right). At weak coupling (vertical solid line) it consists of two operators with bare dimension $2+S$ and $4+S$. As a function of $\xi^2$ the states can be analytically continued from one to another. }
	\label{fig:E0spectrum}
\end{figure}

We can now use \re{E0} to determine the scaling dimension of  
the zero-magnon  operators contributing to the correlation function \re{Gres}. 
Substituting \re{E0} into the relation \eq{forspec} and 
replacing $\chi=(16\pi^2\xi^2)^2$ we find that the scaling dimensions
$\Delta=2+2i\nu$  
satisfy the following quartic equation
\begin{align}\label{eq-A}
(\nu^2+S^2/4) (\nu^2 +(S+2)^2/4)=\xi^4\,,
\end{align}
subject to the additional  condition $\Im \nu<0$.
At finite coupling, this yields the following expressions for the scaling dimensions
\begin{align}\notag\label{D0}
\Delta_2(S) = 2+\sqrt{(S+1)^2+1 -2 \sqrt{ (S+1)^2+4\xi^4}}\,,
\\[2mm] \Delta_4(S) = 2+\sqrt{(S+1)^2+1 +2 \sqrt{ (S+1)^2+4\xi^4}}\,,
\end{align}
The two remaining solutions to \eq{eq-A} are related to \eq{D0} by $\Delta\to 4-\Delta$ and describe shadow operators with ${\rm Re}\Delta<2$.

At weak coupling, for $\xi^2 < 1$, and nonzero Lorentz spin, $S>0$, 
the scaling dimensions \re{D0} look  as
\begin{align}\notag\la{D0t2}
& \Delta_2(S)=2+S-\frac{2 \xi ^4}{S(S+1)}+\frac{2\xi ^8 ((S-1) S-1)}{S^3 (S+1)^3}+{\cal O}\left(\xi ^{12}\right)\,,
\\
& \Delta_4(S)=4+S+\frac{2 \xi ^4}{(S+1)(S+2)}-\frac{2 \xi ^8 (S (S+5)+5)}{(S+1)^3(S+2)^3}+{\cal O}\left(\xi ^{12}\right)\,,
\end{align}
and the corresponding operators can be identified as
twist-2 and twist-4 operators, respectively. \footnote{This explains the meaning of the subscript of $\Delta_t$.} They have the following form
$O_2=\tr(X (n\partial)^S X+\dots$ and $O_4=\tr(\Box X (n\partial)^S X+\dots$ where dots denote similar terms with light-cone derivatives distributed between the fields. Similarly to \cite{Gromov:2017cja}, the twist $4$ operators can be written, due to the equations of motion, as $O_4=\tr X \partial^S \bar Z X Z+\dots$.

Notice that the weak coupling expansion of \re{D0t2} goes in 
powers of $\xi^4$ which is exactly what one expects since each wheel in the graph shown in 
Fig.~\ref{fig:types}(left) is attached to the rest of the diagram through two single-trace vertices.
Something special happens at $S=0$. In this case, we find from \eq{D0}
\begin{align}\label{xi2exp}
\Delta_2(S=0) = 2+\frac{2 i \sqrt{2} \xi ^2}{\sqrt{1+\sqrt{4 \xi ^4+1}}}\,.
\end{align}
and the weak coupling expansion looks as
\begin{align}\label{tw2-weak}
    \Delta_2(S=0) = 2 i \xi ^2- i \xi ^6+\frac{7 i \xi ^{10}}{4}+O\left(\xi ^{14}\right)\,.
\end{align}
Surprisingly, for $S=0$ the weak coupling expansion of the scaling dimension $\Delta_2(S)$ starts from the power \(\xi^2\), instead of the naively expected \(\xi^4\) (the power corresponding to each insertion of the graph-building operator~\eqref{Hc_kernel0}). 

To understand the reason for this we examine the eigenvalue of the zero-magnon kernel \re{E0} for $S=0$
and $\Delta=2+2i\nu$
\begin{align}
E_\ga\Big|_{S=0} = {\pi^4 c^2\over \nu^2(\nu^2+1)} \,.   
\end{align}
We notice that it goes to infinity at small $\nu$. Then, expanding \re{Ga-div} in powers of $\xi^4$ we find that the contribution of the
states with $S=0$ to the correlation function at order $O(\xi^{4n})$ is proportional to
$\int d\nu (E_\ga\big|_{S=0})^n$ and it diverges for $\nu\to 0$. This is in
agreement with our expectations that the contribution of the wheel graphs
is well defined for all states except those with $S=0$. To remove the divergence we have to include the $O(\xi^{4n})$ contribution of double traces.

We observe that, in the resummed expression for the correlation function \re{Ga-div}, the contribution of the states with $S=0$ to $\Ga$ involves the integral $\int d\nu /(1-(16 \pi^2\xi^2)^2E_\ga\big|_{S=0})$ which is convergent for $\nu\to 0$ at finite $\xi^2$ (for the integral to be well-defined we assume that $\xi^2$ has a small imaginary part). It is easy to see that, at weak coupling, the integration over small $\nu$ produces a square-root singularity at the origin, $\Ga \sim \sqrt{\xi^4}$. This explains why the weak coupling expansion of $\Ga$ in powers of $\xi^4$ is divergent. 
At the same time, this also suggests that the double-trace contribution should be essential only in the weak coupling regime whereas at finite coupling it can be safely ignored. We discuss this issue in more detail in Sect.~\ref{sec:doubletrace}.

Arriving at \re{Gres} we have tacitly assumed that the
physical poles \re{forspec} are located away from real $\nu-$axis in \re{Ggen}. As follows from \re{eq-A}, at weak coupling, the two physical poles located at 
$\nu=\pm iS/2 + O(\xi^4)$ pinch the integration contour at the origin for $S\to 0$ and produce a divergent contribution. The role of the double-trace contribution is to subtract this divergence and, thus, make the weak coupling expansion of $G_\ga$ well defined.
Turning the logic around, we can say that the double traces provide a nonvanishing contribution to the scaling dimensions 
because the eigenvalue \re{E0} diverges as ${E}_{\ga}(\nu,0)\sim 1/\nu^2$ for $\nu\to 0$. 
The relation \re{xi2exp} is in a perfect 
agreement with the result of explicit $7-$loop calculation  \cite{Grabner:2017pgm}.

We recall that the correlation function \re{Gres} is given by the sum over the solutions to     \eq{forspec} with ${\rm Re}\,\Delta>2$. In our present case for $S=0$ the solution \eq{xi2exp} satisfy ${\rm Re}\,\Delta = 2$ for real $\xi^2$. As was mentioned above, for the correlation function \re{Gres} to be well-defined $\xi^2$ should have a nonvanishing imaginary part (see Appendix \ref{analGa} for discussion of analytical properties of \re{Gres}). 
The expression \re{xi2exp} satisfies ${\rm Re}\,\Delta = 2$ for $\Im \xi^2<0$. 
For $\Im \xi^2>0$, the scaling dimension is given by the same expression \re{xi2exp} upon replacing $\xi^2\to -\xi^2$.

Let us examine the properties of the scaling dimensions \re{D0}.  The dependence of $\Delta_2$ and $\Delta_4$ on $\xi^4$  
is shown on Fig.~\ref{fig:E0spectrum} for $S=0$ and $S=2$. We observe that the two functions \re{D0} represent in fact two branches of the same analytic function. It has two branch points located at
 \begin{align}
\xi_-^4=-\frac14(S+1)^2\,,\qqqquad \xi_+^4=\frac1{16}(S(S+2))^2\,.
\end{align}
For $\xi^4=\xi_-^4$ the two operators collide, $\Delta_2(S) = \Delta_4(S)$, whereas for
$\xi^4=\xi_+^4$ one of the operators collides with its shadow, $\Delta_2(S) =2$.~\footnote{If the theory were unitary,  the scaling dimensions $\Delta_2$ and $\Delta_4$ would respect the Neumann-Wigner non-crossing rule and remain to be different from each other for any coupling \cite{Korchemsky:2015cyx}}. 
The collision of operators  at $\xi^4=\xi_+^4$ modifies analytic properties of the correlation function ${\cal G}_{\ga}$. 
According to \re{disc2}, the correlation function has the cut for $\xi^4>0$ that starts at $\xi^4=1/\max_{\nu\in \mathbb R} {E}_{\ga}(\nu,S)$.
It is easy to see from \re{E0} that ${E}_{\ga}(\nu,S)$ is a decreasing positive-definite function of $ \nu^2$. Therefore, the cut starts at $\xi^4=\xi_+^4$ (corresponding to $\nu=0$) and goes to infinity along real $\xi^4-$axis. In the vicinity of the branch point, it follows from \re{eq-A} that $\nu^2\sim \xi_+^4-\xi^4$ for $S\neq 0$ leading
to $\Delta_2(S) -2 \sim \sqrt{\xi_+^4-\xi^4}$ \cite{Korchemsky:2015cyx}.

\paragraph{Strong coupling.} At strong coupling, for $\xi^2\to\infty$, the relation \re{eq-A} has four solutions $i\nu = \xi \e^{i\pi k/2} + O(1/\xi)$ (with $k=0,\dots,3$).
 Among them only two satisfy the additional condition $\Im\nu<0$. The corresponding expressions for the scaling dimensions are
 \begin{align}\la{Delt}
\Delta = 2\xi \e^{i\pi k/2}+2 -{S^2+(S+2)^2\over 8 \,\xi \e^{i\pi k/2}} + O(1/\xi^3)\,,
\end{align}
where integer $0\le k\le 3$ satisfies the condition $\Re(\xi \e^{i\pi k/2})>0$ and depends on $\xi$.

\subsubsection{Structure constants for zero-magnon exchange operators}

We apply the general relation \eq{Cstruct}
to find the OPE coefficient for zero-magnon operators~\cite{Grabner:2017pgm} 
\beq\la{C0}
C_{\Delta,S}=\frac{ (S+1) \Gamma (S-\Delta +4) \Gamma
	\left(\frac{1}{2} (S+\Delta -2)\right) \Gamma \left(\frac12(S+\Delta)\right)}{\left((4-\Delta) \Delta +S(S+2)-2\right) 
	\Gamma^2 \left(\frac{1}{2} (S-\Delta +4)\right) \Gamma (S+\Delta -2)}\;.
\eeq
Stricktly speaking, $C_{\Delta,S}$ is given by the product of (properly normalized) $3-$point correlation functions $\vev{O_{\bar X Z} O_{\bar X\bar Z} O_{\Delta,S}}$ and $\vev{O_{XZ} O_{X\bar Z} \bar O_{\Delta,S}}$. In unitary CFT they are complex conjugated to each other and, as a consequence, $C_{\Delta,S}$ is positive definite. This is not the case for the conformal fishnet theory \re{bi-scalarL} and \re{L-dt}. In virtue of the symmetry
\re{2-sym1} the above mentioned 3-point functions coincide and, therefore, $C_{\Delta,S}$ is given by the square of
the 3-point correlation function of two protected  and one unprotected operators
\beq\la{CG0}
C_{\Delta,S}=(C_\ga^{\bullet\circ\circ})^2\;.
\eeq
We will generalize this result to a more complicated structure constant involving two non-protected operators in section~\ref{sec:gt}.
\paragraph{Weak coupling limit.}
Replacing the scaling dimensions in \re{C0} by their expressions \eq{D0}
we can obtain the OPE coefficients at weak coupling. 

First, consider twist-2 operators with the scaling dimension $\Delta_2$ and  non-zero spin $S>0$.
In this case we get
\beq
C_{\Delta_{t=2,S}}=
\frac{\Gamma^2 (S+1)}{\Gamma (2 S+1)}\left[1
+2\xi ^4\frac{ (S+1)  (\psi(2S+1)-\psi(S+1))+1}{S (S+1)^2}
\right]
+{\cal O}\left(\xi ^8\right)\;,
\eeq
where $\psi(x)$ is the Euler polygamma-function. Notice that $C_{\Delta_{t=2,S}}$ becomes singular for $S\to 0$.
Similar to the situation with the scaling dimension $\Delta_2(S)$, this happens because the two limits $S\to 0$
and $\xi^2\to 0$ do not commute.
To get the correct result for $C_{\Delta_{t=2,S=0}}$ at weak coupling, we should first put $S=0$ in \re{C0} and, then, expand it in powers of $\xi^2$. This gives
\beq\label{C-tw2}
C_{\Delta_{t=2,S=0}}=
1-2 i \xi ^2-2 \xi ^4+i\xi ^6 (5 -4  \zeta_3)+\xi ^8 (6-8 \zeta_3)+{\cal O}\left(\xi ^9\right)\;.
\eeq
In analogy with \re{tw2-weak}, the weak coupling expansion starts at order $O(\xi^2)$ indicating that $C_{\Delta_{t=2,S=0}}$ is sensitive to the contribution of the double traces.

For twist$-4$ operators we find from \re{C0} and \re{D0t2}
\beq\label{C-tw4}
C_{\Delta_{t=4},S}=\xi ^4 
\frac{\Gamma^2 (S+2)}{2 (S+1)(S+2)\Gamma (2 S+3)}+{\cal O}\left(\xi ^8\right)\;.
\eeq
In distinction with \re{C-tw2}, the weak coupling expansion of $C_{\Delta_{t=4},S}$ starts at order $O(\xi^4)$ and runs in powers of $\xi^4$. The latter property is in agreement with our expectations that twist$-4$ operators are not affected by double-trace interaction. The twist-4 OPE coefficient \re{C-tw4} is suppressed by the factor of $\xi^4$ as compared with \re{C-tw2}. Due to the equation of
motion, $\Box X=16\pi^2 \xi^2\bar Z X Z $, the corresponding operator takes the form
$O_4=\tr(\Box X (n\partial)^S X)+\dots = 16\pi^2 \xi^2  \tr(\bar Z X Z (n\partial)^S X))+\dots$. The reason for the above-mentioned suppression is that $\vev{O_4 \bar O_4}=O(\xi^4)$
and $\vev{O_4 O_{\bar XZ}O_{\bar X\bar Z}}=O(\xi^4)$ leading to $C_{\Delta_{t=4},S}\sim \vev{O_4 O_{\bar X Z}O_{\bar X\bar Z}}^2/\vev{O_4 \bar O_4} =O(\xi^4)$.
\paragraph{Strong coupling limit.}
At strong coupling the dimension $\Delta$ become large according to \eq{Delt} and we get
\beq\la{CC0strong}
C_{\Delta,S}=2^{5-2 \Delta }\frac{S+1}{\Delta}\tan \left(\pi\frac{\Delta +S}{2}  \right)  \left(1+\frac{3}{2 \Delta }+\frac{4 (S+1)^2+25}{8
   \Delta ^2}+{\cal O}\left(\frac{1}{\Delta^3}\right)\right)\,. 
\eeq
Thus we see that the structure constant at strong coupling is exponentially small since
$\Delta\simeq 2\xi \e^{i\pi k/2} $.

\subsubsection{Zero-magnon 4-point correlation function}

Having determined the conformal data of the zero-magnon operators, we can apply \eq{GGgen} and \eq{Gres} to compute
the four-point correlation function \eq{GGgen} and \eq{Gres}
\beqa\la{G4cg0}
\Ga(x_1,x_2|x_3,x_4)=\frac
{c^2}
{x^2_{12} x^2_{34}}
{\cal G}_{\ga}(u,v)
\eeqa
where we replace the scaling dimensions of the external protected operators by their values, $\Delta_1=\Delta_2=\Delta_3=\Delta_4=1$, and the function ${\cal G}_{\ga}(u,v)$ is given by 
\beqa\label{G0sum}
{\cal G}_{\ga}(u,v)=\sum_{S=0}^\infty\sum_{\Delta=\Delta_{t=2},\Delta_{t=4}}
C_{\Delta,S}\;g_{\Delta,S}(u,v)\,,
\eeqa
where the sum runs over the states with the scaling dimension \eq{D0} and the OPE coefficients \eq{C0}.
Here $g_{\Delta,S}$ is the four-dimensional conformal block defined in \eq{defg} (with $\Delta_i=1$).

The relation \re{G0sum} involves an infinite sum over the conformal blocks and it is not obvious a priori that 
one can find a closed expression for ${\cal G}_{\ga}(u,v)$
even at weak coupling. We show in section~\ref{sect:weak} by
explicit two-loop calculation that ${\cal G}_{\ga}(u,v)$ can be expressed in terms of special functions, the so-called harmonic polylogarithms (HPL). In section~\ref{sec:PT} we extend this result to any 
order of the weak coupling expansion. 
Also in section~\ref{sec:strong} we analyse the strong coupling limit of the expression \re{G0sum}. The analytic properties of ${\cal G}_{\ga}(u,v)$ with respect to the  coupling \(\xi\) are discussed in Appendix~\ref{analGa}.

According to \eqref{GD-def} and \eqref{GA-G},
the correlation functions \(G_{A}\) and \(G_{D}\) 
are given by a linear combination of the zero-magnon functions \(\Ga\)  symmetrized in $x_3\leftrightarrow x_4$. Let us see what effect the exchange of $x_3$ and $x_4$ has on the function $\Ga$. As follows from \re{zbarz}, the cross-ratios transform under the exchange of $x_3$ with $x_4$ as $u \to u/v$ and $v\to 1/v$. The corresponding transformation of the conformal blocks looks as
\beq\la{gsym}
 g_{\Delta,S}(u/v,1/v)=(-1)^Sg_{\Delta,S}(u,v)\,.
\eeq
This relation follows from \eq{Phi-blocks}, it can also be verified directly from the definition \eq{defg}.
Combining together \re{G0sum} and \re{gsym} we conclude that, in the expressions for \(G_{A}\) and \(G_{D}\) the terms in \eq{G0sum} with odd $S$ cancel out whereas those with even $S$ get doubled.

\subsection{One-magnon case and single spiral graphs  ($\Gb$)}\la{sec:G[1]}

\begin{figure}[t!]
        \centering
        $\bea{ccc}
        \includegraphics[scale=0.4]{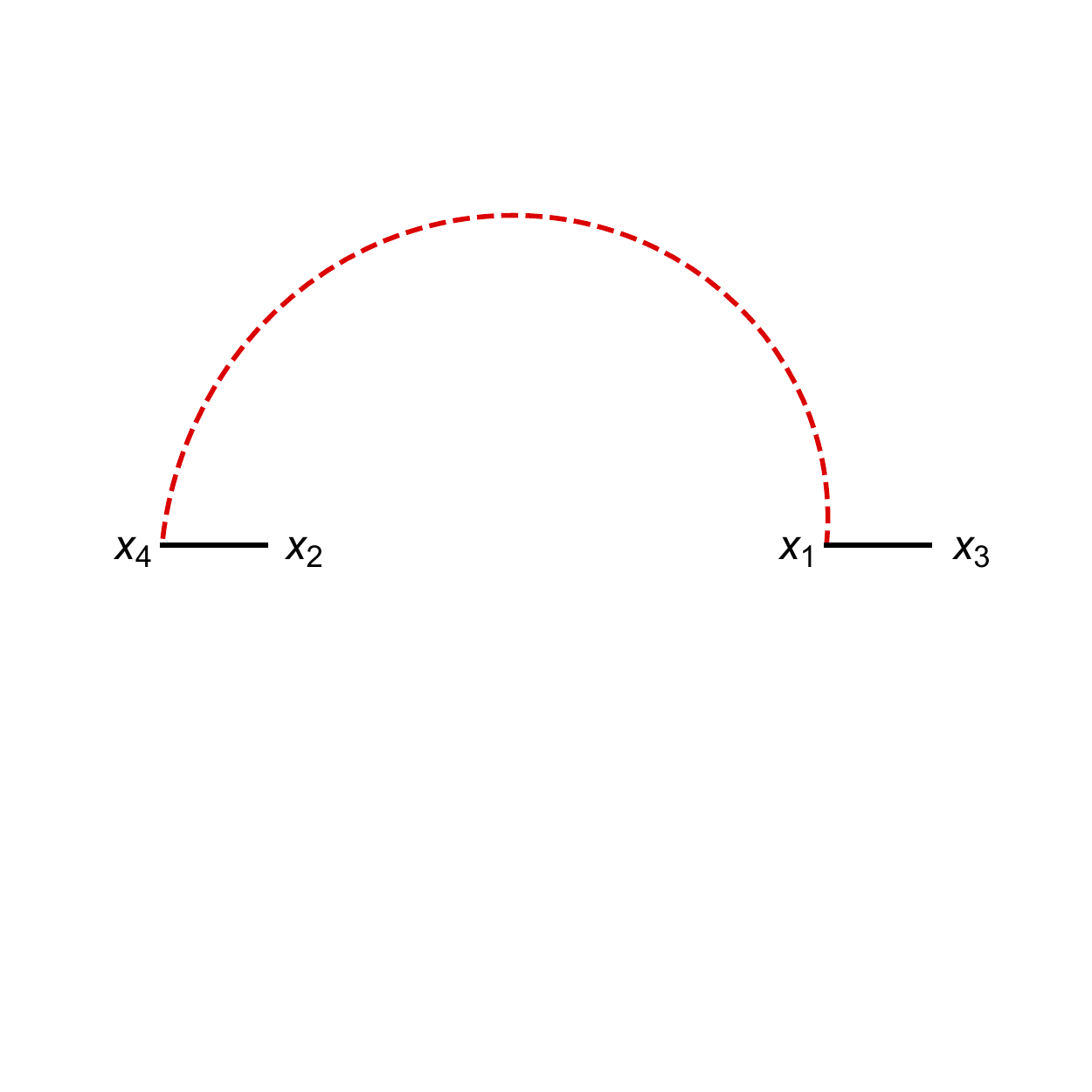}&
        \includegraphics[scale=0.4]{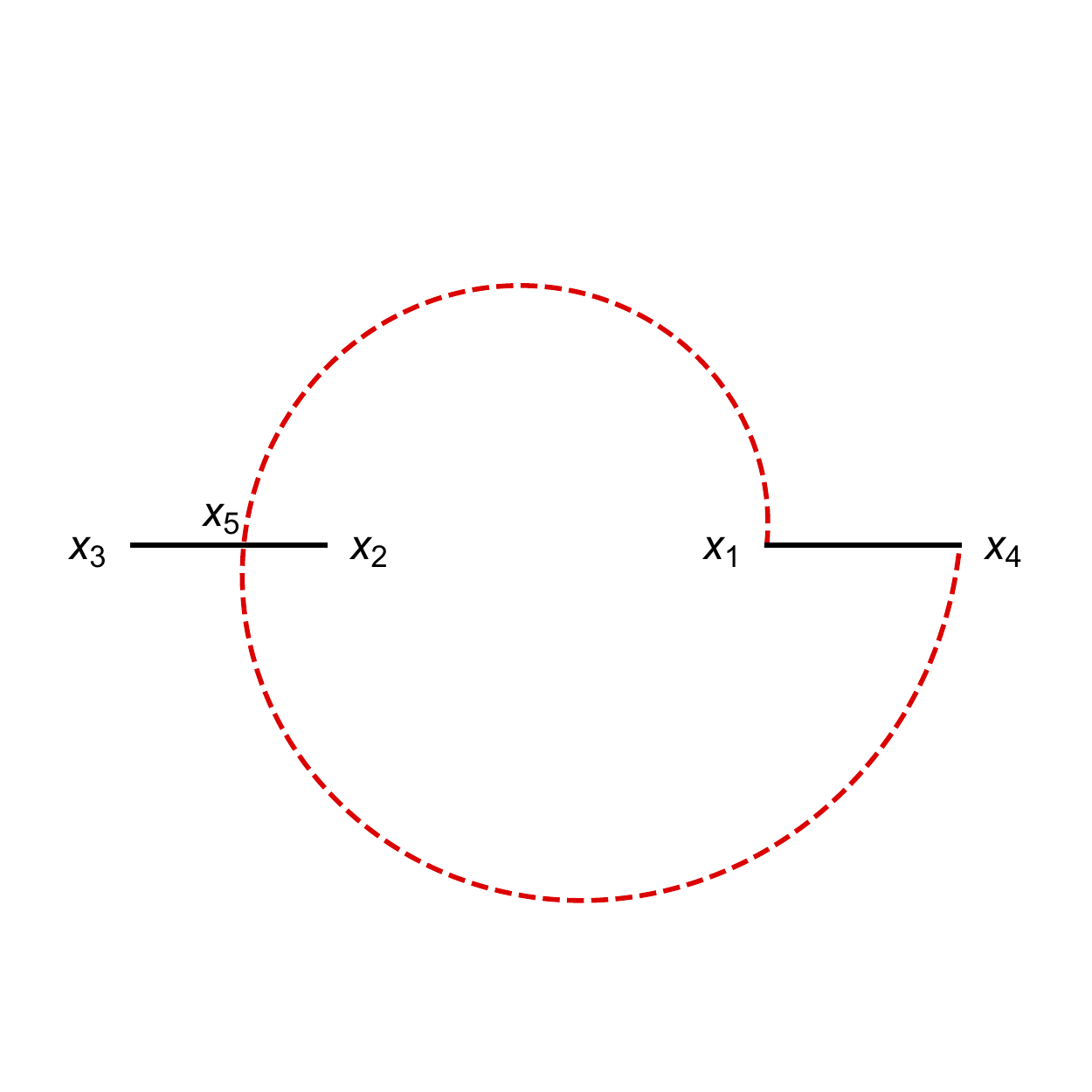}&
        \includegraphics[scale=0.4]{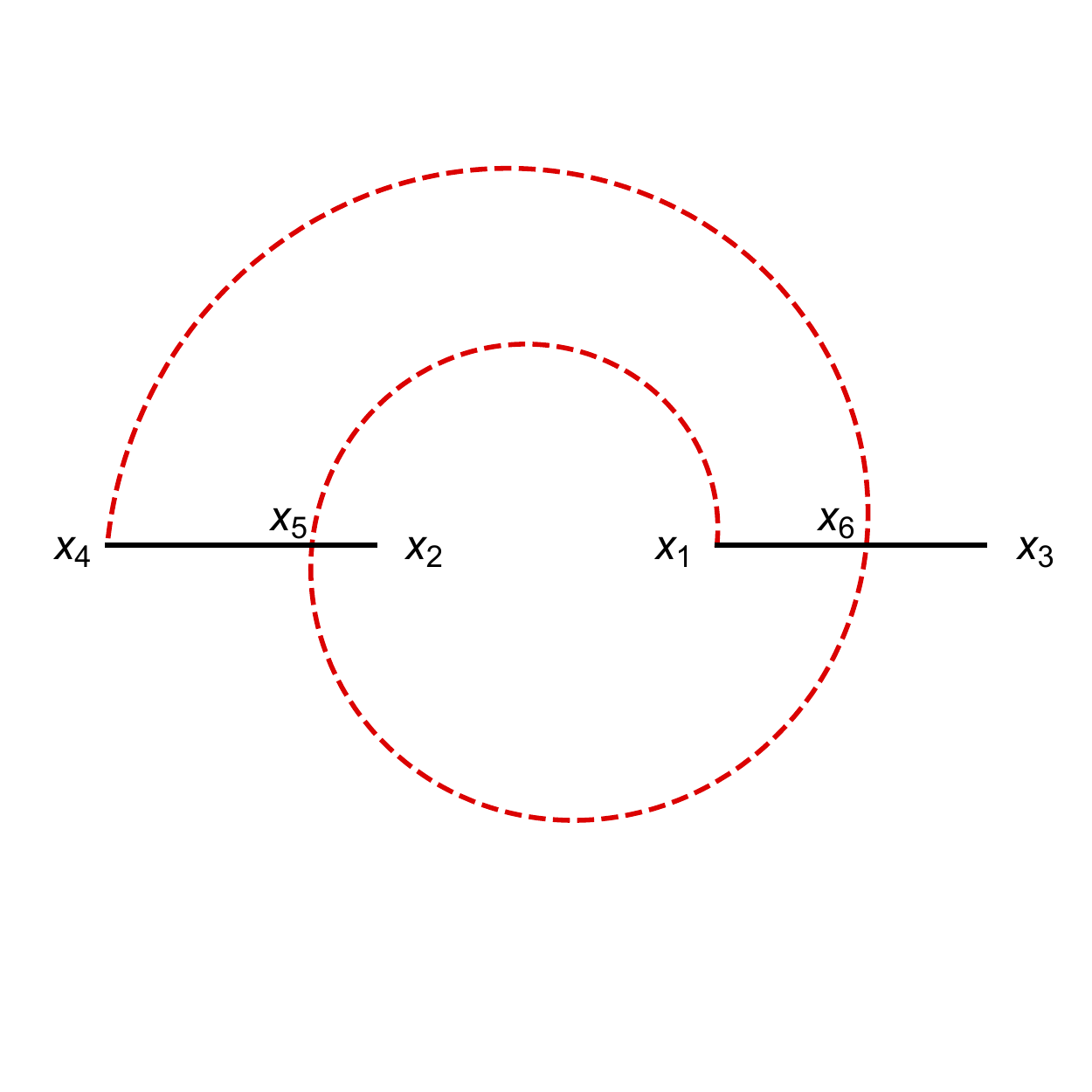}\\
        \xi^0&
        \xi^2&
        \xi^4
        \eea$
        \caption{First $3$ orders contributing to the $\Gb$ correlator. We assume that each of $x_1$ and $x_4$ points are the sources of two propagators, by that reason $x_4$ and $x_3$ are interchanged in the middle picture in comparison to the right and left.}
        \label{fig:G1loops}
\end{figure}

In this subsection, we consider the one-magnon correlation function $\Gb$ described by graphs shown in Fig.~\ref{fig:types}(middle). As we will see shortly, its calculation is simpler than that of $\Ga$ and $\Gc$. 
Since the graph contributing to $\Gb$ have two propagators attached to $x_1$ and $x_4$ and only one to $x_2$ and $x_3$, we identify the scaling dimensions at the external points as
\begin{align}\label{weights-1}
    \Delta_1=\Delta_4=2\,,\qquad\qquad \Delta_2=\Delta_3=1
\end{align}
To identify the graph-building operator $\hat H_\ga$, we
consider the first few terms in the weak coupling expansion of $\Gb$ (see Fig.\ref{fig:G1loops}). The expressions corresponding to the first two diagrams are
\beqa\notag
\la{G10loop}\Gb^{(0)}&=&\frac{c^3}{x_{24}^2 x_{13}^2 x_{14}^2}\;,\\
\la{G11loop}\Gb^{(1)}&=&\int d^4 x_5 \frac{c^5}{x_{25}^2 x_{53}^2 x_{14}^2 x_{15}^2 x_{45}^2}
\;.
\eeqa 
Let us show that these expressions can be represented as $2^{\rm nd}$ and $3^{\rm rd}$ power of the graph building operator with the integral kernel
\begin{align}\label{Hc_kernel1}
{}& H_{\gb}(x_1,x_2|x_3,x_4)=c^2{\delta^{(4)}(x_4-x_1)\over x^{2}_{12}x^{2}_{23}}\;\;.
\end{align}
Indeed, we apply \re{H-conv} to get 
\begin{align}\notag
{}& \langle x_1,x_2|\hat H^2_{\gb}|x_3,x_4\rangle=
    \int d^4 y_5 d^4 y_6 H_{\gb}(x_1,x_2|y_5,y_6)H_{\gb}(y_5,y_6|x_3,x_4)=
\frac{c}{x_{12}^2}\Gb^{(0)}\,,
\\
{}& \langle x_1,x_2|\hat H^3_{\gb}|x_3,x_4\rangle=
\frac{c}{x_{12}^2}\int d^4 y_5 d^4 y_6
G^{(0)}_{\gb}(x_1,x_2|y_5,y_6)H_{\gb}(y_5,y_6|x_3,x_4)
=
\frac{c}{x_{12}^2}G^{(1)}_{\gb}\;.
\end{align}
It is clear from these examples that, in general, $\hat \Gb^{(n)}=4\pi^2x_{12}^2 \hat H_{\gb}^{n+1}$. Thus the correlation function $\Gb$ can be written as
\beq
\hat \Gb=\sum_{\ell=0}^\infty (16\pi^2\xi^2)^\ell\hat \Gb^{(\ell)}=\frac{x_{12}^2}{c}\frac{\hat H_{\gb}^2}{1-(16\pi^2\xi^2)
\hat H_{\gb}}\;.
\eeq
Comparing with the general form
\eq{Ggeneral} we see that $\chi=16\pi^2\xi^2$ and $n=2$.

\subsubsection{Eigenvalue of the one-magnon graph-building operator}

As in the previous case, we can verify that the integral operator $\hat H_\gb$ with kernel given by \re{Hc_kernel1} commutes with the generators of the conformal group acting on the external points $x_i$ and the corresponding conformal weights given by \re{weights-1}. 
As a consequence, its eigenstates are given by \eq{waveFGeneral} with $\Delta_1=2$ and $\Delta_2=1$. The eigenvalue equation \eq{waveGeneral} reduces to an integral
which can be evaluated using the star-triangle identity, as explained in Appendix \ref{app:derE1}. The result turns out to be quite simple
\beq\la{E1}
{E}_{\gb}=(-1)^S\frac{4 \pi ^2 c^2}{(-\Delta +S+3)
        (\Delta +S-1)}\;.
\eeq
It is obviously invariant under $\Delta\to 4-\Delta$ and satisfies \eq{Esym}.
We can use \re{E1} to verify the condition \eq{conditionE0}. 
In the present case, for $\Delta_1=2$ and $\Delta_2=1$, we find from \eq{rn} that
$r_{2n}=0$ for integer $n$ so that the relation \eq{conditionE0} takes the form
\beq
\left.E_{\gb}\right|_{\Delta=3+s+2n+1,S=s}=\left.E_{\gb}\right|_{\Delta=3+s,S=s+2n+1}\;\;, 
\eeq
where $n,s=0,1,2,\dots$. It is easy to see that it holds indeed.
\footnote{Note that the factor \((-1)^S\) in the expression for \(E_{\gb}\) plays an important role for this equation to be satisfied.}

\subsubsection{Spectrum of one-magnon exchange operators}
\begin{figure}[t!]
        \centering
        \includegraphics[scale=0.6]{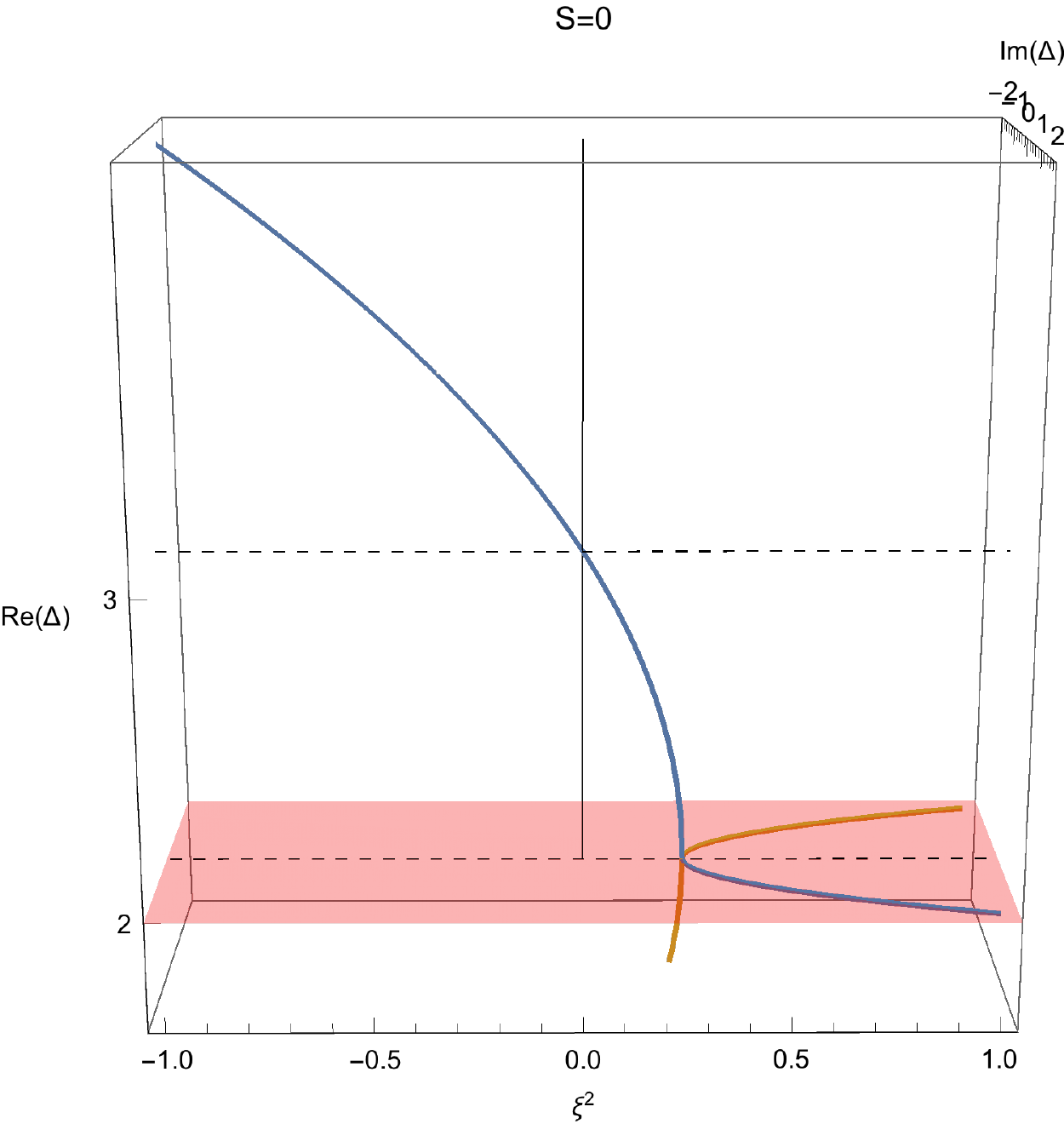}
        \includegraphics[scale=0.6]{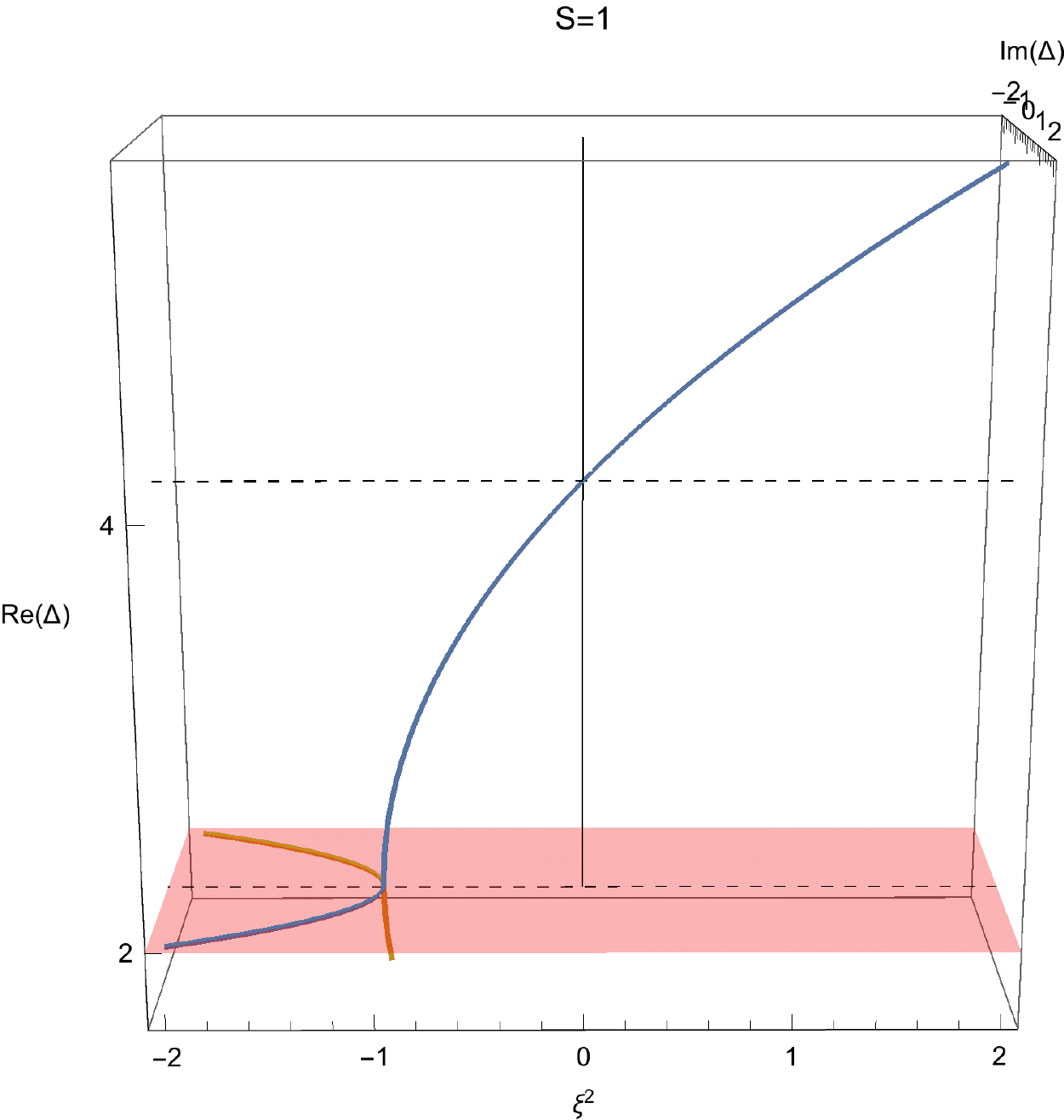}
        \caption{Dependence of $\Delta_{\gb}$ on the coupling constant $\xi^2$ for $S=0$ (left) and $S=1$ (right). At weak coupling $\Delta_{\gb}=3+S+O(\xi^2)$. For $\xi^2=(-1)^S (S+1)^2/4$ the  function $\Delta_{\gb}$ approaches the value $\Delta=2$ and, then,  continues into the complex plane $\Re \Delta=2$ after the collision with the shadow level $4-\Delta_\gb$.}
        \label{fig:E1spectrum}
\end{figure}

Substituting \re{E1} into \eq{forspec} and taking into account that
$\chi=16\pi^2\xi^2$, we determine the scaling dimension of the one-magnon operators $O_\gb=\tr(X Z (n\partial)^S X)+\dots$ contributing to the correlation function \re{Gres} 
\beq\la{D1}
\Delta_\gb =2+\sqrt{(S+1)^2-4(-1)^S \xi ^2}\;.
\eeq
For $S=0$ we have 
\beq\la{Dexact}
\Delta_\gb\big|_{S=0}=2+\sqrt{1-4 \xi ^2}\,.
\eeq
Interestingly, this expression coincides with the asymptotic dispersion relation for the one-magnon state previously found from the double-scaling limit of asymptotic Bethe ansatz (ABA) in Ref.~\cite{Caetano:2016ydc}.
Following the ABA approach, one could have expected that the asymptotic dispersion relation \re{Dexact} should be corrected already at order $O(\xi^4)$ by the wrapping corrections. 
The relation \re{Dexact} 
implies that the wrapping corrections to the one-magnon operator \(\tr(X^2Z)\)  vanish. This is indeed the case for the single wrapping contribution found in \cite{Caetano:2016ydc}. The relation \re{Dexact} is indeed consistent with the ABA, including the known single wrapping correction!

\paragraph{Weak coupling.}
At weak coupling, the scaling dimension \re{D1} of the one-magnon states reads
\beq\label{D1-weak}
\Delta_\gb=3+S-\xi ^2 \frac{2(-1)^S }{S+1}-\xi ^4\frac{2 }{(S+1)^3}-\xi ^6 \frac{4 (-1)^S}{(S+1)^5}+O\left(\xi ^7\right)\;.
\eeq

\paragraph{Strong coupling.}
At strong coupling, we find from \re{D1}
\beq\la{D1str}
\Delta_\gb=2+(-1)^{\frac{S+1}{2}}\left(
2 \xi -\frac{(-1)^S (S+1)^2}{4 \xi}-\frac{  (S+1)^4}{64 \xi ^3}
\right)+{\cal O}\left(\frac{1}{\xi^5 }\right)\;,
\eeq
where 
the branch of $(-1)^{\frac{S+1}{2}}$ is such that ${\rm Re}\,\Delta>2$. Interestingly, at the leading order we get the same coefficient as for the zero-magnon case \eq{Delt}.

\subsubsection{Structure constants with one-magnon exchange operators}
We can also employ the general equation \eq{Cstruct}
to find the OPE coefficient 
\beq\la{C1}
C_{\Delta,S}= (S+1)\frac{ \Gamma^2 \left(\frac{1}{2}
        (S+\Delta -1)\right)\Gamma (S-\Delta +4)}{ \Gamma^2 \left(\frac{1}{2} (S-\Delta
        +5)\right) \Gamma (S+\Delta -1)}\;.
\eeq
As in the previous case, it is given by the square of the $3-$point function of two protected and one unprotected operator, $
C_{\Delta,S}= (C_\gb^{\bullet\circ\circ})^2$.

\paragraph{Weak coupling.} At weak coupling, we obtain from \re{C1} and \re{D1-weak}
\beq
C_{\Delta,S}=
\frac{\Gamma (S+1) \Gamma (S+2)}{\Gamma (2 S+2)}\left[1-2 \xi ^2(-1)^S\frac{  \psi(S+1)-\psi({2 S+2})}{S+1}+{\cal O}\left(\xi ^4\right)\right]\;.
\eeq
In particular for $S=0$ we find
\beq
C_{\Delta,0}=1+2 \xi ^2+6 \xi ^4-4 \xi ^6 (\zeta _3-5)+\xi ^8 (70-20 \zeta_3)+{\cal O}\left(\xi ^{10}\right)\;.
\eeq
Since the first few coefficients do not involve wrapping it should be possible to compare with the perturbative ABA based general expressions from \cite{Escobedo2010,Gromov2012}.

\paragraph{Strong coupling.} At strong coupling  $\Delta$ becomes large \eq{D1str} and we can use it as a large expansion parameter since the expressions are more compact
\beq
C_{\Delta,S}=
 2^{6-2\Delta }\frac{S+1}{\Delta}\cot
   \left(\pi\frac{S-\Delta
   }{2}   \right) \left(1+\frac{5}{2 \Delta }+\frac{4 S^2+8
   S+53}{8 \Delta ^2}+{\cal O}\left(\frac{1}{\Delta^3}\right)\right)\;.
\eeq
The structure constant is again exponentially decaying as it was in the zero-magnon case \eq{CC0strong}.

\subsubsection{One-magnon four-point correlation function}

Substituting \re{weights-1} into \re{GGgen} and \re{Gres}, we obtain the one-magnon 4-point correlation function 
\beqa\la{G4cg}
\Gb(x_1,x_2|x_3,x_4)=c^3\frac
{(x_{13}^2 x_{24}^2)^{1/2}}
{(x^2_{12}  x^2_{34})^{3/2}x^2_{14} }
{\cal G}_{\gb}(u,v)\,,
\eeqa
where
\beqa\la{calG1}
{\cal G}_{\gb}(u,v)=\sum_{S=0,1,2,\dots}C_{\Delta,S}\;g_{\Delta,S}(u,v)
\eeqa
and the sum runs over the states with the scaling dimensions $\Delta$ and the OPE coefficients $C_{\Delta,S}$ given by \eq{D1} and \eq{C1}, respectively. Notice that in the above sum there is only one state for each value of the spin $S$ and spin takes all non-negative integer values. 
In section~\ref{sec:PT} we study \eq{calG1} at weak coupling and compare it with the result of perturbative calculation performed in section~\ref{sect:weak}. 
Also in section~\ref{sec:strong} we analyse the strong coupling limit of \re{calG1}.

Finally, we can use \eqref{G1p-def} and \re{G4cg} to calculate the 4-point correlation functions \(G_{C'}\) and \(G_{C''}\).

\subsection{Two-magnon case and double-spiral graphs  ($\Gc$)}

\begin{figure}[t!]
        \centering
        $\bea{ccc}
        \includegraphics[scale=0.4]{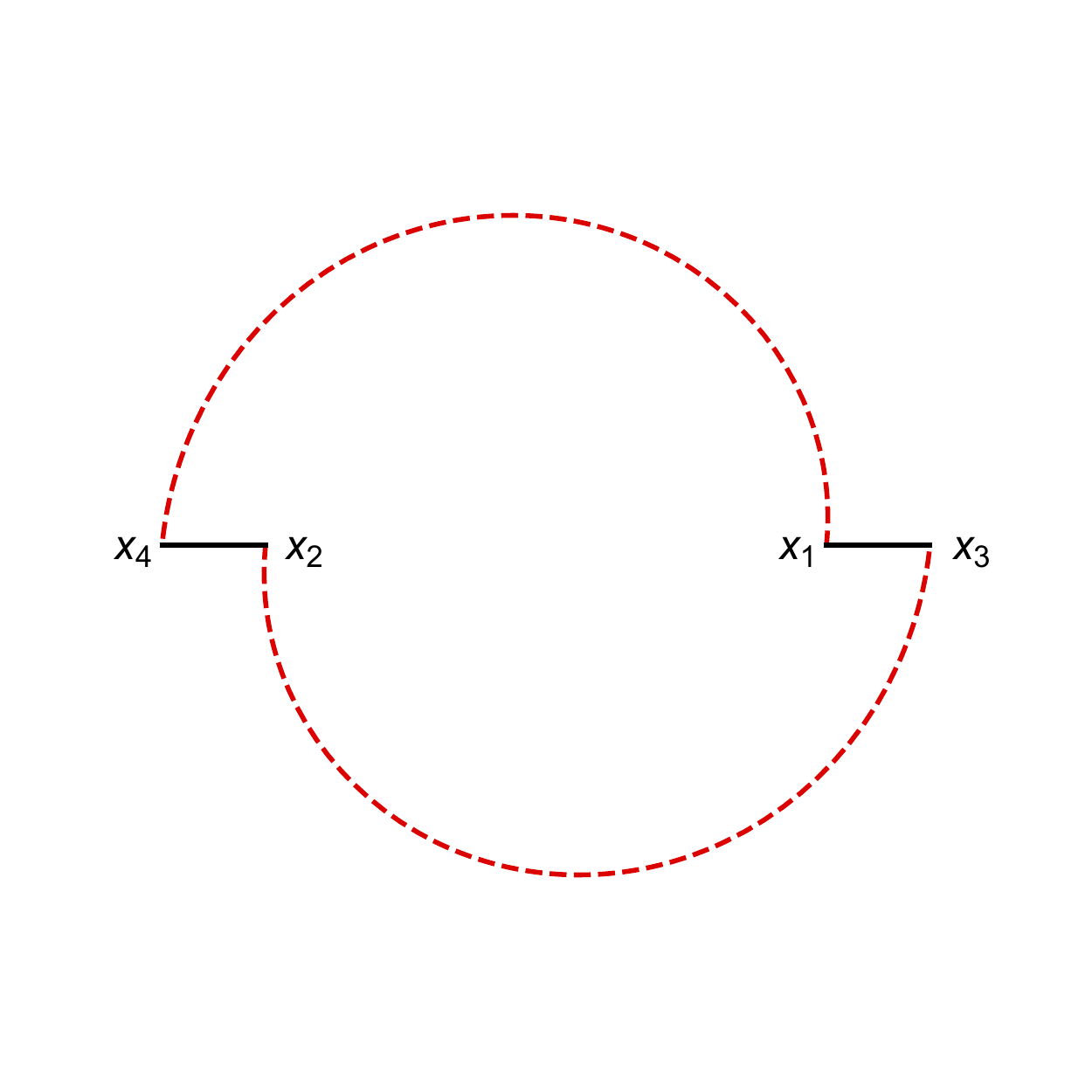}&
        \includegraphics[scale=0.4]{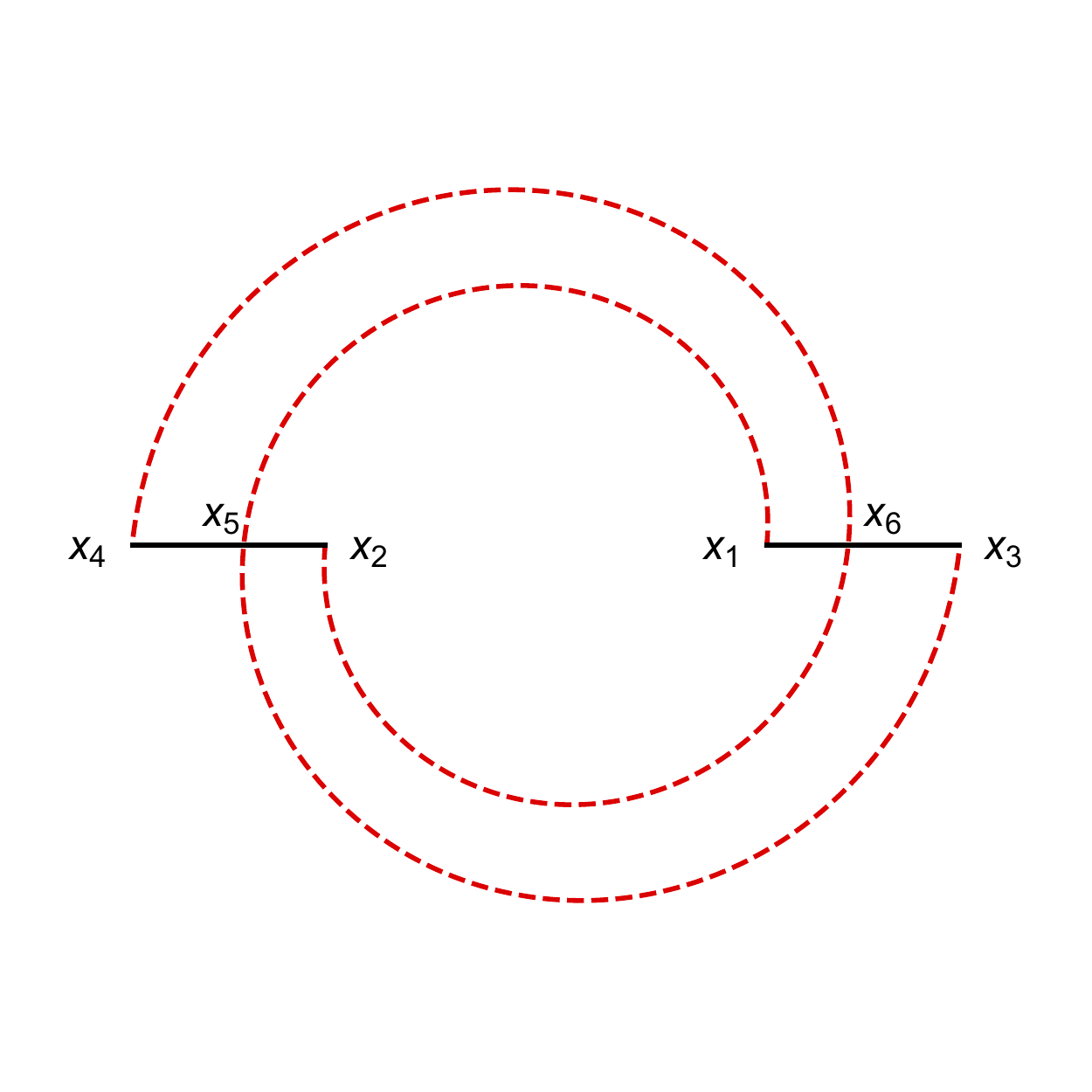}&
        \includegraphics[scale=0.4]{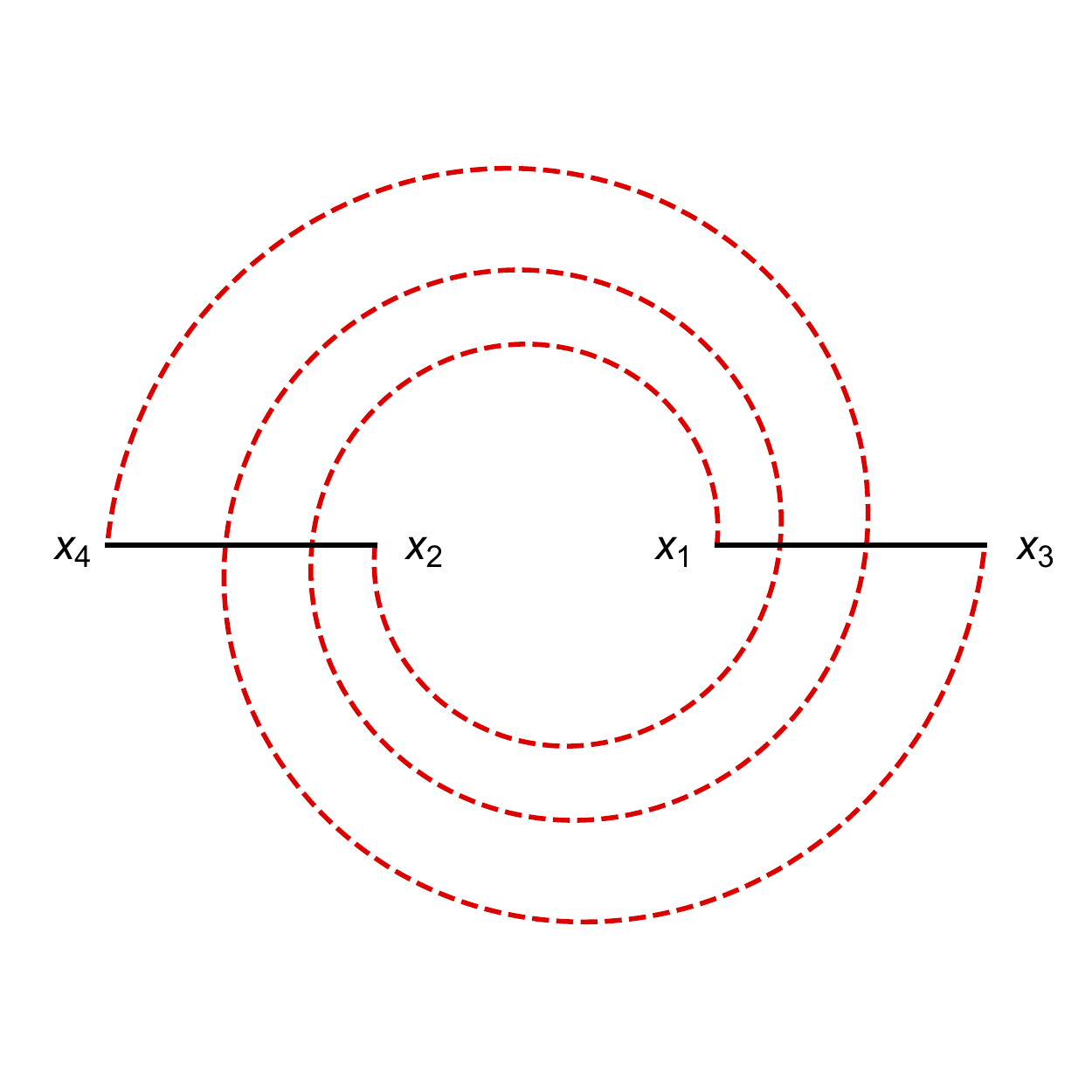}\\
        \xi^0&
        \xi^4&
        \xi^8
        \eea$
        \caption{First $3$ orders contributing to the $\Gc$ correlator. }
        \label{fig:G2loops}
\end{figure}

The two-magnon correlation function $\Gc$ is given by graphs shown in Fig.~\ref{fig:types}(right). 
Since these graphs have two propagators attached to all four external points, we identify the scaling dimensions as
\begin{align}\label{weights-2}
    \Delta_1=\Delta_2=\Delta_3=\Delta_4=2 
\end{align}
As before, in order to construct the graph building operator for the two-magnon case $\hat H_\gc$, we examine the first few orders of the weak coupling expansion of $\Gc$ (see Fig.\ref{fig:G2loops}).  
The expressions corresponding to the first two diagrams on Fig.\ref{fig:G2loops} are 
\beqa\la{diagsG2}\notag
\Gc^{(0)}&=&\frac{c^4}{ x_{13}^2 x_{14}^2 x_{23}^2x_{24}^2}\\
\Gc^{(1)}&=&\int d^4 x_5 d^4 x_6 \frac{c^8}{x_{25}^2 x_{54}^2 x_{15}^2 x_{53}^2  x_{16}^2 x_{63}^2 x_{26}^2 x_{64}^2}
=
\left[\int d^4 x_5 
\frac{c^4}{x_{25}^2 x_{54}^2 x_{15}^2 x_{53}^2  }\right]^2
\;.
\eeqa 
Note that the two-loop integral entering $\Gc^{(1)}$ factorizes into a product of one-loop integrals. This is not the case already at the next
$O(\xi^8)$ order for the right-most diagram in Fig.\ref{fig:G2loops}.

The kernel $H_{\gc}$ generating the two-magnon diagrams is
\begin{align}\label{H2}
 H_{\gc}(x_1,x_2|x_3,x_4)& \equiv   \frac{c^4}{x_{13}^2 x_{14}^2 x_{23}^2 x_{24}^2}\,.
\end{align}
Indeed we verify that the convolution of  $H_{\gc}$ 
reproduces all the diagrams depicted in Fig.\ref{fig:G2loops}. 
\beqa\notag
\Gc^{(0)}&=& H_{\gc}(x_1,x_2|x_3,x_4)\;,\\
\Gc^{(1)}&=&\int d^4 y_5d^4 y_6 H_{\gc}(x_1,x_2|y_5, y_6) H_{\gc}(y_5, y_6|x_3,x_4)\;.
\eeqa
Thus for the sum of all two-magnon diagrams we get 
\begin{align}\label{G2}
\hat \Gc = \sum_{\ell=0}^\infty (16\pi^2\xi^2)^{2\ell} \hat \Gc^{(\ell)}=  {\hat H_{\gc} \over 1- (16\pi^2\xi^2)^2 \hat H_{\gc}}\,.
\end{align}
Comparison with the general expression 
\eq{Ggeneral} shows that $\chi=(16\pi^2\xi^2)^2$ and $n=1$.

\subsubsection{Eigenvalue of the graph-building operator}

To compute the two-magnon correlation function \re{G2} we have to diagonalize the operator $\hat H_{\gc}$. We can use \re{H2} to show that it commutes with the generators of the conformal group. As a consequence,
its eigenstates are given by \eq{waveFGeneral} with $\Delta_1=\Delta_2=2$. To find the corresponding eigenvalue $E_\gc$, we replace the eigenstates in \eq{waveGeneral} by their explicit expressions \eq{waveFGeneral}. This leads to a rather complicated integral on the left-hand side of \eq{waveGeneral}. We can simply its calculation by sending $x_0\to\infty$ on the both sides of \eq{waveGeneral}. In addition, we project
all Lorentz indices on an auxiliary light-like vector $n^\nu$ (with $n^2=0$) and obtain the following representation for $E_\gc$
\begin{align}\label{EB-int}
E_\gc(\Delta,S) = {1\over (4\pi^2)^4}   \int {d^4 x_3 d^4 x_4 \, (nx_{34}) ^S \over x_{13}^2 x_{14}^2 x_{23}^2 x_{24}^2   (x_{34}^2)^{1-i\nu +S/2}} \,,
\end{align}
where $\Delta=2+2i\nu$ and we put $x_{12}^2=(nx_{12})=1$ for convenience. 

Since the integrand of \re{EB-int} acquires the $(-1)^S$ factor under the exchange of the integration points, $x_3\leftrightarrow x_4$, $E_\gc$ vanishes for odd $S$. For even $S$ the calculation of \re{EB-int} yields
 (see appendix~\ref{app:EB} for details)
\begin{align}\la{E2}
{E}_{\gc}\notag
{}& = {1\over 4\pi^4(S+1)}  \sum_{n=0}^\infty \frac{(-1)^n (2 n+S+2)}{(\Delta -2 n-S-4)^2 (\Delta +2 n+S)^2}
\\
{}& =\frac{\psi ^{(1)}\left(\frac{1}{4}
        (S-\Delta +4)\right)-\psi
        ^{(1)}\left(\frac{1}{4} (S-\Delta
        +6)\right)-\psi
        ^{(1)}\left(\frac{1
        }{4}(S+\Delta)\right)+\psi
        ^{(1)}\left(\frac{1}{4} (S+\Delta
        +2)\right)}{(4\pi)^4 (\Delta
        -2) (S+1)}\;,
\end{align}
where $\psi^{(1)}(x)=d \psi(x)/dx$.

The expression for ${E}_{\gc}$ is manifestly invariant under $\Delta\to 4-\Delta$.
Let us verify the condition \eq{conditionE0} for cancelling the spurious poles. 
In the present case, for  $\Delta_1=2$ and $\Delta_2=2$,  it follows from \eq{rn} that $r_{2n+1}=0$ for integer $n$, and the relation \eq{conditionE0} reduces to
\beq
E_{\gc}(\Delta=3+s+2n,S=s)= E_{\gc}({\Delta=3+s,S=s+2n}) \,, \eeq
for $n,s=0,1,2,\dots$.
It is easy to check that it is indeed satisfied. 

\subsubsection{Spectrum}
\begin{figure}[t!]
        \begin{center}
        \includegraphics[scale=0.7]{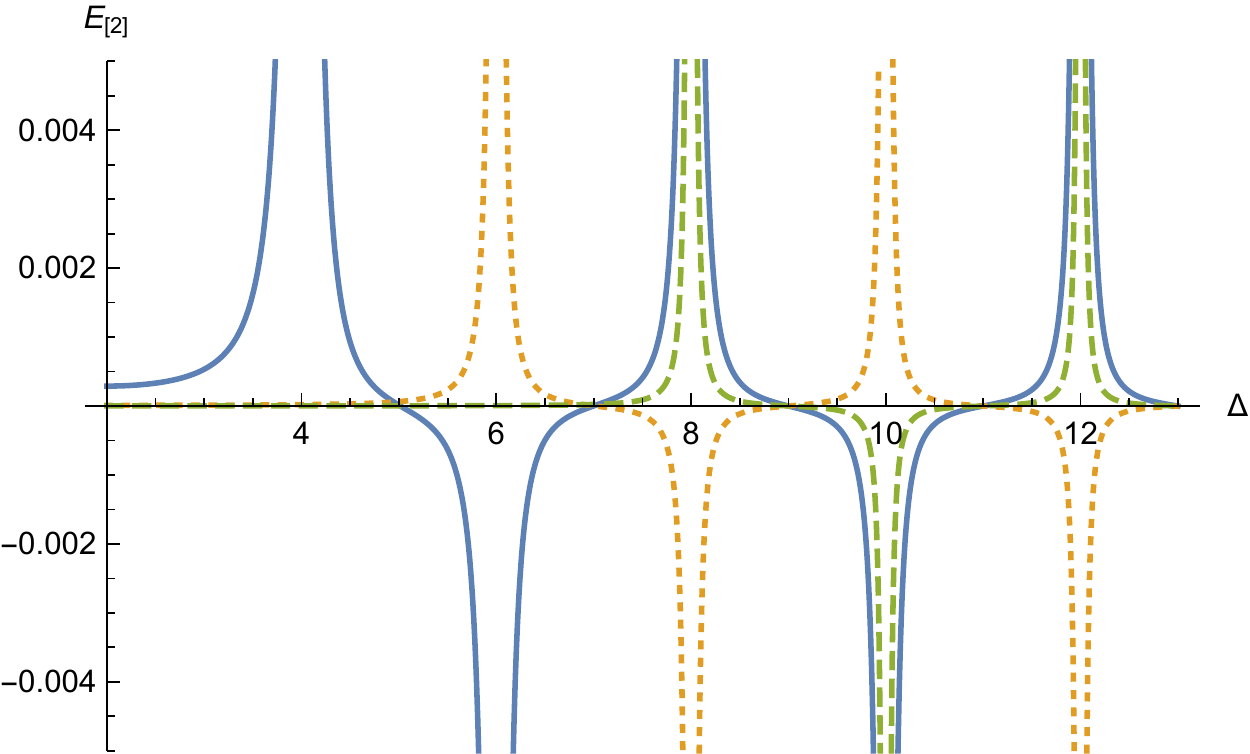}
        \end{center}
        \caption{\la{EB}Function $E_{\gc}(\Delta)$ for $S=0$ (solid line), $S=2$ (dashed line) and $S=4$ (dotted line). At $\Delta=S+4,S+6,\dots$ the function has a double pole. Since we have to impose \eq{E2todelta} at weak coupling $E_{\gc}$ goes to infinity. We see there are always two values of $\Delta$ for which $E_{\gc}$ is large in the vicinity of $\Delta=S+t$ for $t=4,\;6,\;8,\dots$ and thus the tree level spectrum is twice degenerate.}
\end{figure}

The scaling dimensions of the two-magnon operators satisfy the relation
\beq\la{E2todelta}
E_{\gc}(\Delta,S)=\frac{1}{(16\pi^2\xi^2)^2}\;.
\eeq
subject to ${\rm Re}\, \Delta>2$ and $S$ being even nonnegative.

\begin{figure}[t!]
    \centering
    \includegraphics[scale=0.6]{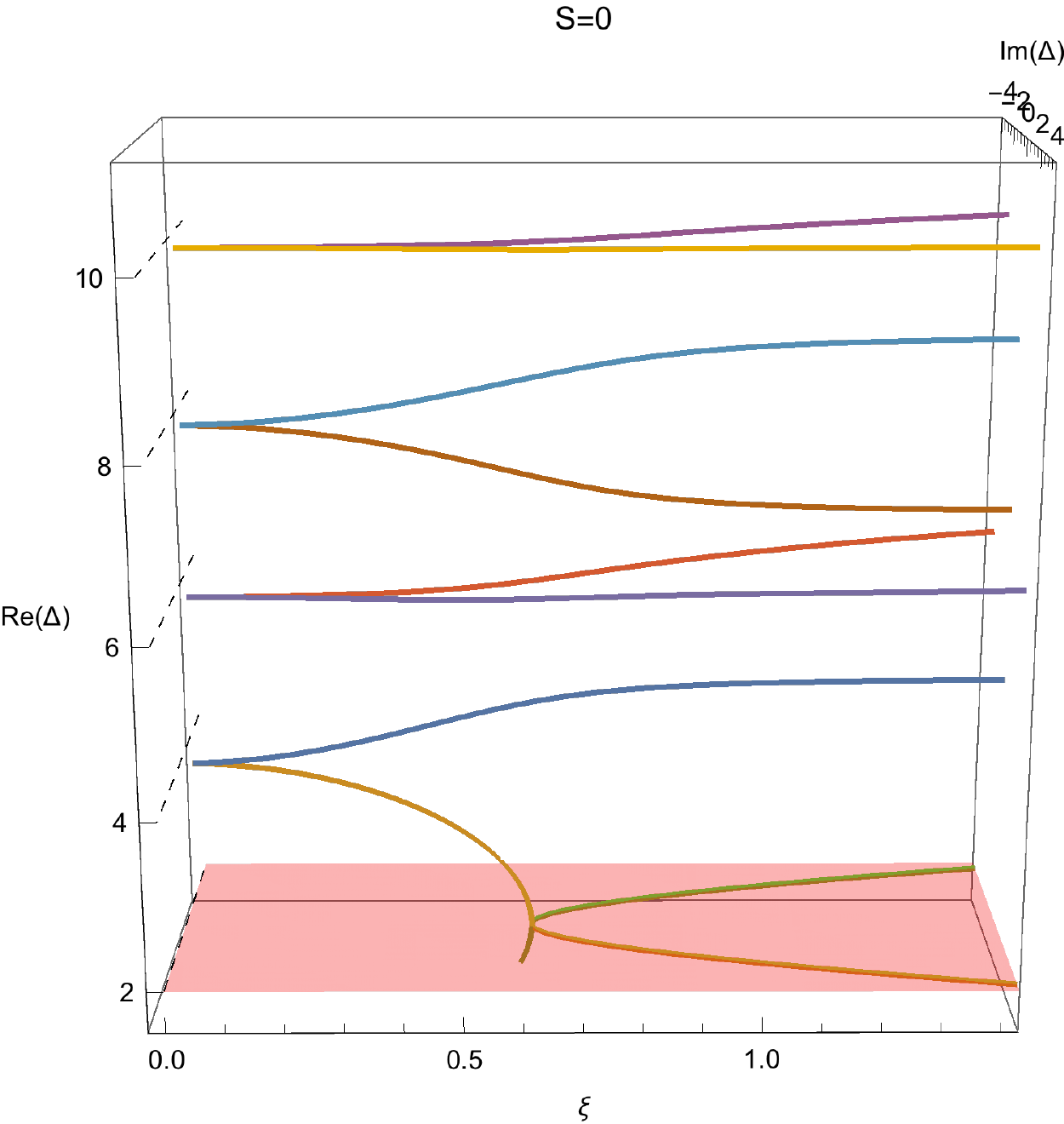}
    \includegraphics[scale=0.6]{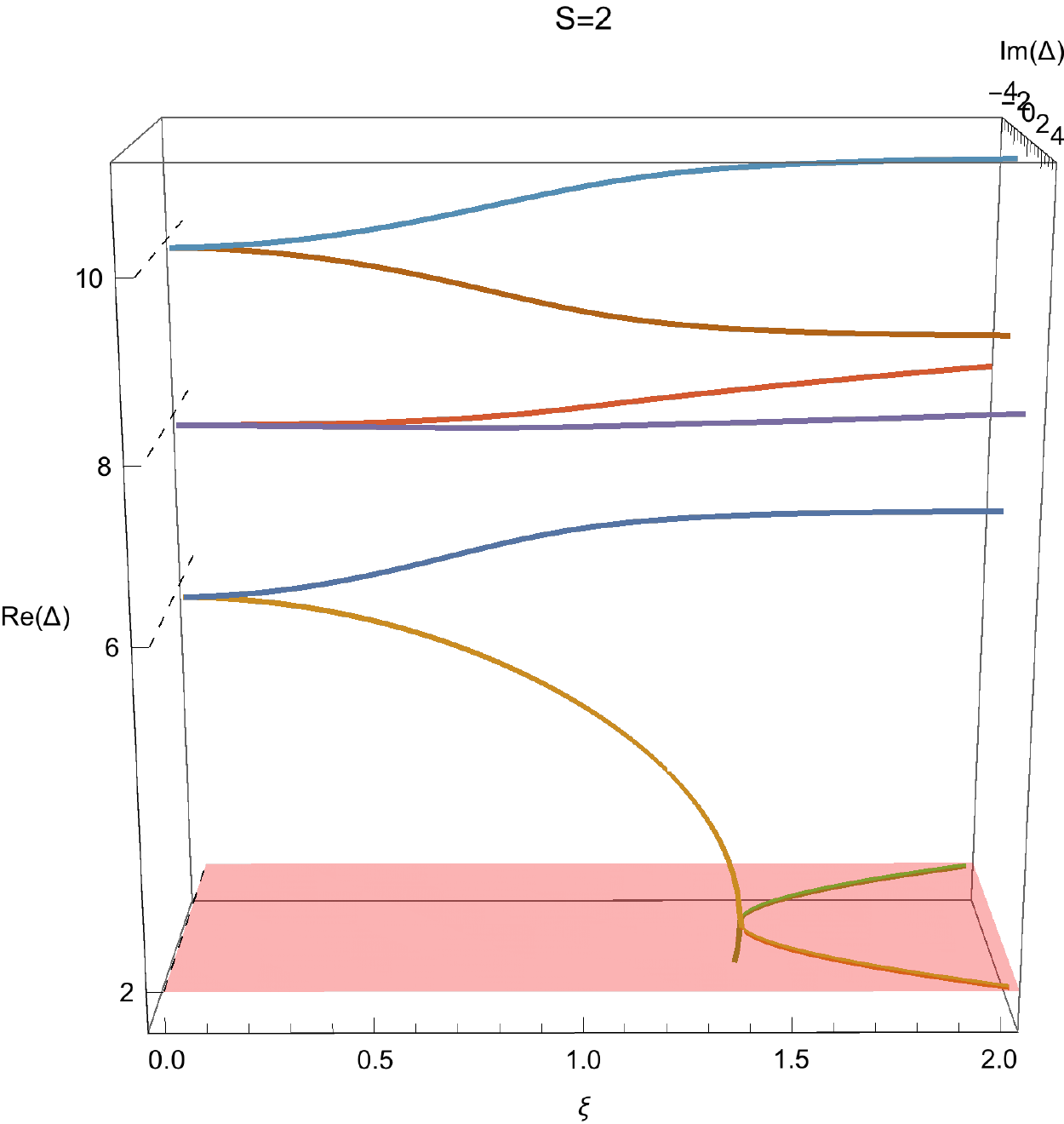}
    \caption{Spectrum of the scaling dimensions 
    $\Delta_{\gc}(\xi)$ of the two-magnon operators  for $S=0$ (left) and $S=2$ (right)}
    \label{fig:EBspectrum}
\end{figure}

This time the spectrum of $\Delta$'s has a rich structure since
for each value of $S$ there are infinitely many solutions to \re{E2todelta}.
Indeed, as follows from the first relation in \re{E2}, the function \eq{E2} has an infinite sequence of double poles at $\Delta=S+t$
for $t=4,6,8\dots$. As a consequence, for small $\xi$ the relation \re{E2todelta}
always has two solutions in the vicinity of $\Delta=S+t$
 describing  operators with 
twist $t$ and even spin $S$.~\footnote{Like in the case of $\Ga$ due to the symmetry of the correlation function under the exchange of points $x_3$ and $x_4$ only even spins contribute to $\Gc$.}
Indeed we see on Fig.\ref{fig:EBspectrum} that all levels are twice degenerate at weak coupling. 

The weak coupling expansion of $\Delta_\gc$ can be found by replacing  $E_{\gc}$ by its expansion 
in the vicinity of the pole. Going through the calculation we obtain
\begin{align}\label{sol4}
\Delta_{\gc}=S+t \pm \xi^2 \gamma_{t,S}^{(0)} + \xi^4  \gamma_{t,S}^{(1)} + O(\xi^6)\;\;,\;\;t=4,6,\dots\,,
\end{align}
with the expansion coefficients given by
\begin{equation}\label{gamma's}
 \gamma_{t,S}^{(0)} =   {4\, i^{t/2} \over\sqrt{  (S+1)(t+S-2)}}
\;\;,\qquad
 \gamma_{t,S}^{(1)} = -{8 (-1)^{t/2}\over (S+1)(t+S-2)^2}\,.
\end{equation}
For each $t$ and $S$, the relations \re{sol4} and \re{gamma's} yield two scaling dimensions that are related to each other through the transformation $\xi^2\to -\xi^2$.
For even $t/2$, the expansion coefficients in \re{sol4} are real. For odd $t/2$, the leading coefficient $\gamma_{t,S}^{(0)}$ is pure imaginary (see Fig.\ref{fig:EBspectrum}).
The relations \re{sol4} and \re{gamma's} describe the scaling dimensions of an infinite set of operators, two per each twist $t=4,6,\dots$ and spin $S=0,2,\dots$. 

In particular, for $S=0$ and $t=4$,  for the two-magnon operators of the form ${\rm tr}(X^2Z^2)+\dots$ the scaling dimensions are given at weak coupling by
\begin{align}\notag
\Delta_{S=0,t=4} {}& =4\pm 2 \sqrt{2} \xi ^2-2 \xi ^4\pm\frac{\xi ^6}{\sqrt{2}}+\left(8-12 \zeta _3\right) \xi
   ^8
\\
{}&
   \pm\left(30 \sqrt{2} \zeta _3-\frac{513}{8 \sqrt{2}}\right) \xi ^{10}+\left(-96
   \zeta _3-60 \zeta _5+168\right) \xi ^{12}\pm O\left(\xi ^{14}\right)
\end{align}
The first 4 terms reproduce the prediction from ABA \cite{Caetano:TBP} including the first L\"uscher correction.

\paragraph{Critical coupling.}

The dependence of the two-magnon scaling dimensions $\Delta_\gc$ on the coupling constant is shown 
on Fig.\ref{fig:EBspectrum}. 
As can be seen from this figure, for each $S$
the lowest level approaches the value $\Delta=2$ at some finite $\xi=\xi_*$.
Expanding $E_{\gc}$ near $\Delta=2$ we find the corresponding value of the coupling constant $\xi_*$
\beq
\xi_*^4=\frac{2 (S+1)}{\psi ^{(2)}\left(\frac{S}{4}+1\right)-\psi
        ^{(2)}\left(\frac{S+2}{4}\right)}\;.
\eeq 
Numerically, for $S=0,2,4,\dots$ we obtain\footnote{For comparison, for a different operators with $J=3$ with zero spin of the type ${\rm tr}\Box^n X^3$ we get a very similar behaviour with the critical points at $0.589884,1.32836,2.02683,2.71805,3.40652,\dots$ \cite{Gromov:2017cja}.}
\beq
\xi_*=0.610212,\quad 1.397,\quad2.20284,\quad 3.02178,\quad 3.84809,\quad\dots\;.
\eeq
For $\xi>\xi_*$, the scaling dimension  $\Delta$ develops  an imaginary part. As we see in a moment, it grows linearly with $\xi$ at strong coupling.

\paragraph{Strong coupling.}
Solving \re{E2todelta}
at strong coupling, we have to identify the values of $\Delta$ at which $E_{\gc}(\Delta,S)$ vanishes. A close examination of \re{E2} shows that $E_{\gc}(\Delta,S)$ has infinitely many zeroes in $\Delta$ for any even of $S$. For example, for $S=0$ the first few zeroes of \re{E2} satisfying ${\rm Re}\,\Delta>2$ can be found numerically as
\beqa
\Delta_{S=0}=5.0145,\;\; 6.6879 \pm 4.08478 i,\;\; 6.99634,\;\; 9.0014,\;\; 10.9993,\;\; 11.0453 +5.05341 i,\dots\notag
\eeqa
Notice that the real part of most of the zeros is close to integer. 
They determine the leading large $\xi$ asymptotics of the scaling dimension of all but two states shown in  Fig.\ref{fig:EBspectrum}.

The remaining two states satisfy $\Re \Delta=2$ and have an imaginary part that grows linearly with $\xi$. They correspond to the solution to \re{E2todelta} with $\Delta\to\infty$. Indeed, the function \re{E2} decays at large $\Delta$ as $E_\gc\sim 1/\Delta^4$ and the relation \re{E2todelta} is automatically satisfied.
By expanding $E_{\gc}$ at large $\Delta$ and solving \eq{E2todelta} we find
\beq\label{DeltaS2m}
\Delta_{\gc}=2\pm i\left(2 \sqrt[4]{2} \xi -\frac{S^2+2 S-2}{4 \sqrt[4]{2} \xi }-\frac{S^4+4 S^3+24 S^2+40
   S-68}{64\ 2^{3/4} \xi ^3}+{\cal O}\left(\frac{1}{\xi^4 }\right)\right)\;.
\eeq

\subsubsection{OPE coefficients} \label{sect:ope}

From \eq{Cstruct} we get for $C_{\Delta,S}$:
\beqa\la{CstructEB}
C_{\Delta,S}=- \frac{\pi/c^4}{c_2(\Delta,S)}\frac{1}{\partial_\Delta (1/E_{\gc}(\Delta,S))}\;.
\eeqa
where $c_2$ is given by \eq{c2general} for
$\Delta_1=\Delta_2=2$. 
Replacing $E_{\gc}$ with \re{E2}, we obtain 
a rather cumbersome expression for $C_{\Delta,S}$, we do not present it here. 

\paragraph{Weak Coupling.}
Expanding the resulting expression for $C_{\Delta,S}$ in powers of $\xi^2$, we get
the OPE coefficients for the operators with twist $t=\left.(\Delta-S)\right|_{\xi=0}=4,6,8,\dots$  and Lorentz spin $S=0,2,4,\dots$
\begin{align}\label{C-weak2}
C_{\Delta,S}  = (-1)^{t/2}\frac{ \Gamma^2 (t/2-1) \Gamma^2 (S+t/2)}{\Gamma (t-3) \Gamma (2
        S+t-1)}\left[ 1+ \xi^2c_{t,S}^{(0)} \gamma_{t,S}^{(0)}+O(\xi^4)\right]\,.
\end{align}
Here $\gamma_{n,S}^{(0)}$ is the one-loop anomalous dimension defined in \re{gamma's} and the coefficient $c_{n,S}^{(0)}$ is given by
\begin{align}
c_{t,S}^{(0)} = \psi({t/2+S})-\psi({t+2S-1}) +\psi({t/2-1})-\psi({t-3})-\frac{1}{2(S+t-2)} \,. 
\end{align}

\paragraph{Strong Coupling.}
To analyse the strong coupling behaviour of \re{CstructEB} it is convenient to rewrite it as
\beq\la{C2fancy}
C_{\Delta,S}=-\frac{\pi/c^4}{ c_2(\Delta,S)}\frac{\partial \Delta}{\partial (4\pi\xi)^4}
\eeq
where we used \eq{E2todelta} to get rid of $E_{\gc}$. Since for most of the states $\Delta$
approaches a constant value at strong coupling, we see from \eq{C2fancy} that $C_{\Delta,S}$ should decay as $1/\xi^8$. For the two remaining states
whose imaginary part grows linearly in $\xi$, 
the OPE coefficient \re{C2fancy} decreases slower as $1/\xi$ for real $\xi$. We recall however that 
in order for the correlation function \eq{Ggen} to be well-defined, $\xi$ should have an imaginary part. 
For $\xi$ with large imaginary part 
the structure constant of the lowest level at each
$S$ decays exponentially 
and thus the OPE expansion at strong coupling should be dominated by the other state, for which $\Delta\to\text{const}$ at $\xi\to\infty$.
This is rather different behaviour to that of two other correlators and needs further investigation. We discuss this issue briefly in section~\ref{sec:strong}.

\subsubsection{4-point correlation function}
Having determined the scaling dimensions and the OPE coefficients we can compute the two-magnon correlation function
\beqa\label{G2Ggen}
{\Gc(x_1,x_2|x_3,x_4)}\equiv
\frac{c^4}{x^4_{12}x^4_{34}} {\cal G}_{\gc}(u,v)\;,
\eeqa
where
\begin{align}\label{GB-sum}
\mathcal \Gc(u,v) = \sum_{t=4,6,\dots} \sum_{S=2,4,\dots} C_{\Delta_+,S}\,  g_{\Delta_+,S}(u,v)+C_{\Delta_-,S}\,  g_{\Delta_-,S}(u,v)\,.
\end{align}
We note that for each spin $S$ and twist $t$
there are two states, as one can see from the weak coupling expansion \eqref{sol4}.

Notice that the weak coupling expansion of \re{sol4} and \re{C-weak2} goes in powers of $\xi^2$. However, due to the symmetry of the spectrum, the two terms in the sum \re{GB-sum} are related to each other through transformation $\xi^2\to -\xi^2$ so that the weak expansion of
$\mathcal \Gc(u,v)$ runs in powers of $\xi^{4}$.

In section~\ref{sec:PT}, we study the equation \eq{GB-sum} at weak coupling and compare the result with the predictions from the perturbation theory of section~\ref{sect:weak}.

\section{Correlation functions at weak coupling from Feynman diagrams}

\label{sec:doubletrace}\label{sect:weak}

In the previous sections, we have derived three different types of four-point correlation functions by applying the operatorial methods. Namely, we have solved the underlying Bethe-Salpeter equations by diagonalizing the corresponding ``graph-building" kernels with a help of the conformal symmetry. Doing so, we have ignored  the double trace counterterms~\eqref{L-dt} which are nessesary for the consistent definition of the bi-scalar model \eqref{bi-scalarL} on the quantum level and for restoring the conformal symmetry of the theory. 

In this section we discuss the role of the double-trace interaction terms \eqref{L-dt}. We show that they are necessary at weak coupling in order to make each order of the 
perturbation expansion of the correlation functions to be finite. At the same time, they do not affect the results for the correlation functions at finite coupling obtained in the previous section.

Let us review  the role of double-trace couplings \(\alpha_1\) and \(\alpha_2\) from the action \eqref{L-dt} in perturbative computations of the correlation functions \(G_{A}, G_{B}, G_{C}\) and \(G_{D}\). 
Below we discuss which of the topologies of the Feynman graphs $\Ga$, $\Gb$ or $\Gc$
have to be completed with the double-trace interactions in order to have meaningful weak coupling expansion of the abovementioned four-point functions.

\subsection{Double-trace contribution to  \(G_{A}, G_{B}\) }

We recall that  the four-point functions  \(G_{A}, G_{B}\) are completely defined, at least for any finite \(\xi\), by the zero-magnon function \(\Ga\). The latter is given by sum over the wheel graphs shown in Fig.~\ref{fig:types2}. As was already mentioned, each wheel in these graphs develops a ultraviolet divergence at short distances. We expect that the double-trace contribution should remove this divergence.  

\begin{figure}[t!] 
\begin{center}	
 \includegraphics[scale=0.35]{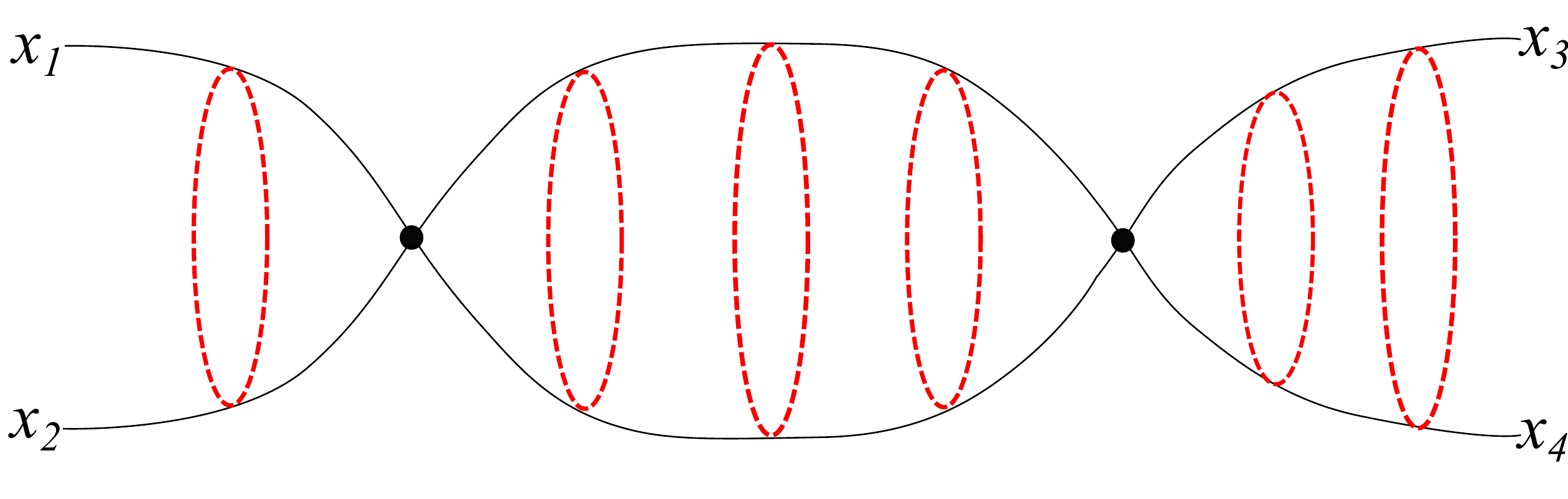}
\end{center}
\caption{Feynman diagrams contributing to the four-point function $ \Ga(x_1,x_2|x_3,x_4)$. Dots denote double-trace $\alpha_1^2$ vertices, all
other quartic vertices describe single-trace $\xi^2$ coupling.}
\label{fig:G-aux}
\end{figure}

The double-trace interaction is described by the action \eqref{L-dt}. 
 It is easy to see that, due to the cylindrical topology of the underlying planar graphs, among four different double-trace interaction terms in \eqref{L-dt} only one term $(4\pi)^2\alpha_1^2 \tr(X^2)\tr(\bar X^2)$ can contribute to \(\Ga\) in the planar limit. It generates a new local quartic scalar vertex. The resulting planar graphs contributing to \(\Ga\) are shown in  
 Fig.\ref{fig:G-aux}. They are obtained by gluing together wheel graphs. 
  Indeed, the insertion of the double-trace vertex \(16\pi^2\alpha_1^2\tr(\bar X)^2\,\tr X^2\) effectively splits the planar wheel Feynman graph into two disconnected parts, with the single trace operators, \(\tr(\bar X)^2\) and \(\tr X^2\), attached to each part.  

\begin{figure}
        \centering
        \includegraphics[scale=0.5]{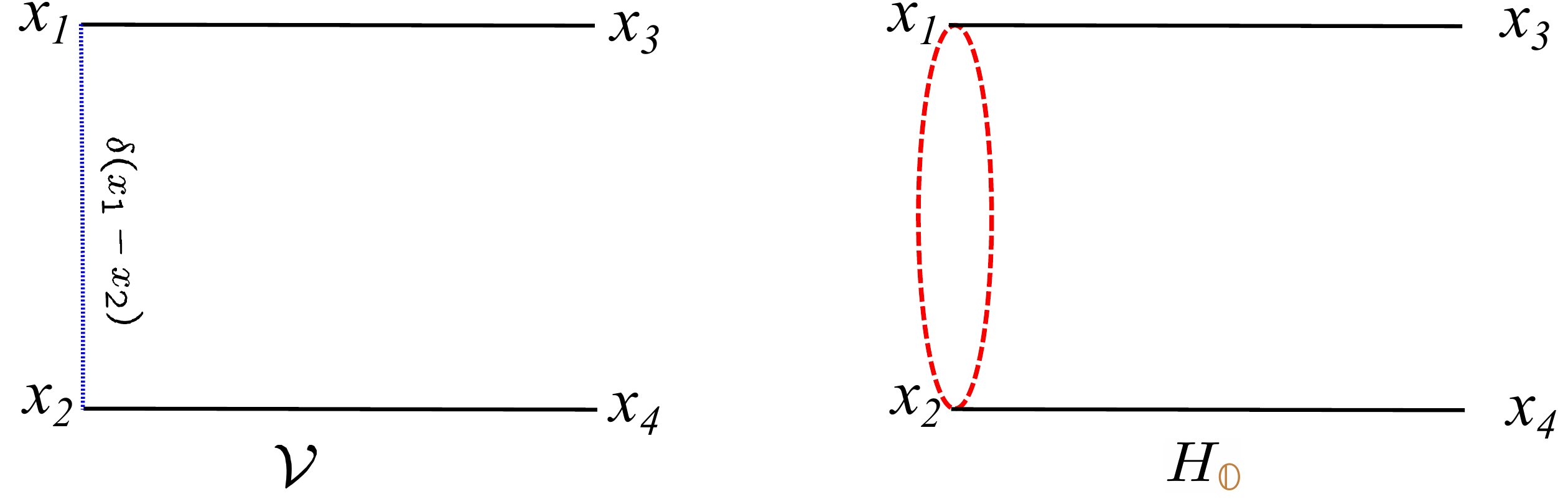}
        \caption{Diagrammatic representation of the graph building operators \({\cal V}\) and \(H_\ga\) defined in \re{ker-def}. Dashed blue line denotes delta-function, black and dashed red lines stand for free scalar propagators.  } \label{fig_HV}
\end{figure}

As we demonstrated in the previous section, the wheel graphs can be summed up by introducing the graph building operators $\hat H_{\ga}$. In the similar manner, we can take into account the graphs shown in 
Fig.~\ref{fig:G-aux} by  replacing the kernel \((16\pi^2\xi^2)^2\hat H_{\ga}\) in the equation \eqref{G0sum} by a linear combination \((16\pi^2\alpha_1^2) \, \mathcal{V} +(16\pi^2\xi^2)^2 \hat H_{\ga}\)     of operators $\mathcal V$ and $\hat H_{\ga}$ generating double-trace vertices and scalar loops, respectively (see Fig.~\ref{fig:G-aux}). Since the contribution to the correlation function of individual diagram shown in Fig.~\ref{fig:G-aux} is divergent, we
introduce dimensional regularization with $d=4-2\epsilon$. Then, the regularized operators $H_{\ga }$ and $\mathcal V$ are defined as
\begin{align}\notag\label{ker-def}
{}& H_{\ga } \, \Phi(x_3,x_4) = c^{4}\int {d^{4-2\epsilon}x_{1}d^{4-2\epsilon}x_{2} \over (x_{13}^2 x_{24}^2 (x_{12}^2)^2)^{1-\epsilon} } \Phi(x_1,x_2)\,,
\\
{}& \mathcal V  \, \Phi(x_3,x_4) =2 c^{2}\int {d^{4-2\epsilon}x_{1}d^{4-2\epsilon}x_{2} \over (x_{13}^2 x_{24}^2)^{1-\epsilon} } \delta^{(4-2\epsilon)}(x_{12}) \Phi(x_1,x_2)\,,
\end{align}
where $\Phi(x_1,x_2)$ is a test function.
They admit a simple diagrammatic representation, see  Fig.~\ref{fig_HV}(right) and (left),
 respectively. 
 We would like to emphasize that the operators \re{ker-def} are well-defined for $\epsilon\neq 0$. 
 
Making use of the operators \re{ker-def} we obtain the following representation for the zero-magnon correlation function
\begin{align}\label{G-repr}
\Ga(x_1,x_2,x_3,x_4) =(4\pi^2)^2 x_{12}^4 \lim_{\epsilon\to 0}  
(x_{1},x_{2}| {1\over 1-(16\pi^2\alpha_1^2 ) \, \mathcal{V} -(16\pi^2\xi^2)^2 \hat H_{\ga}} \hat H_{\ga}|x_{3},x_{4})\,,
\end{align}
where $\alpha^2_1\equiv\alpha_\pm^2(\xi)$ is the double-trace coupling at the fixed point \re{fixed}.
Expanding \re{G-repr} in powers of the couplings $\alpha_1^2$ and $\xi^2$  we find that the first few terms of the weak-coupling expansion of $\Ga$ are given by graphs depicted in Fig.~\ref{fig:G0loopsalpha} below. We compute them later in this section. It is easy to see that 
higher order terms of the weak-coupling expansion of \re{G-repr} produce graphs shown in Fig.~\ref{fig:G-aux}.

Note that for $\epsilon\neq 0$ the conformal symmetry of the correlation function \re{G-repr} is broken. To elucidate the mechanism of restoration of the conformal symmetry of $\Ga$ and the role played by the double traces, we present below the two-loop calculation of the correlation function \re{G-repr}.

Applying the identity $1/(x_{34}^{2})^{2-2\epsilon}=\pi^2 \delta(x_{34})/\epsilon$ as $\epsilon\to 0$, we find from \re{ker-def} that 
$ H_{\ga } \, \Phi(x_1,x_2)\sim \epsilon^{-1}\int d^4 x_3 \Phi(x_3,x_3)/(x_{13}^2 x_{23}^2)$. This means that the operator $ H_{\ga }$ is singular on the space of functions $\Phi(x_3,x_4)$ that do not vanish at short distances $x_{34}\to 0$. Examining the expression for the eigenfunctions \re{waveFGeneral}, we find they scale at short distances as $\Phi_{\nu,S,x_0}(x_3,x_4)\sim 1/x_{34}^{\Delta_1+\Delta_2-\Delta} = 1/x_{34}^{-2i\nu}$ for $\Delta_1=\Delta_2=1$ and $\Delta=2+2i\nu$. We recall that computing the correlation functions in the previous section we deformed the integration contour over $\nu$ into the lower half-plane. It is easy to see that in this case $\Phi_{\nu,S,x_0}(x_3,x_4)$ vanishes for $x_{34}\to 0$ and,
as a consequence, the operator $H_{\ga }$ does not develop UV divergences. Moreover, as follows from the definition \re{ker-def}, the double-trace operator $\mathcal V$
annihilates the eigenstates $\Phi_{\nu,S,x_0}(x_3,x_4)$ with $\Im \nu<0$ and,
therefore, does not contribute. This explains why the double-trace interaction can be neglected when computing the four-point function \(\Ga\) by the Bethe-Salpeter method. The appearance of UV divergences at weak coupling is a manifestation of analytic properties of $\Ga$. As a function of $\xi^4$, it  has a square-root cut at the origin so that its pertubative expansion runs in powers of $(-\xi^4)^{1/2}$. 
We have already observed this phenomenon on the example of the scaling dimension \eqref{tw2-weak}.

\subsection{Double-trace contributions to  \(G_{B}\) and \(G_{C}\) }

\begin{figure}[t!]
    \begin{center}
    \includegraphics[scale=0.25]{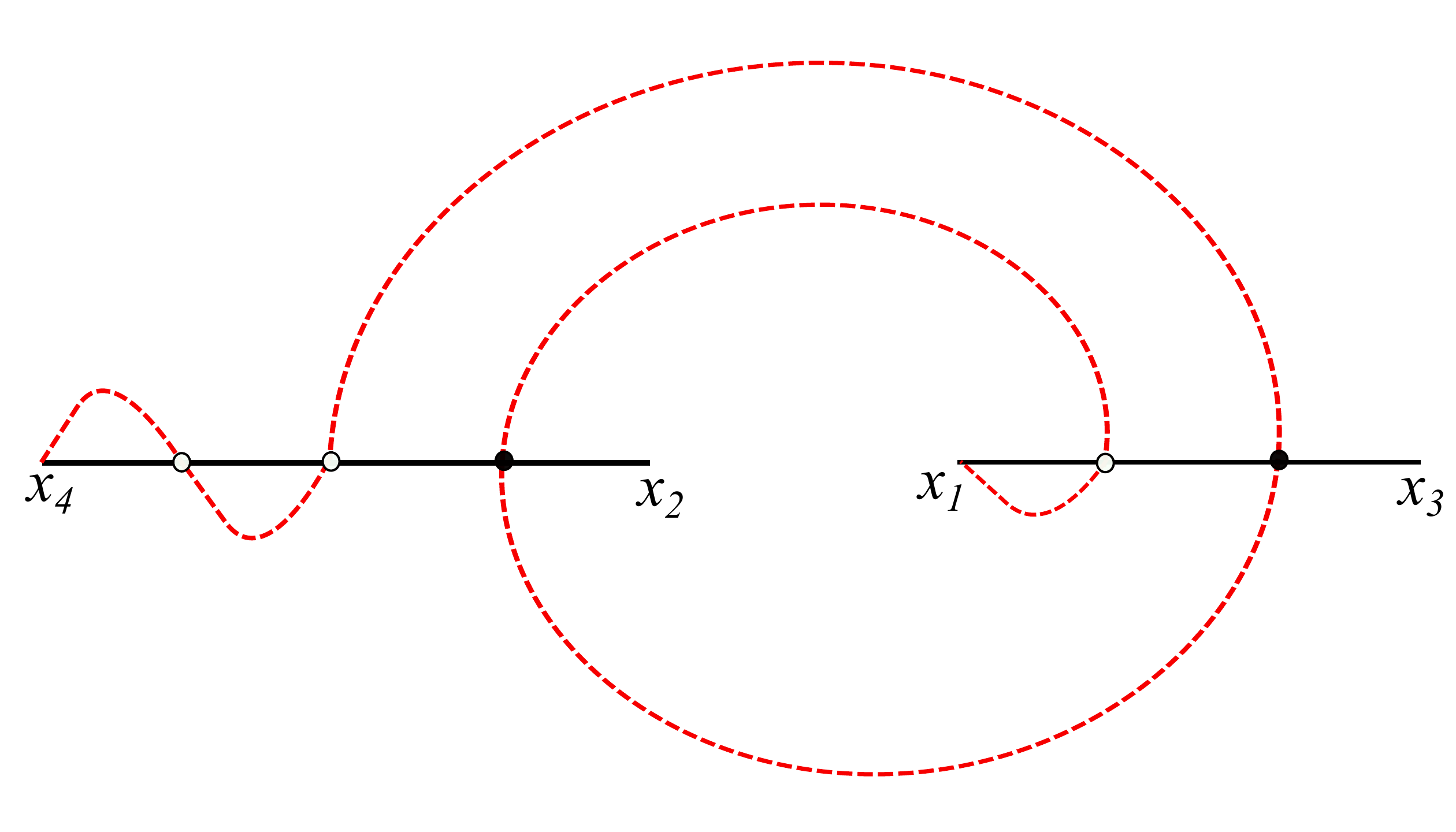}
    \end{center}
    \caption{Typical planar graphs producing UV divergent contribution to the four-point correlation function \(G_{C}\). The black blobs denote the single-trace vertices with the coupling \(\xi^2\). The white blobs denote both the single-trace  \(\xi^2-\)vertices and the double-trace \(\alpha_2^2-\)vertices. Their contribution is proportional to $(\xi^2-\alpha_2^2)$ and it vanishes at the fixed point. Similarly, for the correlation function \(G_{B}\) the contribution of  analogous UV divergent planar graphs vanishes at the fixed point through the same mechanism.      }
    \label{fig:2trace_graphs}
\end{figure}

 The inspection of Feynman graphs defining \(G_{B}\) and   \( G_{C}\) shows that the double-trace interactions with the coupling \(\alpha_1\) do not contribute in the planar limit. On the other hand, the interactions  with the double-trace  coupling \(\alpha_2\) do contribute to both correlation functions through the graphs of the type shown on Fig.\ref{fig:2trace_graphs} on the example of    \( G_{C}\). Each vertex depicted by white blobes on Fig.\ref{fig:2trace_graphs} describes both single- and double-trace couplings. The contribution of each such vertex to $G_C$ is UV divergent and it is proportional to $(\xi^2-\alpha_2^2)$. As a result, it vanishes at the fixed point \re{fixed},
 so that we are left only with the sums over UV finite single spiral graphs summed up by the UV finite structure function  \( \Gb\). This property is not surprising given the fact that the correlation function $G_{C}$
has to be a finite function of $\xi^2$ whereas the Feynman diagram in  Fig.\ref{fig:2trace_graphs} involves the  ultraviolet divergent 
scalar loops.     

The function $G_{C}$ is regular  at \(\xi^2\to 0\) and its weak-coupling expansion runs in powers of \(\xi^2\). The expression for $G_C$ at arbitrary coupling has been derived in section~\ref{sec:G[1]} using the Bethe-Salpeter equation in the form of conformal partial wave expansion, Eqs.~\re{G1p-def}, \re{G4cg} and \re{calG1}.

The correlation function \(G_{B}\) also receives UV divergent contribution from planar graphs similar to those shown in Fig.~\ref{fig:2trace_graphs}. Their contribution vanishes at the fixed point through the same mechanism as in the previous case. This means that  \(G_{B}\) is defined by the two-magnon function  \(\Gc\), see Eq.~\re{GB-def}. 
Since two-magnon graphs contributing to \(\Gc\) contain even number of single-trace vertices, the weak coupling expansion
of \(\Gc\) runs in powers of $\xi^4$. At arbitrary coupling, \(\Gc\)
is given by the conformal partial wave expansion \re{G2Ggen} and \re{GB-sum}.

In the rest of this section, we compute the first few terms of the weak coupling expansion of the 4-point correlation functions and,  then, compare them with the exact expressions obtained in section~\ref{sec:GBO}.

\begin{figure}[t]
	\centering
	$\bea{cccc}
	\includegraphics[scale=0.3]{mag0loop0.pdf}&	
	\includegraphics[scale=0.3]{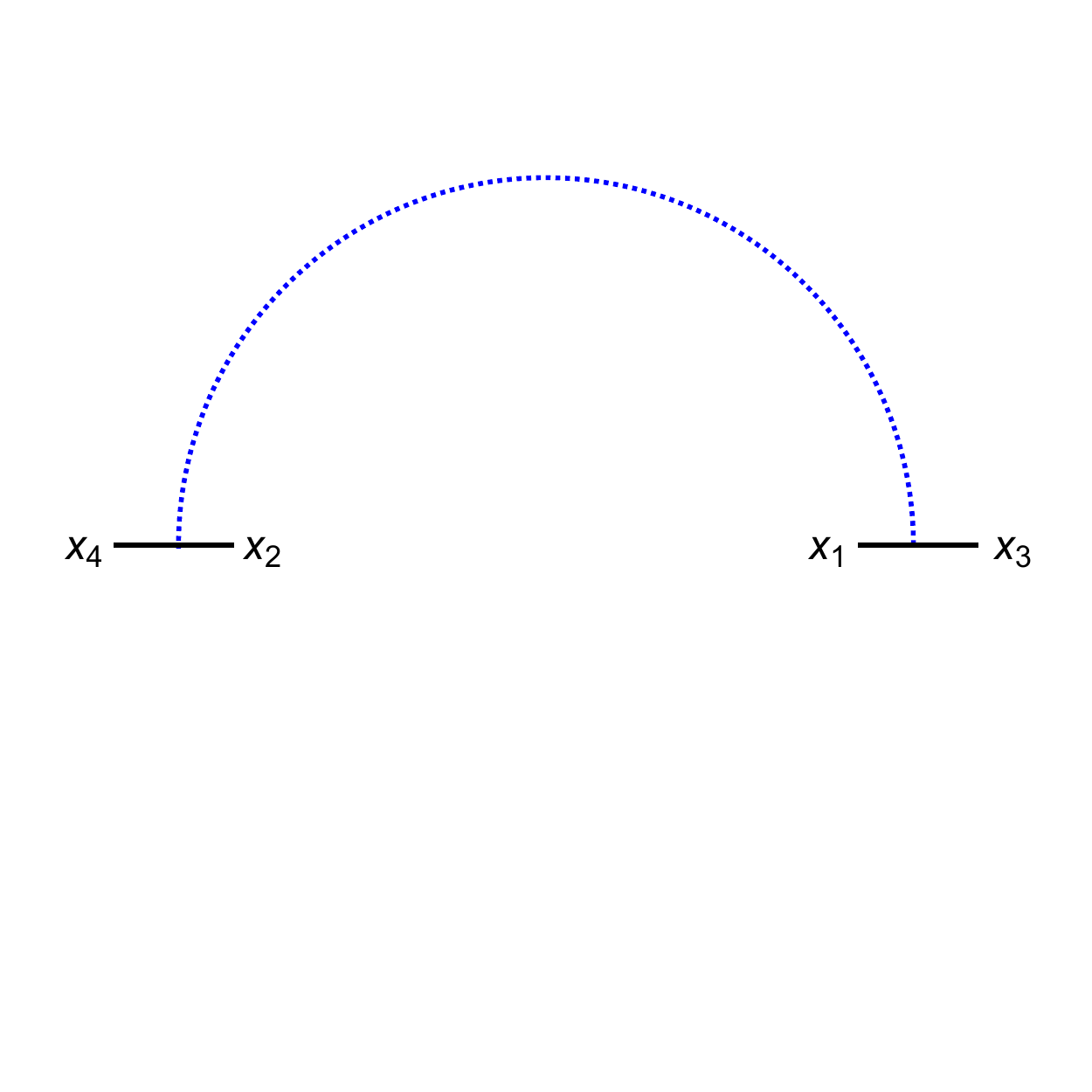}&
	\includegraphics[scale=0.3]{mag0loop1.pdf}&
	\includegraphics[scale=0.3]{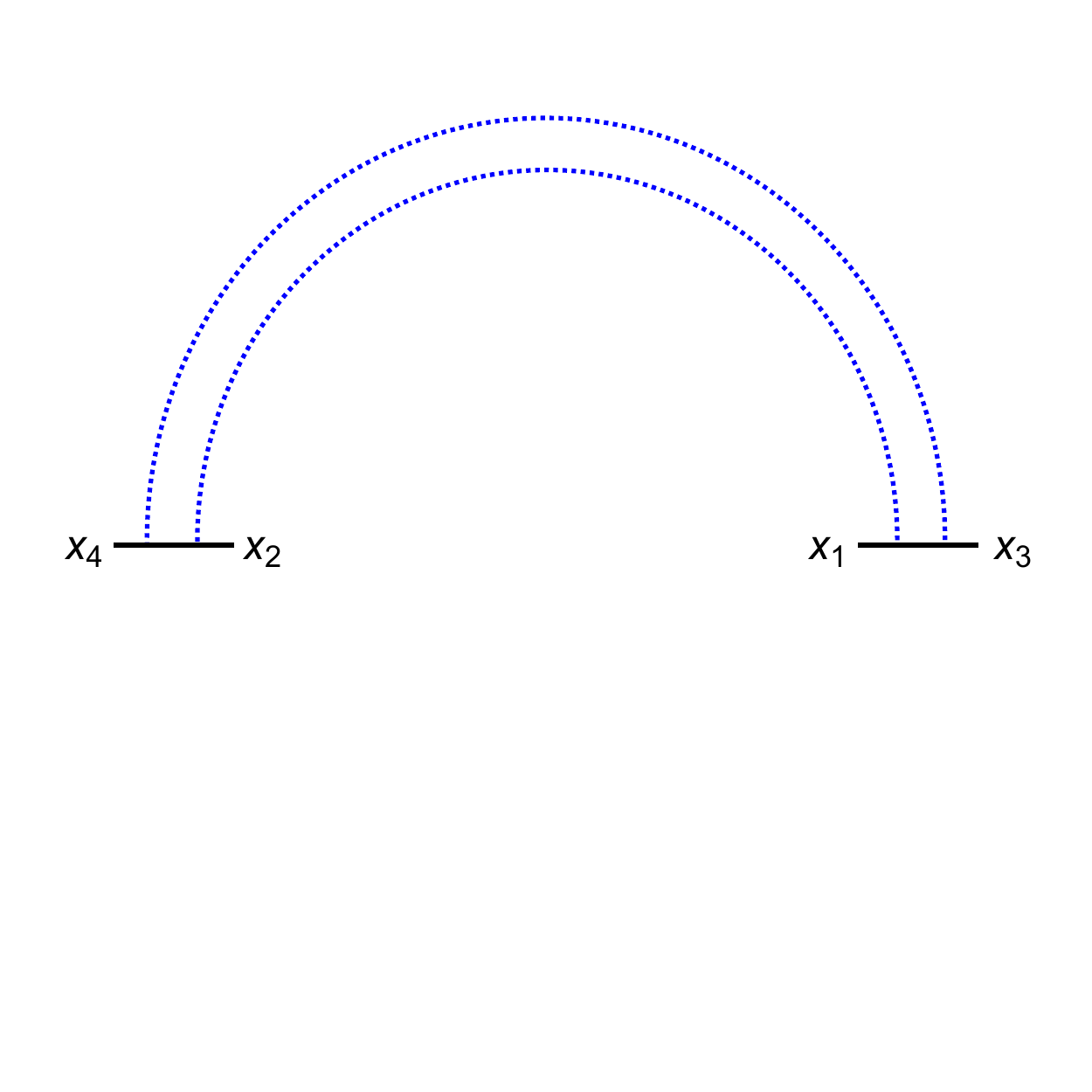}\\
	G^{(0,0)}&
	\alpha_1^2\, G^{(1,0)}&
	\xi^4\, G^{(0,2)}&
	\alpha_1^4\, G^{(2,0)}
	\eea$
	\caption{Feynman diagrams defining the first few terms of the weak-coupling expansion of $\Ga$. $G^{(n,k)}$ stands for diagram with $k$ single-trace vertices and $n$ double-trace vertices. Dashed blue line denotes delta-function, black and dashed red lines denote free scalar propagators.}
	\label{fig:G0loopsalpha}
\end{figure}

\subsection{Type $G_D$}\label{sect:conf}

In the previous sections, \(G_D\) was defined in terms of the function $\Ga$ by \eqref{GD-def}, and $\Ga$ is given by the wheel diagrams shown in Fig.\ref{fig:G0loops}. As we already discussed, this definition is not complete and has to be supplimented with the double-trace contributions.
The reason for this is that each diagram in Fig.\ref{fig:G0loops} is UV
divergent and the extra double trace contribution is needed to make each
term  of the weak coupling expansion to be UV finite.

To illustrate this, we examine the first few terms of the weak coupling expansion of $\Ga$. They are given by Feynman diagrams shown 
in Fig.~\ref{fig:G0loopsalpha}
\begin{align}\label{G-def1}
 \Ga = G^{(0,0)} + 16\pi^2\alpha_1^2 G^{(1,0)} + (16\pi^2\xi^2)^2 G^{(0,2)} + (16\pi^2\alpha_1^2)^2  G^{(2,0)} + {\cal O}(\xi^6) \,,
\end{align}
where $G^{(n,k)}$ denotes the contribution of diagram with $n$ double-trace vertices and $k$ single-trace vertices. We recall that $\Ga$ depends on only one double-trace coupling $\alpha_1^2$ whose value is given by \re{fixed} at the fixed point.

The first term on the right-hand side of \re{G-def1} has been previously defined in 
\eq{G0firsttwo}, $G^{(0,0)}=G^{(0)}_{\ga}$.
The $O(\alpha_1^2)$ correction to \re{G-def1}
can be computed from Feynman graphs contributing to \(G_D\) and  is given by a finite four-dimensional ``cross" integral (see Fig.~\ref{fig:G0loopsalpha})
\begin{align}\label{G01}
 16\pi^2\alpha_1^2G^{(1,0)} = \frac{1}{2}\times 4\times ({16\pi^2}{\alpha_1^2})  \int {d^4 x_5 c^4\over x_{15}^2 x_{25}^2 x_{35}^2 x_{45}^2}
  = \frac{ c^2}{x_{12}^2 x_{34}^2}
 2\alpha_1^2 {u  D(u,v)}
  \,,
\end{align}
where the factor \(\frac{1}{2}\) comes from the relation \eqref{GD-def} between \(G_D\) and \(\Ga\), $4$ is the symmetry factor,
$c=1/(4\pi^2)$ and $D(u,v)$ has a simple form when expressed in terms of
auxiliary variables defined in \eq{zbarz}
\begin{align}\label{Phi}
 D(u,v)={1\over z-\bar z}\left[2 \text{Li}_2(z)-2 \text{Li}_2({\bar z})+\ln \left(\frac{1-z}{1-{\bar z}}\right) \ln
   (z {\bar z})\right]\,.
\end{align}
The function inside the brackets is known as the Bloch-Wigner function.

The two-loop corrections $G^{(0,2)}$ and $G^{(2,0)}$ come from last two Feynman diagrams shown in Fig.~\ref{fig:G0loopsalpha}. The corresponding Feynman integrals are divergent and require regularization. In dimensional regularization with $d=4-2\epsilon$ we
have 
\begin{align}\notag
G^{(0,2)} =  c^6 I(x_1,x_3|x_2,x_4)
\,,\qqqquad
G^{(2,0)} = 4c^6  I(x_1,x_2|x_3,x_4)\,,
\end{align}
where the notation was introduced for
\begin{align}
I(x_1,x_2|x_3,x_4) =   \int {d^{4-2\epsilon} x_5 d^{4-2\epsilon} x_6 \over [x_{15}^2 x_{25}^2 x_{36}^2x_{46}^2 (x_{56}^2)^2]^{1-\epsilon}}\,.
\end{align}
The integral on the right-hand side has a UV divergence coming from integration at short distances $x_{56}^2\to 0$. 
Applying the identity $1/(x_{56}^2)^{2-2\epsilon} = \pi^2 \delta^{(4-2\epsilon)}(x_{56})/\epsilon+ O(\epsilon^0)$ we find that the
residue of $I(x_1,x_2|x_3,x_4)$ at the pole $1/\epsilon$ is proportional to the same one-loop integral that enters \re{G01}.
The same  is true for the function $I(x_1,x_3|x_2,x_4)$.  As a consequence, the divergent
part of two loop correction to \re{G-def1} takes the form   
\begin{align}
(16 \pi^2\xi^2)^2 G^{(0,2)} + (16 \pi^2\alpha_1^2)^2 G^{(2,0)} =  c^2{4\alpha_1^4+\xi^4\over\epsilon}{u D(u,v)\over x_{12}^2x_{34}^2}
+O(\epsilon^0)\,.
\end{align}
Since $\alpha_1^4=-\xi^4/4 + O(\xi^6)$ at the fixed point \re{fixed}, UV divergences cancel in the sum the single- and double-trace contributions. We conclude that the two-loop correction to \re{G-def1} is UV finite as it should be. 

Going through the calculation of a finite part of the two-loop contribution we find that it factorizes into a product of one-loop correction \re{G01} and a logarithm of the cross-ratio 
\begin{align}\la{born}
(16\pi^2\xi^2)^2 G^{(0,2)} + (16\pi^2\alpha_1^2)^2 G^{(2,0)} = {c^2\xi^4 \over 2x_{12}^2 x_{34}^2} u D(u,v)\ln u\,.
\end{align}
Combining this relation with \re{G01} and \re{born} we obtain that the correlation function \re{G-def1} takes the expected form \re{G4cg0} with 
the function $\mathcal G(u,v)$ given at two loops by
\begin{align}\label{calG-res2}
\mathcal  \Ga (u,v) = u - {i\xi^2}\, uD(u,v) + \xi^4 \, uD(u,v)\lr{\ft12\ln u-1} + O(\xi^6)\,.
\end{align}
Here we replaced the double-trace coupling by its value \re{fixed} at the fixed point $\alpha_1^2=\alpha_-^2$.\footnote{This choice is syncronized with the sign convention in \eq{xi2exp}.} For $\alpha_1^2=\alpha_+^2$, the function
$\mathcal \Ga (u,v)$ is given by the same expression with $\xi^2$ replaced with $-\xi^2$.

Notice that the $O(\xi^2)$ correction to \re{calG-res2} comes entirely from the double-trace contribution. In the next section we show that \re{calG-res2} is in a perfect agreement with the exact expression for $\Ga$, Eq.~\eq{G0sum}, which was obtained by
resumming the wheel graphs shown in Fig.~\ref{fig:G0loops}. This is in agreement with our expectations that the double-trace contribution to $\Ga$ can be ignored at finite coupling.

\subsection{Type $\Gb$}

The weak coupling expansion of the one-magnon correlation function $\Gb$ is defined by Feynman diagrams shown in Fig.~\ref{fig:G1loops}. 
The contribution of the first two diagrams is given by \eq{G10loop}.
In distinction from the previous case, the corresponding integrals are well-defined in $D=4$ dimensions and do not require regularization.

In particular, the one-loop correction $\Gb^{(1)}$ defined in \eq{G10loop}
can be expressed in terms of the ``cross" integral \eq{G01} 
\beqa
16\pi^2 \xi^2 \Gb^{(1)}&=& c^3 \xi^2 \frac{ u D(u,v)}{x_{12}^2x_{14}^2 x_{34}^2}=
c^3 \xi^2
\frac
{(x_{13}^2 x_{24}^2)^{1/2}}
{(x^2_{12}  x^2_{34})^{3/2}x^2_{14} } { u^{3/2} D(u,v)}
\;.
\eeqa
Then, the one-magnon correlation function $\Gb$ takes the expected form
\re{G4cg} with ${\cal G}_{\gb}(u,v)$ given at weak coupling by
\beq\la{G1pert}
{\cal G}_{\gb}(u,v)=u^{3/2}+\xi^2  u^{3/2} D(u,v)+{\cal O}(\xi^4)\;.
\eeq
In the next section, we reproduce this expansion from the exact expression \eq{calG1} for $\Gb$ and also produce explicit expressions for higher order terms.

\subsection{Type $\Gc$}

The two-magnon correlation function $\Gc$ is defined by Feynman diagrams shown in Fig.~\ref{fig:G2loops}. 
Like in the previous case, the corresponding integrals are well-defined in $D=4$ dimensions and do not require regularization.

The weak coupling expansion of $\Gc$ runs in powers of $\xi^4$ and first two terms are given by \eq{diagsG2}. 
As was mentioned before,  
the $O(\xi^4)$ contribution to $\Gc$ is given by a two-loop Feynman integral \re{diagsG2} that factorizes into the product of one-loop 
integrals. The latter take the form \re{G01} and, as a consequence, it can be expressed in terms of  Bloch-Wigner function 
\beqa\la{diagsG21}
(16\pi^2\xi^2)^2 \Gc^{(1)}=\frac{\xi^4}{x_{12}^4 x_{34}^4}\left[ {c^2}  u D(u,v) \right]^2
\;.
\eeqa
The resulting expression for the two-magnon correlation function takes the expected form \re{G2Ggen} with 
\begin{align}\label{calG-B}
\mathcal \Gc(u,v) ={u^2\over v} +   \xi^4u^2 D^2(u,v) + O(\xi^8)\,.
\end{align}
We managed to reproduce both terms of the expansion \eq{calG-B} from its expansion over conformal blocks \eq{GB-sum} by expanding it at weak coupling. For that we followed the procedure explained in the next section \ref{sec:PT}.

\section{Prediction for the 4-point correlation functions at weak coupling from OPE data}\la{sec:PT}

In section  \ref{sec:GBO}, we determined the conformal data of the operators that appear in the conformal partial wave expansion \eq{Gres} of the correlation functions at finite coupling. This expansion  
 takes the form of double infinite sums over spins and dimensions. 
It is not obvious a priori that these sums can be
evaluated in a closed form. We demonstrate in this section that, at weak coupling, 
these sums can be computed order by order in perturbation theory.   
For zero- and one-magnon functions, $\Ga$ and $\Gb$, respectively, the result can be written in terms of special class of iterated integrals known as
single valued harmonic polylogarithm functions (SVHPL) \cite{Brown:2004ugm,Dixon:2012yy}. The two-magnon function $\Gc$ has a more complicated form and it can be expressed in terms of elliptic polylogarithms.

\subsection{Zero-magnon case (Type $\Ga$)} 
As we already emphasized before, an interesting fact about $\Ga$ is that it receives corrections from the double trace interaction. 
As a result, its expansion goes in powers of $\xi^2$ rather than in $\xi^4$. 

In the previous section we computed the first three terms of the weak-coupling expansion of the function $\mathcal \Ga$ defined in \eq{G0sum}. Taking into account \eq{calG-res2} as well as the explicit expression \eq{Phi} for the function $D(u,v)$, it is natural to look for a general expression for $\mathcal \Ga$ in the form
\beq\label{sum-G}
{\cal G}_{\ga}=\frac{z\bar z}{z-\bar z}\sum_{n=0}^\infty (i\xi^2)^n {\cal G}_{\ga}^{(n)}(z,\bar z)\;,
\eeq
where $z$ and $\bar z$ are defined in \re{zbarz}.
The goal of this section is to compute ${\cal G}_{\ga}^{(n)}$ explicitly starting from the OPE expansion \eq{G0sum}.

Note that the dependence on the coupling constant $\xi^2$ enters into \eq{G0sum} only through the scaling dimensions $\Delta_2(S)$ and $\Delta_4(S)$ given by \eq{D0}. 
For general $S$ their weak coupling expansion only involves powers of $\xi^4$.
We recall  however, that, due to non-comutativety of the limits $\xi\to 0$ and $S\to 0$, for $S=0$ the weak coupling expansion of $\Delta_2(S=0)$ does contain powers $\xi^2$ powers. This means that all terms on the right-hand side of \re{sum-G} with odd powers of $\xi^2$ come entirely from the contribution of the
twist-2 operator with zero spin. 
We would like to emphasize that the scaling dimension of this operator  has to
satisfy the condition ${\rm Re}\;\Delta_2(0)>2$. Together with \eq{xi2exp}, 
this implies that $\xi^2$ should have a nonzero imaginary part ${\rm Im}\, \xi^2<0$. For ${\rm Im}\, \xi^2>0$ we have to exchange the operator with its shadow
whose scaling dimension is given by $4-\Delta_2(S)$ and it can be obtained from
\eq{xi2exp} through the transformation $\xi^2\to - \xi^2$.
This ambiguity exactly corresponds to the ambiguity of choosing the fixed point
\re{fixed}. In what follows we assume that ${\rm Im}\, \xi^2<0$ and apply \eq{xi2exp}.

In order to find the explicit expressions for the coefficient functions ${\cal G}_{\ga}^{(n)}(z,\bar z)$ we match \re{sum-G} into the OPE expansion \eq{G0sum}
in the short distance limit, $u\to 0$ and $v\to 1$, or equivalently $z\,,\bar z\to 0$. In this limit, the conformal blocks scale as $g_{\Delta,S}(u,v)\sim (z\bar z)^{(\Delta-S)/2}$ and their expansion in powers of the coupling generates terms 
of the form $z^n \bar z^m \log^k (z\bar z)$ with $k$ not exceeding the order in
the coupling. The small $z\,,\bar z$ expansion of ${\cal G}_{\ga}^{(n)}(z,\bar z)$
involves the same terms and their coefficients can be computed for any finite $n+m$
using the exact expressions for the conformal data of the operators. 

A nontrivial property of ${\cal G}_{\ga}^{(n)}(z,\bar z)$ is that for $n\ge 1$ they can be expanded over the basis of special iterated integrals, the so-called harmonic polylogarithms (HPL) (see \cite{Remiddi:1999ew,Maitre:2005uu}), schematically
\begin{align}\label{G-HPL0}
{\cal G}_{\ga}^{(n)}(z,\bar z) = \sum_{\vec{a} ,\vec{b}} C_{ \vec{a} ,\vec{b}} \,  H_{a_1,a_2,\dots}(z) H_{b_1,b_2,\dots}(\bar z)\,,
\end{align}
where the sum runs over the two sets of indices (including empty sets) $\vec a= (a_1,a_2,\dots)$ and $\vec b= (b_1,b_2,\dots)$ with $a_i$ and $b_j$ taking  the values $\{0,1\}$. Most importantly, the number of terms on the right-hand side of \re{G-HPL} is finite
for any $n$ thus allowing us to find the expansion coefficients $C_{ \vec{a} ,\vec{b}}$ by matching the small $z,\bar z$ expansion of ${\cal G}_{\ga}^{(n)}(z,\bar z)$ into the corresponding expansion of the basis of HPL functions. Namely, comparing the coefficients in front of $z^n \bar z^m \log^k (z\bar z)$ terms on the both sides of \re{G-HPL0} with sufficiently large $n+m$ we obtain an overdetermined system
of linear equations for $C_{ \vec{a} ,\vec{b}}$. It is remarkable that for any finite $n$ this system has a unique solution. 

For the first few coefficient functions we find
\begin{align}\notag\label{G-HPL}
\mathcal G_{\ga}^{(0)}&= z-\bar z
\\ \notag
\mathcal G_{\ga}^{(1)}&= 
H_{1,0}-\bar{H}_{1,0}+H_1 \bar{H}_0-H_0 \bar{H}_1+\bar{H}_{0,1}-H_{0,1}
\\[1mm]\notag 
\mathcal G_{\ga}^{(2)}&= H_1 \bar{H}_{0,0}-\bar{H}_1 H_{0,0}+\bar{H}_0 H_{1,0}
-H_0\bar{H}_{1,0}
+\bar{H}_{1,0}-H_{1,0}
-\bar{H}_{1,0,0}+H_{1,0,0}
\\
& -H_1 \bar{H}_0+H_0
\bar{H}_1-\bar{H}_{0,1}+\bar{H}_{0,0,1}+H_{0,1}-H_{0,0,1}
 \,,
\end{align}
where we introduced a short-hand notation for $H_{a_1,a_2,\dots}=H_{a_1,a_2,\dots}(z)$ and $\bar H_{a_1,a_2,\dots}=H_{a_1,a_2,\dots}(\bar z)$. 
The same expressions can be rewritten in terms of classical (di)logarithm functions
as
\\\beqa\notag
{\cal G}_{\ga}^{(1)}&=&-2 \text{Li}_2(z)+2 \text{Li}_2\left(\bar{z}\right)-\log \left(\frac{1-z}{1-\bar{z}}\right) \log
\left(z \bar{z}\right)\\
{\cal G}_{\ga}^{(2)}&=&-\frac{1}{2} \left(\log \left(z \bar{z}\right)-2\right) \left(2 \text{Li}_2(z)-2
\text{Li}_2\left(\bar{z}\right)+\log \left(\frac{1-z}{1-\bar{z}}\right) \log
\left(z \bar{z}\right)\right)\,.
\eeqa
We verify that these expressions coincide with the result of the two-loop
calculation \eq{calG-res2}.

It is convenient to assign to each term on the right-hand side of \re{G-HPL0} and \re{G-HPL} 
the weight equal to the total length of the sets $\vec a \cup \vec b$. Then, 
${\cal G}_{\ga}^{(1)}$ is given by a linear combination of weight$-2$ functions
whereas expansion of ${\cal G}_{\ga}^{(2)}$ contains both weight$-2$ and weight$-3$ functions. 
We see that the maximal weight of  increases by one at each loop order.
Correspondingly, the dimension of the basis of the functions increases rapidly at higher loops. The number of terms in the expression for $\mathcal G_\ga^{(n)} $ at $n$ loops is given by:
\begin{table}[h!] 
\centering
 \begin{tabular}{llllll}
\hline
\multicolumn{1}{c}{\text{loops}} & \multicolumn{1}{c}{3} & \multicolumn{1}{c}{4} & \multicolumn{1}{c}{5} & \multicolumn{1}{c}{6} & \multicolumn{1}{c}{7} 
\\ \hline
\multicolumn{1}{c}{\text{dimension}} &\multicolumn{1}{c}{38} & \multicolumn{1}{c}{48} & \multicolumn{1}{c}{154} & \multicolumn{1}{c}{244} & \multicolumn{1}{c}{508} 
\\ \hline
\end{tabular}
\end{table}

\noindent
The resulting expressions for $\mathcal G_\ga^{(n)}$ in terms of HPLs are rather lengthy and we do not present them here.

The coefficient functions $\mathcal G^{(n)}(z,\bar z)$ should be single-valued functions of $z$ for $\bar z=z^*$. This property is not obvious from \re{G-HPL0} since HPL
functions have, in general, branch cuts that start at $z=0$ and $z=1$.
We can use this information to restrict the basis of possible functions. The 
HPLs can enter \re{G-HPL} through special linear combinations which are free from the branch cuts.  They 
are known as single-valued harmonic polylogarithms $\mathcal L_{a_1,a_2,\dots}$,  their definition can be found in \cite{Brown:2004ugm,Dixon:2012yy}. The resulting expressions for $\mathcal G_{\ga}^{(n)}$  in terms of SVHPLs are
\footnote{We attach to the arXiv submission an auxilary file containing the expressions for SVHPLs of the weight up to $8$ in terms of the HPLs.}
\beqa
{\cal G}_{\ga}^{(1)}&=&-( \mathcal{L}_{01}- \mathcal{L}_{10})\\
\nn{\cal G}_{\ga}^{(2)}&=&
 ( \mathcal{L}_{0 1}- \mathcal{L}_{1 0})- \mathcal{L}_{0 0 1}+ \mathcal{L}_{1 0 0}\\
\nn{\cal G}_{\ga}^{(3)}&=&
\frac{3}{2}\left(  \mathcal{L}_{0 1}- \mathcal{L}_{1 0}\right)
- \mathcal{L}_{0 0 0 1}
+ \mathcal{L}_{0 0 1 0}
- \mathcal{L}_{0 1 0 0}
- \mathcal{L}_{0 1 0 1}
+ \mathcal{L}_{1 0 0 0}
+ \mathcal{L}_{1 0 1 0}+4
\zeta _3  \mathcal{L}_1\\
\nn{\cal G}_{\ga}^{(4)}&=&
(3-4\zeta_3)  (\mathcal{L}_{0 1}- \mathcal{L}_{1 0})
- \mathcal{L}_{0 0 0 0 1}+ \mathcal{L}_{0 0 0 1 0}- \mathcal{L}_{0 1 0 0
	 0}- \mathcal{L}_{0 1 0 0 1}+ \mathcal{L}_{1 0 0 0 0}+ \mathcal{L}_{1 0 0 1 0}\\
\nn{\cal G}_{\ga}^{(5)}&=&
\frac{49}{8} 
\left(\mathcal{L}_{0 1}-\mathcal{L}_{1 0}\right)
+\frac{1}{2} 
\left(\mathcal{L}_{0 0 0 1}-\mathcal{L}_{0 0 1 0}+\mathcal{L}_{0 1 0 0}+\mathcal{L}_{0 1 0 1}-\mathcal{L}_{1 0 0 0}-\mathcal{L}_{1 0 1 0}\right)-2 \left(\zeta _3-6 \zeta
_5\right)  \mathcal{L}_1\\
\nn &+&4
\zeta _3 
\left(\mathcal{L}_{0 0 1}-\mathcal{L}_{0 1 0}+\mathcal{L}_{1 0 0}+\mathcal
{L}_{1 0 1}\right)
-\mathcal{L}_{0 0 0 0 0 1}+\mathcal{L}_{0 0 0 0 1 0}-\mathcal{L}_{0 0
	 0 1 0 0}-\mathcal{L}_{0 0 0 1 0 1}+\mathcal{L}_{0 0 1 0 0 0}\\
\nn &+&\mathcal{L}_{
	0 0 1 0 1 0}-\mathcal{L}_{0 1 0 0 0 0}-\mathcal{L}_{0 1 0 0 0 1}-\mathcal{L}_{0 1 0 1 0 0}-\mathcal{L}_{0 1 0 1 0 1}+\mathcal{L}_{1 0 0 0 0 0}+\mathcal{L}_{1 0 0 0 1 0}+\mathcal{L}_{1 0 1 0 0 0}+\mathcal{L}_{1 0 1 0 1 0}\;.
\eeqa
Similar expressions up to  $7$ loops can be found in a Mathematica file attached to this submission.

\subsection{One-magnon case (Type $\Gb$)}

Going along the same lines as in the previous case, we were able to expand the sum \eq{calG1} in terms of the SVHPLs. The weak-coupling result \eq{G1pert} 
suggests to look for ${\cal G}_{\gb}$ in the form
\beq
{\cal G}_{\gb}=\frac{(z\bar z)^{3/2}}{z-\bar z}\sum_{n=0}^\infty \xi^{2n}{\cal G}_{\gb}^{(n)}\;.
\eeq
The explicit expressions for the first few coefficient functions are
\beqa\notag
&&{\cal G}_{\gb}^{(0)}=z-\bar z
\\\notag
&&{\cal G}_{\gb}^{(1)}=-{\cal L}_{10}+{\cal L}_{01}
\\\notag
\la{Gcg2}&&{\cal G}_{\gb}^{(2)}={\cal L}_{0100}-{\cal L}_{0010}
\\
\nn&&{\cal G}_{\gb}^{(3)}=
+{\cal L}_{0 0 0 1 0 0}
+{\cal L}_{0 0 0 1 1 0}
-{\cal L}_{0 0 1 0 0 0}
-{\cal L}_{0 0 1 0 1 0}
+{\cal L}_{0
        1 0 1 0 0}
+{\cal L}_{0 1 0 1 1 0}
-{\cal L}_{0 1 1 0 0 0}
-{\cal L}_{0 1 1 0 1 0}
-4 \zeta _3 ({\cal L}_{001}+   {\cal L}_{0 1 1})\\
\nn&&{\cal G}_{\gb}^{(4)}=
- {\cal L}_{0 0 0 
        0 1 0 0 0}- {\cal L}_{0 0 0 0 1 0 1 0}+ {\cal L}_{0 0 0 1 0 0 0 0}+ {\cal L}_{0 0 0 1 0 0 1 0}- {\cal L}_{0 1 
        0 0 1 0 0 0}- {\cal L}_{0 1 0 0 1 0 1 0}+ {\cal L}_{0 1 0 1 0 0 0 0}+ {\cal L}_{0 1 0 1 0 0 1 0}\\
 &&-4 \zeta _3
\left( {\cal L}_{0 0 0 0 1}- {\cal L}_{0 0 0 1 0}+ {\cal L}_{0 1 0 0 1}- {\cal L}_{0 1 0 1 0}\right)\;.
\eeqa
We verify that the first two terms reproduce correctly the perturbation theory result \eq{G1pert}.

\subsection{Two-magnon case (Type $\Gc$)}
In this case we only managed to reproduce the tree level and $2$-loop perturbation theory result \eq{calG-B}.
We found that it is not possilbe to express the $4$-loop expression in terms of SVHPLs. We found that it is given by elliptic function. In small $z$ limit the underlying elliptic curve degenerates and the correlator can be written in terms of HPL's of $\bar z$ multiplied by powers and logarithms of $z$.
\footnote{We would like to thank F.~Aprile, J.~Bourjaily, J.~Drummond, P.~Heslop and O.~Gurdogan for very useful discussion on related issue.}

\section{Classical (Strong Coupling) Limit of the 4-point Correlators}\la{sec:strong}
In this section, we investigate the strong coupling limit of the 4-point correlators.
Even though the world-sheet description of this theory is still not known, it was shown  in \cite{Gromov:2017cja}  that there is a classical limit of the underlying integrability construction for the spectrum where
it reduces to an algebraic curve reminiscent of that of the classical strings in $AdS^5\times S_5$ in the full theory. 
Similarly, we will see that the leading strong coupling  \(\xi\to\infty\)\ asymptotics of the correlation function 
is saturated by one state with large \(\Delta,S\sim\xi \) and the corresponding result scales as \(e^{-\xi A(z,\bar z)}\), where \(A(z,\bar z)\) is a certain function of cross-ratios. 
Moreover, since \(\xi=g e^{-i\gamma_3/2}\) this classical asymptotics reminds the behavior of the three point correlation functions for short operators  in 
strongly coupled  \({\cal N}=4\)~SYM theory in planar limit (see for example \cite{Gromov:2011jh,Kazama:2012is}).
Whereas it would be premature to conclude that this points to the dual string description for the bi-scalar model, the observed behaviour looks very much like an action evaluated on some classical solution. 
All that constitutes an evidence towards existence of the classical limit of the fishnet theory at strong coupling \(\xi\to\infty\). 
The possible relation to the recently proposed AdS sigma model description of the ground state of the fishnet theory for infinite $L$ by \cite{Basso:2018agi} is also yet to be understood.
In this section we assume that $z$ takes a generic value in the range $0<|z|<1$\footnote{As below we find square roots and logariphms for definiteness we also assume in the calculation that $0<\arg z<\pi/4$.}. The convergence is not uniform and the limits $z\to 0$ or $z\to 1$ have to be taken with extra care. We also assume that $\xi=e^{-i\phi}\xi_0$ where $\xi_0$ is large and real and $\phi$ is a small positive phase. This assumption is necessary as the correlator has poles at real values of $\xi$ which accumulate at infinity and the limit is not defined.

\subsection{Correlation function for zero-magnon case (wheel-graphs)}

First we consider the correlation function $G_{D}$. It is obtained from the zero-magnon function $\Ga$ given by \eq{G0sum} via \eq{GD-def}. As we discussed above, this results in  dropping terms with odd $S$ in the sum \eq{G0sum} and doubling the terms with even $S$. In this subsection, we compute \eq{G0sum} in the limit when $\xi\to\infty$. 

Our main assumption, which is backed by intensive numerical analysis, is that for $\xi\to\infty$ the sum in  \eq{G0sum} is saturated by large spins $S\sim \xi$.
Then, replacing  the conformal blocks in \eq{G0sum} by their 
asymptotic behavior at large $S$ and $\Delta$, we can
evaluate the sum over $S$ by the saddle-point method. In what follows  we only evaluate the leading exponential factor. The pre-exponent and the subleading terms can be computed by the same method, we leave it to future studies.

Before we begin we notice the following property of the structure constant \re{C0}
\beq
C_{S,\Delta}=-C_{-2-S,\Delta}\,.
\eeq
It allows us to write the 4-point function \eq{G0sum} in the following way
\beqa\label{G0sumway}
{\cal G}_{\ga}(u,v)=
\sum_{\Delta=\Delta_{t=2},\Delta_{t=4}}
\sum_{S=-\infty}^\infty
C_{\Delta,S}\;
(-1)^S \frac{z\bar  z}{
 z - \bar z}  k(\Delta + S, z) k(\Delta - S - 2, \bar z)
\,,
\eeqa
where $k(\Delta,z)$ is given by the hypergeometric function in \eq{defk}. This expression considerably simplifies our analysis as it allows to replace the sum over $S$ by an integral with exponential precision at large $\xi$ and the evaluate it by the saddle point method.

The asymptotic behavior  of the conformal block can be found
by using a series representation of the hypergeometric functions in \eq{defk} 
\beq
\, _2F_1\left(\frac{\Delta +S}{2},\frac{\Delta +S}{2};\Delta +S;z\right)=\sum_{k=0}^\infty
\frac{z^k 2^{\Delta +S-1} \Gamma \left(\frac{S}{2}+\frac{\Delta
        }{2}+\frac{1}{2}\right) \Gamma \left(k+\frac{S}{2}+\frac{\Delta
        }{2}\right)^2}{\sqrt{\pi } \Gamma (k+1) \Gamma \left(\frac{S}{2}+\frac{\Delta
        }{2}\right) \Gamma (k+S+\Delta )}\,.
\eeq
Then rescaling the variables as
\beq\label{scal}
\Delta= \xi\; d\,,\qquad S=\xi\; s\,,\qquad k=\xi\; w\,,
\eeq
we expand the expression under the sum at large $\xi$ and extremize in $w$ to obtain
\beq
\, _2F_1\left(\frac{\Delta +S}{2},\frac{\Delta +S}{2};\Delta +S;z\right)\sim \exp\left[ \frac{\Delta+S}{2} \log \left(\frac{4}{z}\frac{1-\sqrt{1-z}}
{1+\sqrt{1-z}}\right)\right]\;,
\eeq 
 in agreement with \cite{Hogervorst:2013sma}. 
 
Similarly we replace the structure constants \re{C0} by their leading asymptotic behaviour at large $S$ and $\Delta$ to get from \eq{G0sumway}
and \eq{GD-def}
\beq\label{GD-sum}
{\cal G}_{D}(z,\bar z)\sim \sum_{\Delta,S}\exp\left[
\frac{\Delta}{2}  \log \left(\frac{\left(\sqrt{1-z}-1\right)
        \left(\sqrt{1-\bar{z}}-1\right)}{\left(\sqrt{1-z}+1\right)
        \left(\sqrt{1-\bar{z}}+1\right)}\right)+\frac{S}{2}  \log
\left(\frac{\left(\sqrt{1-z}-1\right)
        \left(\sqrt{1-\bar{z}}+1\right)}{\left(\sqrt{1-z}+1\right)
        \left(\sqrt{1-\bar{z}}-1\right)}\right)
\right]\;.
\eeq
where the sum runs over even spins $S$, both positive and negative.
We recall that for each $S$ there are only two values of $\Delta$ that contribute to the sum, $\Delta_2(S)$ and $\Delta_4(S)$, given by \re{D0}. At strong coupling, we apply \re{scal} to find that they have different dependence on the spin
\beqa\label{delta's}
\Delta_{t=2}\simeq \xi  \sqrt{s^2-4}\,,\qquad\quad
\Delta_{t=4}\simeq \xi  \sqrt{s^2+4}\;.
\eeqa
Note that for $s<2$ we get purely imaginary $\Delta_{t=2}/\xi$. In order to get a consistent strong coupling limit we should take $\xi$ to have slightly negative phase, so that ${\rm Re}\;\Delta_{t=2}\to+\infty$ at strong coupling.
After that we substitute \re{scal} into \re{GD-sum} and
extremize the expression in the exponent over $s$ to get 
\beq\la{Gcl}
{\cal G}_{D}(z,\bar z)\sim e^{-\xi A_{t=2}(z,\bar z)}+e^{-\xi A_{t=4}(z,\bar z)}\;,
\eeq
where the two states \re{delta's} produce two different exponents
\beq\la{A2A4}
A_{t=2}=i A_{t=4}\,,\qquad\quad
A_{t=4}(z,\bar z)\equiv -i\sqrt{\log
        \left(\frac{\sqrt{1-z}-1}{\sqrt{1-z}+1}\right)^2
        \log
        \left(\frac{\sqrt{1-\bar z}+1}{\sqrt{1-\bar z}-1}\right)^2}\;,
\eeq
where $\bar z=z^*$.
Notice that $A_{t=4}$ is positive whereas $A_{t=2}$ is purely imaginary.
As a consequence, $-\xi A_{t=2}$ has a large negative real part 
and, therefore, the correlation function \re{A2A4} is exponentially suppressed at strong coupling,
as it is expected for the tunneling processes in the classical limit.

Let us also point out that the state which saturates the sum over $S$ in the sum \re{GD-sum} has the following scaling dimension and spin
\begin{align}\la{de}\notag
& \Delta^{cl}_{t}=\pm\frac{2\xi}{A_{t}}  \log \left(\frac{\left(\sqrt{1-z}-1\right)
        \left(\sqrt{1-\bar{z}}-1\right)}{\left(\sqrt{1-z}+1\right)
        \left(\sqrt{1-\bar{z}}+1\right)}\right) \,,
\\
& S^{cl}_{t}=
\mp\frac{2\xi}{A_{t}}  \log \left(\frac{\left(\sqrt{1-z}-1\right)
        \left(\sqrt{1-\bar{z}}+1\right)}{\left(\sqrt{1-z}+1\right)
        \left(\sqrt{1-\bar{z}}-1\right)}\right)\,,
\end{align}
with the upper sign for $t=2$ and lower for $t=4$.
It would be interesting to find a classical model which  reproduces these results.\footnote{In analogy with $sl(2)$ spin chain we expect that to be some variation of the classical Toda spin chain.}

As a test of our result we consider the small $z$ limit of \re{Gcl}.
In this limit,  we expect that the leading contribution to the correlation function should come from the states with $S=0$ leading to
${\cal G}_{D}(z,\bar z)\sim (z\bar z)^{\Delta_t(S=0)/2}$.  
Expanding the relations \eq{A2A4} and \eq{de} at small $z$ we find
\begin{align}\notag
& e^{-\xi A_4}\simeq \left(\frac{z\bar z}{16}\right)^\xi\,, && \hspace*{-20mm} \Delta_4^{cl}\simeq 2\xi\,, && \hspace*{-20mm}  S_4^{cl}\simeq 0
\\
& e^{-\xi A_2}\simeq \left(\frac{z\bar z}{16}\right)^{i\xi}\,, && \hspace*{-20mm}  \Delta_2^{cl}\simeq 2i\xi\,, &&\hspace*{-20mm}   S_2^{cl}\simeq 0\;,
\end{align}
in a perfect agreement with  our expectation.

To further test our result, we computed the 4-point function \re{GD-sum} numerically for  $\xi=n e^{-i\pi/6}$ (with $n=100,105,110\dots,200$) and $z=e^{i\pi/5}/8$. Fitting the data
we obtained the following result for the contribution of states with $\Delta_{t=2}$ and $\Delta_{t=4}$ 
\beqa\label{pres}\nn
{\rm Re}\log{\cal G}_{D}\Big|_{t=2}&=&
-3.479085434\; n+(0.5000000000 \log
   n-4.602673564)-\frac{0.3908689558}{n}+\dots\\
{\rm Re}\log{\cal G}_{D}\Big|_{t=4}&=&
-6.025952735\; n+(0.5000000000 \log
   n-4.602673564)-\frac{0.6770048906}{n}+\dots\;.
\eeqa
The leading term on the right-hand side agrees with analytic result
\re{A2A4} for ${\rm Re}(-\xi A_{t=2}(z,\bar z))$
and ${\rm Re}(-\xi A_{t=4}(z,\bar z))$
up to $27$ significant digits (i.e. within the fit precision~\footnote{For presentation purposes we lowered the precision in \re{pres}}). The relation \re{pres} clearly indicates that the first subleading correction to $\log{\cal G}_{D}$ is the same for the two states. It scales as $\ln \xi/2$ and generates $\sqrt\xi$ in pre-exponents on the right-hand side of \re{Gcl}. This factor
could come from some zero modes in the semiclassical analysis (see very similar prefactors in the expectation values of a circular Wilson loop in ${\cal N}=4$ SYM  \cite{Medina-Rincon:2018wjs}). 

\subsection{Correlation function for one-magnon case (single spiral graphs) }

The strong coupling limit of the correlation function $\Gb$ happens to be very similar to the previous case. We immediately observe using \re{D1} that the states contributing to $\Gb$ have exactly the same strong coupling asymptotics as $t=2$ states in $\Ga$ from the previous section, namely, for even $S$ we have
\beq\la{Deltaxi}
\Delta\simeq \xi\sqrt{s^2-4}
\eeq
where $s=S/\xi$ and the odd $S$ have to be considered separately and their contribution can be obtained by replacing $\xi$ by $-i\xi$.
From \eq{Deltaxi}, it is not surprising that the leading asymptotics of the $4-$point correlator appears to be also the same as that of  ${\cal G}_D$, Eq.~(\ref{Gcl})
\beq\label{pre}
{\cal G}_{\gb}(z,\bar z) \sim e^{-\xi A_{t=2}(z,
        \bar z)}+e^{-\xi A_{t=4}(z,
        \bar z)}\;.
\eeq
where $A_{t=2}$ comes from even spins and $A_{t=4}$ from odd spins.
Note, however,
that  the pre-exponential factors in (\ref{Gcl}) and (\ref{pre}) are different.

To verify our result we computed ${\cal G}_{\gb}(z,\bar z)$ numerically for fixed values of $\xi=n e^{-i\pi/6},\;n=100,105,110\dots,200$ and for fixed $z=\frac{1}{8}e^{i\pi/5}$. Fitting this data
with $n,\log n,1,1/n,1/n^2,\dots,1/n^{18}$ we obtain the following result
\beqa
\left.{\rm Re}\log{\cal G}_{\gb}\right|_{{\rm even\; }S}&=&
-3.479085434\; n+(0.5 \log
   n-4.705957686)+\frac{0.2721446760}{n}+\dots\\
\left.{\rm Re}\log{\cal G}_{\gb}\right|_{{\rm odd\; }S}&=&
-6.025952735\; n+(0.5 \log
   n-4.705957686)+\frac{0.4713684059}{n}+\dots
\eeqa
The leading coefficients agrees with our analytic result $-{\rm Re}(e^{-i\pi/6}A_{t=2}(z,\bar z))$ and $-{\rm Re}(e^{-i\pi/6}A_{t=4}(z,\bar z))$ with $27$ digits (i.e. within the fit precision). 

\centerline{*\qquad*\qquad *}

The analysis of the strong coupling behavior of the two-magnon correlator (double spiral graphs) is more complicated and we leave it for the future studies. 
It would be interesting to guess the  ``world-sheet" degrees of freedom leading to these classical asymptotics, similarly to what was done \(AdS^5\times S_5\) in the full \(AdS_5\times S^5\) duality.  A possible way to further elucidate whether the bi-scalar model has a string dual description is probably to study the classical behavior of the long operators with many magnons.  The  finite gap description of these states in the model can help to identify the string degrees of freedom, in analogy with the finite gap construction    for the full \(AdS_5\times S^5\) string~\cite{Kazakov2004a,Beisert2005,Kazakov2004}.

\section{Gluing triangles: general structure constants}\la{sec:gt}
\begin{figure}[t!]
	\begin{center}
		\includegraphics[scale=0.5]{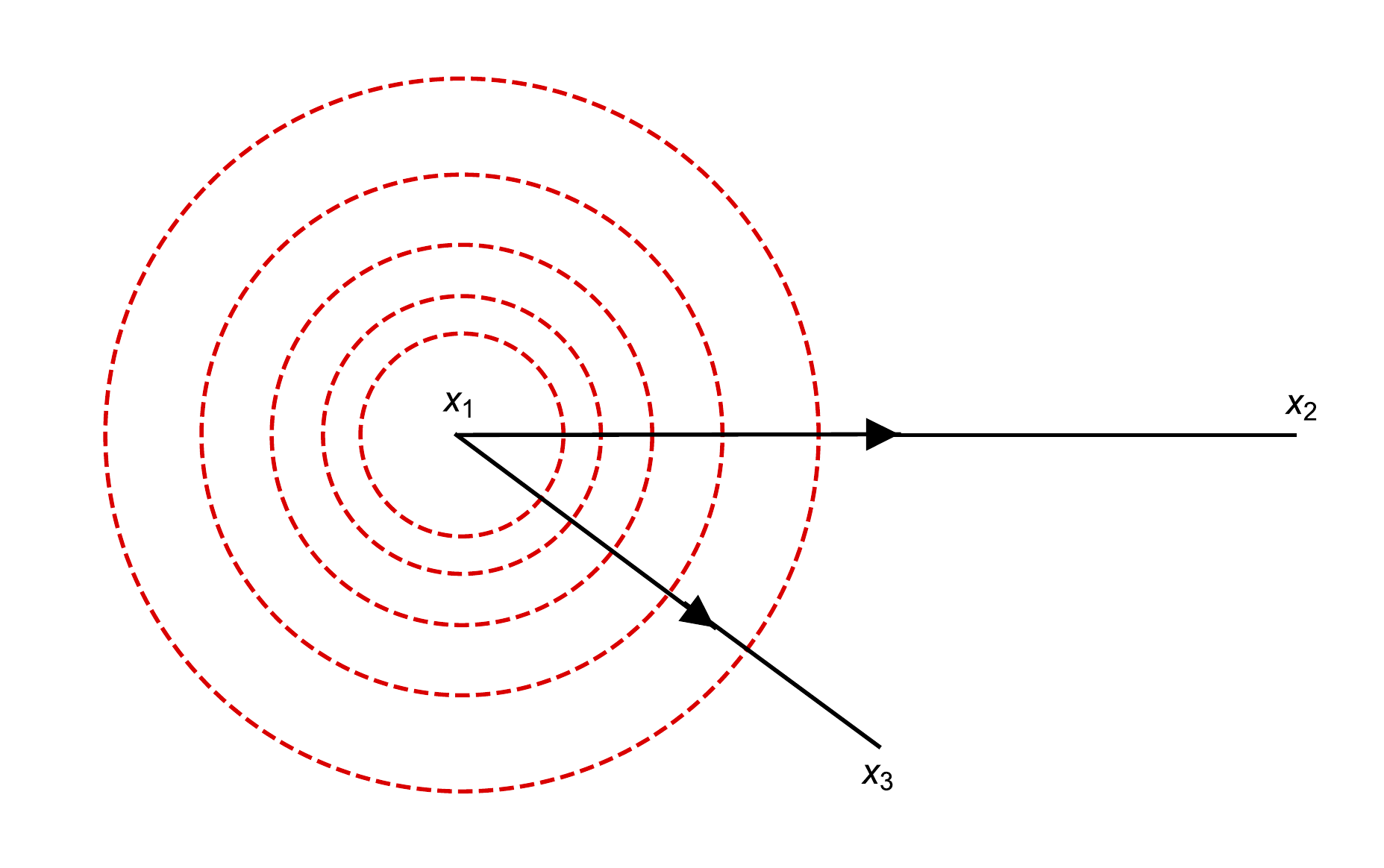}
	\end{center}
	\caption{\la{fig:gluing0}Three-point correlation function of one unprotected operator located at point $x_1$ and two protected operators at points $x_2$ and $x_3$. }
\end{figure}	

In the previous section, we computed the OPE coefficients $C_{\Delta,S}^{\bullet\circ\circ}$ for the
two protected and one unprotected zero-magnon operator, see  \eq{CG0}. 
In this section we generalize these results to the OPE coefficients of one protected and two unprotected zero-magnon operators $C_{\Delta_1,S_1,\Delta_2,S_2}^{\bullet\bullet\circ}$. 

We show below that these OPE coefficients can be easily obtained by gluing together the three-point correlation functions defined by $C_{\Delta,S}^{\bullet\circ\circ}$, schematically
\footnote{The factor $1+(-1)^S$ on the r.h.s. comes from symmetrization in $x_2$ and $x_3$.}
\beq\la{corel0}
\langle 
{\cal O}_S^{n}(x_1)
\tr[\bar X(x_2)\bar X(x_3)]\rangle
=C_{\Delta,S}^{\bullet\circ\circ}\frac{(1+(-1)^S) c^2 }{x_{12}^{\Delta-S} x_{13}^{\Delta-S} x_{23}^{2-\Delta+S}} 
\left(
{2(n\cdot x_{13})\over x_{13}^2}
-
{2( n\cdot x_{12})\over x_{12}^2}
\right)^S
\eeq
where $n$ is an auxiliary light-like vector and
\beq\label{O-n}
{\cal O}_S^{n}(x)={\cal N}\Tr\left[X(x)(n\cdot\partial)^S X(x)+\dots\right]\,,
\eeq
with dots denoting terms with derivatives acting on both fields. As usual, the normalization factor ${\cal N}$ is fixed by the two-point function, ${\cal N}^2\sim 1/\vev{{\cal O}_S^{n}\bar{\cal O}_S^{n}}$.
The stucture constant $C_{\Delta,S}^{\bullet\circ\circ}$ is given by \eq{C0} and \eq{CG0} and the scaling dimension $\Delta=\Delta_2(S)$ is defined in \re{D0}.

\begin{figure}[t!]
	\begin{center}
		\includegraphics[scale=0.5]{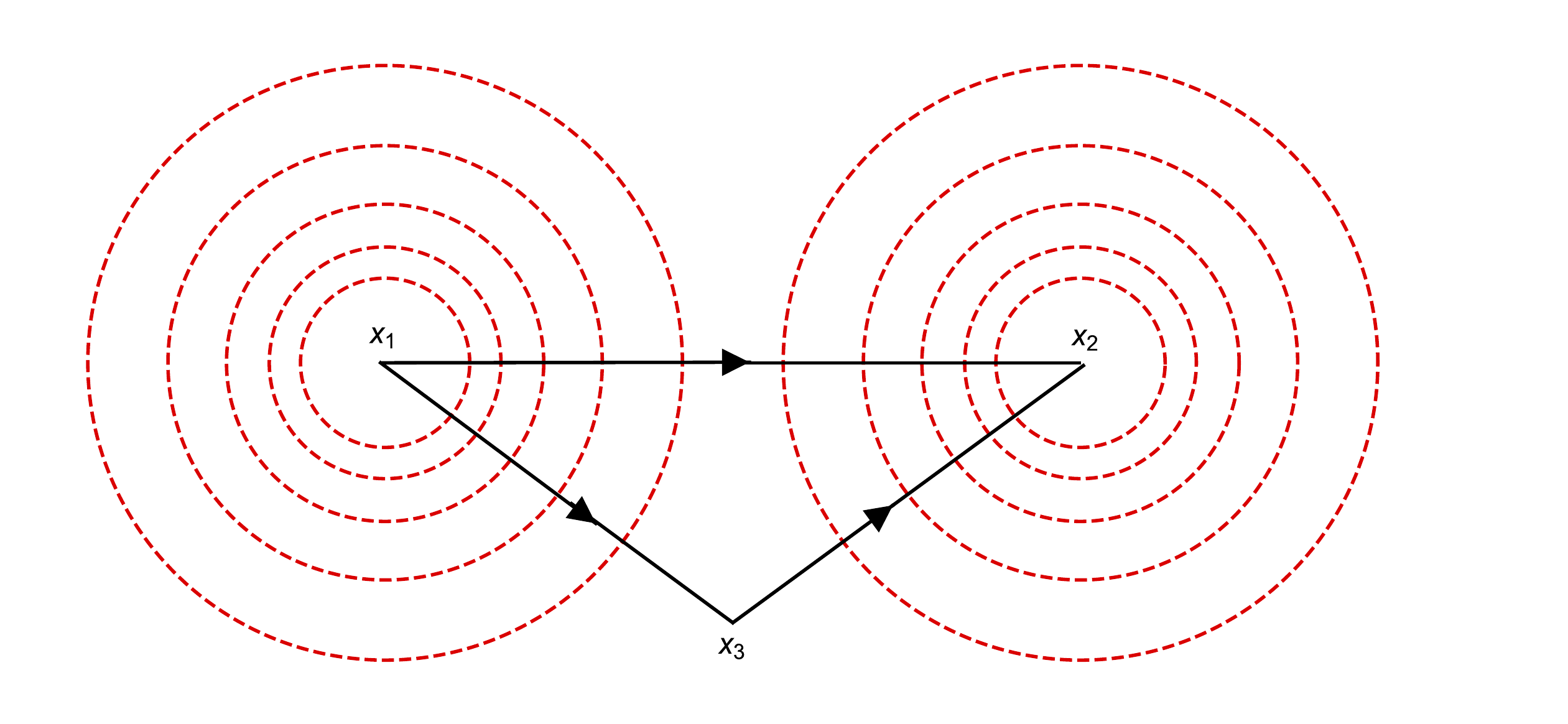}
	\end{center}
\caption{\la{fig:gluing} Three-point correlation function of two unprotected   twist-2 operators and one protected operator ${\rm tr}(X\bar X)$. It can be obtained by gluing together 
two wheel diagrams shown in Fig.~\ref{fig:gluing0} through the common external points.
}
\label{fig:joinedtriangle}\end{figure}

The three-point function \re{corel0} resums the wheel diagrams shown in Fig.\ref{fig:gluing0}. They contain two scalar lines connecting $x_1$ with the two external
points $x_2$ and $x_3$ dressed by an abritrary number of wheels incircling $x_1$. We can use this result as a building block for a more complicated three-point correlation function, involving two twist-2 operators with an arbitrary spin and one protected operator
\beq\label{T-def}
T_{S_1,S_2}(x_1,x_2,x_3)=\langle 
{\cal O}_{S_1}^{n_1}(x_1)
\bar {\cal O}_{S_2}^{n_2}(x_2) 
\tr[\bar X(x_3)X(x_3)]\rangle\,.
\eeq
where $\bar {\cal O}_{S_2}^{n_2}$ is given by \re{O-n} with $X$ replaced with $\bar X$ and $n_i$ (with $i=1,2$) being light-like vectors. In the planar limit, this correlation function receives contribution from the diagrams shown on Fig.~\ref{fig:joinedtriangle}. The main observation is that they can obtained 
by merging together two wheels depicted in Fig.~\ref{fig:gluing0}.

To exemplify the idea we first consider \re{T-def} for $S_1=S_2=0$. In this case,
we have
\beq\label{T-example}
T_{0,0}(x_1,x_2,x_3)=2 c^4\int d^4 x_0 \frac{C^{\bullet\circ\circ}_{\Delta_1,0}}{(x_{10}^2)^{\Delta_1/2}(x_{13}^2)^{\Delta_1/2}(x_{03}^2)^{1-\Delta_1/2}}
(-\Box_0)\frac{C^{\bullet\circ\circ}_{\Delta_2,0}}{(x_{20}^2)^{\Delta_2/2}(x_{23}^2)^{\Delta_2/2}(x_{03}^2)^{1-\Delta_2/2}}
\eeq
where the operator $(-\Box_0)$ amputates an extra propagator and the integration over $x_0$ glues two sets of wheels together. Here $\Delta_1=\Delta_2$ are the scaling dimensions of the operators $\tr(X^2)$ and $\tr(\bar X^2)$ given by \re{xi2exp} but it is convenient to keep $\Delta_1$ and $\Delta_2$ to be arbitrary.~\footnote{Note that we can in principle take different  couplings of wheels around $x_1$ and around $x_2$ (similarly to \cite{Cavaglia:2018lxi}), i.e. $\xi_1$ and $\xi_2$, making also $\Delta_1$ and $\Delta_2$ unequal.} Then,
after differentiation in $x_0$ the integral becomes
\beq\label{T-int}
T_{0,0}(x_1,x_2,x_3)=2c^4\int d^4 x_0 \frac{C^{\bullet\circ\circ}_{\Delta_1,0}
C^{\bullet\circ\circ}_{\Delta_2,0}
\left(2-\Delta _2\right) \Delta _2 
}{
x_{01}^{\Delta _1} x_{02}^{\Delta
   _2+2} x_{03}^{6-\Delta _1-\Delta _2} x_{13}^{\Delta _1}
   x_{23}^{\Delta _2-2}
 }\;.
\eeq
We can employ inversions to verify that the integral transforms under the conformal transformation as a three-point correlation function of scalar operators. This property fixes the form of $T_{0,0}(x_1,x_2,x_3)$
up to a structure constant
\beq
T_{0,0}(x_1,x_2,x_3)=2c^3\frac{ C^{\bullet\bullet\circ}_{\Delta_1,0;\Delta_2,0}
}{
	x_{12}^{\Delta _1+\Delta _2-2} x_{13}^{\Delta _1-\Delta _2+2}
	x_{23}^{-\Delta _1+\Delta _2+2}
}\;.\la{T00int1}
\eeq
The integral in \re{T-int} can be computed immediately using the star-triangle identity leading to
\beq
 C^{\bullet\bullet\circ}_{\Delta_1,0;\Delta_2,0}=  C^{\bullet\circ\circ}_{\Delta_1,0}
C^{\bullet\circ\circ}_{\Delta_2,0} 
\frac{\Gamma \left(2-\frac{\Delta _1}{2}\right) \Gamma
	\left(2-\frac{\Delta _2}{2}\right) \Gamma \left(\frac{\Delta_1}{2}+\frac{\Delta_2}{2}-1\right)}{\Gamma \left(\frac{\Delta _1}{2}\right)\Gamma
	\left(\frac{\Delta _2}{2}\right)
	\Gamma \left(-\frac{\Delta _1}{2}-\frac{\Delta _2}{2}+3\right) }\,.
\la{T00int}
\eeq
Replacing $C_{\Delta,S}^{\bullet\circ\circ}$ with \eq{CG0} we obtain the following expression
\beqa\la{Cbbc}
(C^{\bullet\bullet\circ}_{\Delta_1,0;\Delta_2,0})^2=
\frac{4 \Gamma \left(4-\Delta _1\right) \Gamma \left(4-\Delta _2\right)
	\Gamma^2 \left(\frac{\Delta_1}{2}+\frac{\Delta _2}{2}-1\right)}{\left(\left(\Delta _1-4\right) \Delta
	_1+2\right) \left(\left(\Delta_2-4\right) \Delta _2+2\right) \Gamma
	\left(\Delta _1-1\right)\Gamma \left(\Delta _2-1\right) \Gamma^2 \left(-\frac{\Delta _1}{2}-\frac{\Delta
		_2}{2}+3\right) }\;.
\eeqa
which is symmetric in $\Delta_1$ and $\Delta_2$ as it should be.

So far, we treated $\Delta_1$ and $\Delta_2$ as arbitrary parameters. We expect that, in the limit $\Delta_2\to 2$, when the correlator $T_{0,0}$
looses all its wheels around $x_2$, it should reduce to the wheel correlation function  \eq{corel0}
multiplied by a free scalar propagator connecting $x_2$ and $x_3$. Indeed, it is easy to see that $C^{\bullet\bullet\circ}_{\Delta_1,0;2,0}= C^{\bullet\circ\circ}_{\Delta_1,0}$.

In application to the conformal fishnet theory, we have to replace the scaling dimensions $\Delta_1=\Delta_2$ in \re{Cbbc} with their exact expression \re{xi2exp}. In this case, the expression \eq{Cbbc} simplifies to
\beq
(C^{\bullet\bullet\circ}_{\Delta_1,0;\Delta_1,0})^2=\frac{4 \left(\Delta _1-3\right){}^2}{\left(\left(\Delta _1-4\right)
	\Delta _1+2\right){}^2}\;.
\eeq
\paragraph{Non-zero spins.} 
It is straightforward to generalize \re{T-example} to the operators with non-zero spins
\begin{align}\label{T-spins}
    T_{S_1,S_2}(x_1,x_2,x_3)=2 c^4 C^{\bullet\circ\circ}_{\Delta_1,S_1}
C^{\bullet\circ\circ}_{\Delta_2,S_2}\int d^4 x_0\Psi_{\Delta_1,1,1}^{S_1,n}(x_1,x_0,x_3)(-\Box_0)\Psi_{\Delta_2,1,1}^{S_2,n_2}(x_2,x_0,x_3)
\end{align}
where the notation was introduced for the so-called conformal triangle function
\beq
\Psi_{\Delta_1,\Delta_2,\Delta_3}^{S_1,n}(x_1,x_2,x_3)\equiv
\frac{ 1
}{
	x_{12}^{\Delta _1+\Delta _2-\Delta_3-S_1} x_{13}^{\Delta _1-\Delta _2+\Delta_3-S_1}
	x_{23}^{-\Delta _1+\Delta _2+\Delta_3+S_1}
}
\left(
\frac{2 (n x_{13})}{x_{13}^2}
-\frac{2 (n x_{12})}{x_{12}^2} 
\right)^{S_1}
\;\la{Psidef}
\eeq
In order to perform the $x_0-$integration in \re{T-spins}, we make use of the following
identity
\beqa\label{Psi-D}
\Psi_{\Delta_1,\Delta_2,\Delta_3}^{S_1,n}(x_1,x_0,x_3)\equiv
f_{\Delta_1,\Delta_2,\Delta_3}^{S_1}
\left[{\cal D}_{\Delta_1,n}\right]^{S_1}\frac{ 1
}{
	x_{10}^{\Delta _1+\Delta _2-\Delta_3-S_1} x_{13}^{\Delta _1-\Delta _2+\Delta_3-S_1}
	x_{03}^{-\Delta _1+\Delta _2+\Delta_3+S_1}
}
\eeqa
where ${\cal D}_{\Delta_1,n}$ is a differential operator acting on $x_1$\footnote{A similar operator also appeared in \cite{Karateev:2017jgd}.}
\beqa
{\cal D}_{\Delta_1,n}= (n\partial_{x_1})+
2 \left(\Delta _1-1\right) \frac{(nx_{13})}{x_{13}^2}
\eeqa
and the normalization constant is given by 
\beqa
f_{\Delta_1,\Delta_2,\Delta_3}^{S_1}\equiv
\frac{
\Gamma \left(\frac{1}{2} \left(-S_1+\Delta _1+\Delta _2-\Delta
	_3\right)\right)}{\Gamma \left(\frac{1}{2} \left(S_1+\Delta _1+\Delta
	_2-\Delta _3\right)\right)}\,.
\eeqa
Replacing the $\Psi-$functions in \re{T-spins} with \re{Psi-D},
we can pull out the differential operator ${\cal D}_{\Delta_1,n}^S$ outside the $x_0-$integral to obtain
\beqa\notag
&& T_{S_1,S_2}(x_1,x_2,x_3) =
\\f_{\Delta_1,1,1}^{S_1}
f_{\Delta_2,1,1}^{S_2}
\\
&& \times \left[{\cal D}_{\Delta_1,n_1}\right]^{S_1}
\left[{\cal D}_{\Delta_2,n_2}\right]^{S_2}
\int d^4 x_0
\frac{ \left(\Delta _2-S_2\right) \left(S_2+2-\Delta _2\right)C^{\bullet\circ\circ}_{\Delta_1,S_1}
	C^{\bullet\circ\circ}_{\Delta_2,S_2}
}{
x_{0,1}^{\Delta_1-S_1} x_{0,2}^{\Delta _2-S_2+2} x_{0,3}^{S_1+S_2-\Delta_1-\Delta _2+6} x_{1,3}^{\Delta _1-S_1} x_{2,3}^{\Delta
	_2-S_2-2}
}
\eeqa
where the $x_0-$integral can be again computed using the star-triangle identity. 

In this way, we arrive at the following expression
\beqa\la{T11}
&& T_{S_1,S_2}(x_1,x_2,x_3) =
2c^3 C^{\bullet\bullet\circ}_{\Delta_1,S_1;\Delta_2,S_2} \Psi^{S_1,n_1,S_2,n_2}_{\Delta_1,\Delta_2}(x_1,x_2,x_3)
\eeqa
where the dependence on $x_i$ is carried by the function
\begin{align}\label{Psi123}
\Psi^{S_1,n_1,S_2,n_2}_{\Delta_1,\Delta_2} =  \left[{\cal D}_{\Delta_1,n_1}\right]^{S_1}
\left[{\cal D}_{\Delta_2,n_2}\right]^{S_2}
\frac{ f_{\Delta_1,\Delta_2,2}^{S_1+S_2}  
}{
	x_{12}^{\Delta _1+\Delta _2-2-S_1-S_2} x_{13}^{\Delta _1-\Delta _2+2-S_1+S_2}
	x_{23}^{-\Delta _1+\Delta _2+2+S_1-S_2}
}
\end{align}
The structure constant $C^{\bullet\bullet\circ}_{\Delta_1,S_1;\Delta_2,S_2}$ 
depends on the scaling dimensions and spins of the two unprotected operators and reads
\beqa\la{Cgeneral}\nn
\left(C^{\bullet\bullet\circ}_{\Delta_1,S_1;\Delta_2,S_2}\right)^2&=&\frac{4  \left(S_1+1\right) \left(S_2+1\right) }{\left(-\left(\Delta _1-4\right) \Delta _1+S_1
	\left(S_1+2\right)-2\right) \left(-\left(\Delta _2-4\right) \Delta
	_2+S_2 \left(S_2+2\right)-2\right) }\\
&\times&\frac{
\Gamma \left(S_1-\Delta
_1+4\right) \Gamma \left(S_2-\Delta _2+4\right) \Gamma^2 \left(\frac{1}{2}
\left(S_1+S_2+\Delta _1+\Delta
_2-2\right)\right)
}{
\Gamma \left(S_1+\Delta _1-1\right)
\Gamma \left(S_2+\Delta _2-1\right)
\Gamma^2 \left(\frac{1}{2} \left(S_1+S_2-\Delta _1-\Delta
_2+6\right)\right) 
}\;.
\eeqa
This relation constitutes the main result of this section.
We can check again that in the limit $S_2\to 0$ and $\Delta_2\to 2$ the general expression \eq{Cgeneral} reduces to $(C_{\Delta_1,S_1}^{\bullet\circ\circ})^2$ defined in \eq{C0}.
We note that there is a striking similarity of \re{Cgeneral} with the expressions for the cusp structure constants of \cite{Cavaglia:2018lxi,Kim:2017sju}.

The function $\Psi^{S_1,n_1,S_2,n_2}_{\Delta_1,\Delta_2,2}$ describes
the $x_i-$dependence of the three-point correlation function \re{T-def}. Based on the conformal symmetry, we expect that it should have the following general form
\cite{Costa:2011mg} 
\begin{align}\label{ans}
{}&\Psi^{S_1,n_1,S_2,n_2}_{\Delta_1,\Delta_2}
= {1\over (x_{12}^2)^{\alpha_{12}}(x_{23}^2)^{\alpha_{23}}(x_{13}^2)^{\alpha_{13}}}
\sum_{n=0}^{\min(S_1,S_2)} c_n \, H_{12}^n (V_{1,23})^{S_1-n} (V_{2,13})^{S_2-n}
\end{align}
where
\begin{align}\notag
{}&\alpha_{12} = \frac12(S_1+S_2+\Delta_1+\Delta_2-2)\,,
\\\notag
{}&\alpha_{13} = \frac12(S_1-S_2+\Delta_1-\Delta_2+2)\,,
\\
{}&\alpha_{23} = \frac12(-S_1+S_2-\Delta_1+\Delta_2+2)\,,
\end{align}
and the sum runs over the conformal tensors projected onto light-like vectors
$n_1$ and $n_2$
\begin{align}\notag
{}& H_{12} = 2(n_1n_2) x_{12}^2 - 4(n_1 x_{12}) (n_2 x_{12})\,,
\\[2mm]\notag
{}& V_{1,23} ={x_{12}^2 x_{13}^2\over x_{23}^2}\left[ {2(n_1 x_{13})\over x_{13}^2} - {2(n_1 x_{12})\over x_{12}^2}\right] \,,
\\
{}& V_{2,13} ={x_{23}^2 x_{21}^2\over x_{31}^2}\left[ {2(n_2 x_{23})\over x_{23}^2} - {2(n_2 x_{21})\over x_{21}^2}\right] \,.
\end{align}
The expansion coefficients $c_n$ can be found by matching \re{ans} into \re{Psi123} 
\begin{align}\nn
c_n&=
\frac{\Gamma \left(S_1+1\right) \Gamma \left(S_2+1\right) }{\Gamma (n+1) \Gamma \left(-n+S_1+1\right) \Gamma
   \left(-n+S_2+1\right) }\\
   &\times\frac{\Gamma
   \left(\frac{1}{2} \left(S_1-S_2+\Delta _1+\Delta _2-2\right)\right)
   \Gamma \left(\frac{1}{2} \left(-S_1+S_2+\Delta _1+\Delta
   _2-2\right)\right)}{\Gamma \left(\frac{1}{2} \left(S_1+S_2+\Delta
   _1+\Delta _2-2\right)\right) \Gamma \left(\frac{1}{2} \left(2
   n-S_1-S_2+\Delta _1+\Delta _2-2\right)\right)}\;.
\end{align}

Applying \re{Cgeneral} we have to replace the scaling dimensions $\Delta_1$ and $\Delta_2$ 
by their explicit expressions. For the twist-two operators \re{O-n} they are given by the function $\Delta_2(S)$, Eq.~\re{D0}, evaluated for spin $S=S_1$ and $S=S_2$, respectively.
We recall that the scaling dimensions of twist-2 and twist-4 operators, $\Delta_2(S)$ and
$\Delta_4(S)$ are two branches of the same function of the (complexified) coupling constant $\xi^2$. This suggests that depending on the choice of the branch
of the functions $\Delta_1(\xi^2)$  and $\Delta_2(\xi^2)$, the relation \eq{T11} should also describe the three-point correlation function of the protected operator with two unprotected operators each having twist$-2$ or twist$-4$.

\section{Generalization to any  dimension}\label{sec:anyd}

Many of the results obtained in the previous sections for the bi-scalar theory
\eqref{bi-scalarL} in  \(d=4\) dimensions
can be easily generalized to its \(d-\)dimensional version  proposed in~\cite{Kazakov:2018qbr}. In particular,  the zero-magnon four-point correlation function \(G_D\) was computed there for any \(d\), generalizing the \(d=4\) results of~\cite{Grabner:2017pgm}.

The Lagrangian of \(d-\)dimensional bi-scalar model is non-local
\begin{equation}\label{bi-scalarL-D}
    {\cal L}_{d}=  N_c\,\tr
    [\bar X \,\, (-\p_\mu \p^\mu)^{\frac{d}{4}-\omega}\,X + \bar Z \,\, (-\p_\mu \p^\mu)^{\frac{d}{4}+\omega}\,Z 
+(4 \pi)^{\frac{d}{2}} \xi^2 \bar X \bar Z X Z].
  \end{equation} 
where the differential operator in an arbitrary power is defined in a standard way, as an integral operator. For the particular ``isotropic" case \(\omega=0\), the action \eqref{bi-scalarL-D} should be supplemented with the same double-trace counterterms \eqref{L-dt}. As before, the theory has two fixed points with the corresponding values of the double-trace couplings depending on \(\xi^2\) and  computable at least at weak coupling. 
Since the \(d\)-dimensional theory \re{bi-scalarL-D} has the same chiral interaction vertex 
as \eqref{bi-scalarL}, 
we can consider the correlation functions the same types as those shown on Fig.\ref{fig:types2}. An important difference with the $d=4$ case is that free scalar propagators are now given by   \(\ {c}/{(x_{12}^2)^{d/4}}\) with \(c={1}/{(2\pi)^{d/2}}\).

To compute zero- and one-magnon  four-point  correlation functions,  as it is done in  4 dimensions in the previous sections, we have to find, at any \(d\),  the eigenvalues of the graph-building operators $E_{\ga}$ and $E_{\gb}$. 
They are computed in appendix~\ref{app:Es} leading to (see also \cite{Giombi:2017dtl,Kazakov:2018qbr})
\beqa
&&E_{\ga}=
c^4 \pi ^d\frac{ \Gamma \left(\frac{d}{4}+\frac{S}{2}-\frac{\Delta
        }{2}\right) \Gamma \left(-\frac{d}{4}+\frac{S}{2}+\frac{\Delta
        }{2}\right)}{\Gamma \left(\frac{3 d}{4}+\frac{S}{2}-\frac{\Delta
        }{2}\right) \Gamma \left(\frac{d}{4}+\frac{S}{2}+\frac{\Delta
        }{2}\right)}\\
&&E_{\gb}=c^2 \pi ^{d/2} (-1)^S\frac{ \Gamma \left(\frac{3
                d}{8}+\frac{S}{2}-\frac{\Delta }{2}\right) \Gamma
        \left(-\frac{d}{8}+\frac{S}{2}+\frac{\Delta }{2}\right)}{\Gamma
        \left(\frac{5 d}{8}+\frac{S}{2}-\frac{\Delta }{2}\right) \Gamma
        \left(\frac{d}{8}+\frac{S}{2}+\frac{\Delta }{2}\right)}\;.
\eeqa
In addition, we also need the expression for the kinematical factor
$c_2$ defined in \eq{c2general}. It is given in
by \eq{c2general}, where we should take  $\Delta_1=\Delta_2=d/4$ for the zero-magnon case
and $\Delta_1=2d/4,\;\Delta_2=d/4$ for the one-magnon case. Assuming that all the intermediate steps are still valid for general $d$, we  find the the following expression for the OPE coefficients
\beqa\la{CstructD}
C_{\Delta,S}=\frac{4\pi }{c^d}{\rm res}_{\Delta}\left(\frac{1}{c_2(\nu,S)}\frac{E_{\Delta,S}^n}{1-\chi E_{\Delta,S}}\right)\;.
\eeqa
where $\chi_{\ga}=(4\pi)^{d}\xi^4$, $n=1$ and $\chi_{\gb}=(4\pi)^{d/2}\xi^2$, $n=2$ for zero-magnon and one-magnon cases, respectively. The dimension \(\Delta\) of operator appearing at the corresponding pole is related to the representation label \(\nu\) in the following way: \(\Delta=\frac{d}{2}+2i\nu\).
For \(d=2\), the zero-magnon spectrum looks particularly simple~\cite{Kazakov:2018qbr}
\begin{equation}
    \Delta_{\ga}=1+\sqrt{S^2-4\xi^2}\,.
\end{equation}
Interestingly   \cite{Kazakov:2018qbr}, the particular case of $d=2$ with \(\omega\to 1/2\) in \eqref{bi-scalarL-D} is  relevant for the BFKL approximation in high-energy QCD.

For general $d$ the structure of the poles in $\nu$ could change. In particular the poles coming from $E_{\Delta,S}=1/\chi$ will have a different number of solutions for different \(d\) (actually, this number is  infinite for odd dimensions, see \cite{Kazakov:2018qbr} for the analysis of possible exchange states in zero-magnon case), and those have to be taken into account when computing the correlation function. The computations are particularly simple in dimensions \(d=4k\), where \(k\) is integer. Then the spectral equations become polynomial in \(\Delta\). For example, for one-magnon case, the spectral equation for \(d=8\) looks as  
\begin{equation}
   \frac{(-1)^S}{16} (-\Delta +S+6) (-\Delta +S+8) (\Delta +S-2) (\Delta +S)=\xi^2\,,
\end{equation}  etc.

We leave the computations of a more complicated two-magnon correlation function, as well as the  detailed study of the general $d$ case, for future investigation. 

\section{Discussion and Conclusions}

In this work, we computed exactly and explicitly  certain 4-point correlation functions in bi-scalar CFT  
\eqref{bi-scalarL} 
which represent a specific double scaling limit (combining weak coupling and strong imaginary twist) of \(\gamma\)-deformed \({\cal N}=4\) SYM theory in 't Hooft approximation~\cite{Gurdogan:2015csr}. Although this theory obeys the integrability properties, related to the integrability of fishnet Feynman
graphs~\cite{Zamolodchikov:1980mb} dominating in its planar perturbation theory, we concentrate here on the physical quantities which can be computed without any appeal to the integrability and utilizing only the conformal symmetry properties. The three types of such 4-point functions are named as zero-magnon, one-magnon and two-magnon cases,
referring to the exchange operators which are of a  type \(\tr(Z^2)\), \(\tr(Z^2X)\) and \(\tr(Z^2X^2)+\text{perm.}\), as well as their non-zero spin cousins. The last operator is the analogue of the famous Konishi operator in \({\cal N}=4 \text{ SYM theory.}\) It is quite remarkable that we managed to compute explicitly  the all-loop anomalous dimensions of all these operators, as well as their structure constants with the external protected operators. These  formulas give a host of very non-trivial non-perturbative OPE data, with a rich analytic structure. To our knowledge,  this are the first examples of such non-perturbatively and  explicitly computed 4-point correlation functions in an interacting CFT in \(d>2\) dimensions.
  
The results for 4-point functions are presented in a standard form, as explicit OPE expansions over conformal blocks. We studied these functions by the weak coupling expansion and found that the results for zero- and one-magnon cases can be presented, in each order of weak-coupling expansion, in terms of special class of functions called Harmonic Poly Logariphms (HPL), thus facilitating the study of these functions in the cross channels. As for the more complicated two-magnon case, we observed that the HPL representation is only possible in the light-cone limit $z\to 0$, or equivalently $u\to 0$ with $v$ fixed.

Many interesting questions related to the study of these quantities are left out of the scope of this paper. In particular, we left for the future  the study of the operator content  and the OPE in cross-channels of all three 4-point functions, where the analytic structure looks more complicated than in the original channel. It would be also interesting to generalize these 4-point functions to the case of external operators with spins.  The correlation functions we computed could serve as a building block for more complicated $n-$point functions by gluining them together. Similar approach was recently developed in SYK context in~\cite{Gross:2017aos}\footnote{see also \cite{Rosenhaus:2018dtp} for a recent review.}. To demonstrate the procedure we show in section~\ref{sec:gt} how to obtain more complicated structure constants starting from the elementary blocks.

Notice that the main common feature of the 4-point functions, which allowed for the explicit computations using  the combination of Bethe-Salpeter techniques and the conformal symmetry, was the presence of not more than two  scalar fields of each species, \(Z\) or \(X\),  in the exchange operators.   If we want to compute the multi-point correlation functions with more than two scalars of any of the species in exchange operators, we have to appeal to the integrability methods, based on the non-compact conformal Heisenberg type spin chains. All computations become then much more sophisticated, though in many cases possible.      
Already the computations of anomalous dimensions of \(\tr(Z^3)\) operator and its cousins with the same \(R\)-charge, necessitated the extraction of  the results from the  quantum spectral curve of the full \(\gamma\)-deformed \({\cal N}=4\) SYM theory, in the corresponding double scaling limit~\cite{Gromov:2017cja}. The alternative method for computations of anomalous
dimensions, is based on the integrable quantum conformal spin chain~\cite{WithDavidToAppear}.  For the 3-point and 4-point correlation functions in bi-scalar theory, this last method seems to be
particularly promising.
Recently, in a very similar setup it was shown that the separation of variables (SOV) approach could be used to get compact expessions for the correlation function in~\cite{Cavaglia:2018lxi}. The expressions for the structure constants obtained there are very similar to our results, which indicates that a similar SOV-based approach could work here.   
There are also some other options, such as the form factor approach~\cite{Basso:2015zoa}, successfully applied in~\cite{Basso:2017jwq} for the computation of  certain 4-point correlation function based on fishnet graphs with disc topology. 

Interestingly, all our current results for 4-point correlation functions in \(d=4\) dimensions appear to be directly generalizable to the bi-scalar CFT at any \(d\) which was recently  formulated in~\cite{Kazakov:2018qbr}.  The \(d=2\) case is closely related to the BFKL model of reggeized gluons describable by the \(SL(2,C)\) Heisenberg spin chain~\cite{Lipa:1993pmr,Faddeev:1994zg,Derkachov:2001yn,DeVega:2001pu}.  It is worth noticing that the computations of 4- and 6-point correlation functions of BFKL pomeron light-ray operators presented in~\cite{Balitsky:2013npa,Balitsky:2015oux,Balitsky:2015tca}  look closely related to our current computations. It would be interesting to compare those results to our current results. 

The most interesting, physical questions about the bi-scalar CFT  still remain to be answered. It still remains to be understood whether this theory has a dual string description. The availability of explicit all-loop results, such as presented in this paper, should allow for some guesses in this direction. 
  Quite intriguingly, the strong coupling limit of the $4$ point correlators studied in this paper exhibits typical classical exponential scaling with the coupling, suggesting existence of the dual strong coupling classical description. 

\paragraph{Note added:} while this paper was in preparation a paper \cite{Liu:2018jhs} appeared about SYK model, which may however have some overlap with our results.

\section*{Acknowledgements}

We are thankful to F.~Aprile, B. Basso, J. Caetano,
A.~Cavaglia, S.~Derkachev, J.~Drummond, D.~Grabner,  O.~Gurdogan, P.~Heslop,  D.~Kosower,
F.~Levkovich-Maslyuk, E.~Olivucci, M.~Preti, V.~Rosenhaus, E.~Sobko    for
discussions. The work of V.K.  was supported   by the European Research Council (Programme
``Ideas" ERC-2012-AdG 320769 AdS-CFT-solvable). N.G. and V.K. are grateful to Humboldt
University (Berlin) for the hospitality and financial support of this work in the framework of the ``Kosmos" programme. N.G. wishes to thank STFC for support from Consolidated grant number ST/J002798/1. The work of G.K. was supported by
the French National Agency for Research grant ANR-17-
CE31-0001-01. G.K. would like to thank the Galileo Galilei Institute for
Theoretical Physics for the hospitality, the INFN and Simons Foundation for partial support
during the completion of this work.

\appendix

\section{Definitions and relations}
In the main text we use the following definition of the four-dimensional conformal block \cite{Dolan:2000ut}
\begin{align}\la{defg}
&g_{\Delta,S}=(-1)^S \frac{z\bar  z}{
 z - \bar z} \left[ k(\Delta + S, z) k(\Delta - S - 2, \bar z) - 
   k(\Delta + S, \bar z) k(\Delta - S - 2, z)\right],\\
&\text{where} \quad k(\beta_, x) = x^{\beta/2}
  \,\,_2F_1\left(\frac{\beta - (\Delta_1 -\Delta_2)}{2}, \frac{\beta + (\Delta_3 - \Delta_4)}{2}, \beta, 
   x\right)\la{defk}\;.
\end{align}
We also use the following formulas for normalization of the conformal triangles in any dimension $d$ \begin{align}\la{norm}
&\int \frac{d^d x_1d^d x_2}{(x^2_{12})^{d-\Delta_1-\Delta_2}}{\Phi_{-\nu,S,x_0}}(x_1,x_2)
\Phi_{\nu',S',x_0'}(x_2,x_1)=\\ 
\nn &= 
 (-1)^S c_1(\nu,S) \delta(\nu-\nu')\,\delta_{S,S'}\delta^{(d)}( x_{00'})( n n')^S
+ (-1)^S c_2(\nu,S) {\delta(\nu+\nu')\delta_{S,S'}} \frac{Y^S(x_{00'})}{ (x_{00'}^2)^{{d/2-S-2i\nu}}},
\end{align} where $Y(x_{00'})\!=\!(n \p_{x_0} )(n'\p_{x_{0'}})\ln x_{00'}^2 $.
The coefficients $c_1$ and $c_2$ enter into the general relations we discuss in the text.  They  are given by~\cite{Tod:1977harm}
\begin{align}\la{c1general}
&c_1= \frac{ 2^{{S}+1}\, S!\,\Gamma(+2 i \nu)\Gamma(-2 i \nu)  \left( 4 \nu ^2
   +(\frac{d}{2}+S-1)^2\right)^{-1}}{\pi ^{-(3\times d /2  + 1)} \Gamma\left(\frac{d}{2}-1+2i\nu\right)\Gamma\left(\frac{d}{2}-1-2i\nu\right) \Gamma(\frac{d}{2}+S)},
 \\ 
\\&
\label{c2general}c_2=\frac{2 \pi ^{d+1} (-1)^S S! \Gamma \left(\Delta -\frac{d}{2}\right) \Gamma (\Delta
        +S-1) \Gamma \left(\frac{1}{2} \left(d-\Delta +\Delta _1-\Delta
        _2+S\right)\right) \Gamma \left(\frac{1}{2} \left(d-\Delta -\Delta _1+\Delta
        _2+S\right)\right)}{\Gamma (\Delta -1) \Gamma \left(\frac{d}{2}+S\right) \Gamma
        \left(\frac{1}{2} \left(\Delta +\Delta _1-\Delta _2+S\right)\right) \Gamma
        \left(\frac{1}{2} \left(\Delta -\Delta _1+\Delta _2+S\right)\right)
        (d-\Delta +S)}\;.
\end{align}
The functions $\Phi_{\nu,S,x_0}$ form an orthogonal basis for $\nu>0$. This implies the following resolution of identity
\beq\la{identity}
\delta^{(d)}(x_1-x_3)\delta^{(d)}(x_2-x_4)=
\sum_{S=0}^\infty\frac{(-1)^S}{
{(x^2_{12})^{\Delta_1+\Delta_2-d}}
}\int_0^\infty\frac{d\nu}{c_1(\nu,S)}\int d^d x_0
{{\Phi^{\mu_1\dots\mu_S}_{-\nu,S,x_0}}(x_1,x_2)
{\Phi^{\mu_1\dots\mu_S}_{\nu,S,x_0}}(x_4,x_3)
}{}\;.
\eeq
when projected to the functional space spanned by $\Phi_{\nu,S,x_0}$.

\section{Cancellation of the spurious poles}\la{App:cancel}
In this appendix we analyse additional possible contributions
in \eq{Cstruct} due to the extra poles in $g_{\Delta,S}(u,v)$ and the measure factor $1/c_2$. We will see that these contributions cancel each other if an additional condition is imposed on the eigenvalue $E_{\Delta,S}$.

The conformal block $ g_{\Delta,S}(u,v)$ from \eq{defg} has simple poles at $\Delta_{S-n}=S+3-n$ (with $n=1,2,\dots,S$), or equivalently
$2i\nu_n=S+1-n$.  Its residue at the  pole $\nu=\nu_n$ is given by
$r_n\, g_{S+3,S-n}(u,v)$ where (see for example Appendix B in \cite{Simmons-Duffin:2017nub}):
\begin{align}\label{rn}
r_n=(-1)^n \frac{i n \Gamma^2 \left(\frac{1}{2} \left(n-\Delta _{1}+\Delta_2+1\right)\right)
}{2 \Gamma
        (n+1)^2 \Gamma^2 \left(\frac{1}{2} \left(-n-\Delta _{1}+\Delta_2+1\right)\right)
}\,.
\end{align}
This results in the following extra contribution to \eq{Cstruct}:
\beq
R^{g}_{S,m}=\left(\frac{r_m}{c_2(\Delta_{S-m},S)}\frac{E_{\Delta_{S-m},S}^n}{1-\chi E_{\Delta_{S-m},S}}\right)g_{S+3,S-m}(u,v)\;\;,\;\;1\leq n\leq S < \infty\;.
\eeq
The sum of the above contributions in general is not zero, however, there are also poles at $\Delta=S + 3 + k,\;k=0,1,2,\dots$ coming from $1/c_2(
\Delta,S)$ factor in \eq{Ggen}. They can be also expressed in terms of $r_n$ as follows
\beq
\frac{1}{c_2(S+3+k+2i\epsilon,S)}\simeq-\frac{1}{\epsilon}
\frac{r_k}{c_2(S+3,S+k)}
\eeq
resulting in another contribution 
\beq
R^{c_2}_{S,k}=-\left(\frac{r_k}{c_2(\Delta_{S},S+k)}
\frac{E_{\Delta_{S+k},S}^n}{1-\chi E_{\Delta_{S+k},S}}\right)g_{S+3+k,S}(u,v)
\;\;,\;\;0\leq S,k <\infty\;.
\eeq
Let us  show that these poles cancel under certain condition on $E_{\Delta,S}$. The sum of the contributions of $R^g$ can be written as
\beq
\sum_{S=0}^\infty \sum_{m=0}^S R^{g}_{S,m}=
\sum_{m=0}^\infty \sum_{S=m}^\infty R^{g}_{S,m}=
\sum_{k=0}^\infty \sum_{S=0}^\infty R^{g}_{S+k,k}
\eeq
where we use that $r_0=0$ and in the last sum we shift the summation index by $m$ and renamed $m$ to $k$. Since the other contribution has the form
\beq
\sum_{k=0}^\infty \sum_{S=0}^\infty R_{S,k}^{c_2}
\eeq
we have to consider
\beq
R_{S,k}^{c_2}+R_{S+k,k}^g=
-\left(\frac{r_k}{c_2(\Delta_{S},S+k)}
\left[\frac{E_{\Delta_{S+k},S}^n}{1-\chi E_{\Delta_{S+k},S}}-\frac{E_{\Delta_{S},S+k}^n}{1-\chi E_{\Delta_{S},S+k}}\right]\right)g_{S+3+k,S}(u,v)\;.
\eeq
We see that there will be no additional terms in \eqref{Gres}  if we require
\beq\la{conditionE}
r_k (E_{3+S+k,S}-E_{3+S,S+k})=0\;\;,\;\;k=0,1,2,\dots\;.
\eeq 
We verify in the main text that this requirement is indeed satisfied for each case considered.

\section{Eigenvalues of the graphs-building operators}\label{app:Es}

In this appendix we give technical details of the derivation of the eigenvalues of the graphs-building operators $H_\ga$, $H_\gb$ and $H_\gc$. 

\subsection{Star-triangle identity}

The calculation can be simplified by applying  the so-called star-triangle identity
\beq\la{starttriangle1}
\int d^d x_0 \;x_{01}^a x_{02}^b x_{03}^c = 
\pi ^{d/2}\frac{ \Gamma \left(\frac{a}{2}+\frac{d}{2}\right) \Gamma
   \left(\frac{b}{2}+\frac{d}{2}\right) \Gamma
   \left(\frac{c}{2}+\frac{d}{2}\right) }{\Gamma \left(-\frac{a}{2}\right) \Gamma
   \left(-\frac{b}{2}\right) \Gamma \left(-\frac{c}{2}\right)}
   x_{23}^{-a-d} x_{13}^{-b-d}
   x_{12}^{-c-d}
\eeq
where the exponents $a$, $b$ and $c$ satisfy  $a+b+c=-2d$. A particular case of it is
\beq\la{starttriangle2}
\int d^d x_0 \;x_{01}^a x_{02}^b = 
\pi ^{d/2}\frac{ \Gamma \left(\frac{a}{2}+\frac{d}{2}\right) \Gamma
   \left(\frac{b}{2}+\frac{d}{2}\right) \Gamma
   \left(-\frac{a}{2}-\frac{b}{2}-\frac{d}{2}\right) }{\Gamma
   \left(-\frac{a}{2}\right) \Gamma \left(-\frac{b}{2}\right) \Gamma
   \left(\frac{a}{2}+\frac{b}{2}+d\right)}x_{12}^{a+b+d}
\eeq
which can be obtained from \eq{starttriangle1} by sendind point $x_3$ to infinity.

\subsection{Eigenvalue $E_{\ga}$}\label{app:E0}

Here we compute the eigenvalue of the graph-building operator $H_\ga$ corresponding to the correlation function $\Ga$ in the \(d\)-dimensional version of the  bi-scalar theory. 

In virtue of conformal symmetry, the eigenstate of $H_\ga$ is given by \re{waveFGeneral}
with $\Delta_1=\Delta_2=d/4$ being scaling dimension of scalars in $d-$dimensions
\begin{align*}
\Phi_{\nu,S,x_0}(x_1,x_2)=
x_{01}^{-\Delta+S} x_{02}^{-\Delta+S} x_{12}^{\Delta-S -d/2}
\left(
\frac{2 
( n x_{02})}{x_{02}^2}
-\frac{2 
( n x_{01})}{x_{01}^2}
\right)^S   \,.
\end{align*} 
The calculation of the eigenvalue of $H_\ga$  reduces to the following integral
\begin{align}\la{waveGeneral0}
I=\int d^d x_1 d^d x_2\Phi_{\nu,S,x_0}(x_1,x_2)\frac{c^4}{x_{12}^d x_{13}^{d/2} x_{24}^{d/2}}={E_{\ga}}
\Phi_{\nu,S,x_0}(x_3,x_4)\;,
\end{align}
where we used \eq{Hc_kernel0}, generalised to any $d$.

Firstly, we observe that the integrand simplifies by the change of variables $x_a$ (with $a=1,2,3,4$)
\beq
x_{0a}^\mu =\frac{x_{0\bar a}^\mu}{x_{0\bar a}^2}\;\;,\;\;x_{0a}=\frac{1}{x_{0\bar a}}\;\;,\;\;x_{ab}=\frac{x_{\bar a\bar b}}{x_{0\bar a}x_{0\bar b}}\;\;,\;\;d^d x_a = \frac{d^d x_{\bar a}}{x_{0\bar a}^{2d}}
\eeq 
corresponding to the inversion around $x_0$.
This results in the following integral
\begin{align}\notag
I {}& =c^4 2^S\int d^d x_{\bar 1} d^d x_{\bar 2}
 \left(\frac{x_{0\bar{3}} x_{0\bar{4}}}{
x_{\bar{1}\bar{3}}
   x_{\bar{2}\bar{4}}
}\right)^{d/2}    x_{\bar{1}\bar{2}}^{-\frac{3 d}{2}+\Delta -S} (n\cdot x_{\bar{2}\bar{1}}  )^S
\\
{}& =c^4\int d^d x_{\bar 1} d^d x_{\bar 2} \left(\frac{x_{0\bar{3}} x_{0\bar{4}}}{x_{\bar{1}\bar{3}}
   x_{\bar{2}\bar{4}}}\right)^{d/2} \frac{\Gamma \left(\frac{3
   d}{4}-\frac{S}{2}-\frac{\Delta }{2}\right) }{\Gamma \left(\frac{3 d}{4}+\frac{S}{2}-\frac{\Delta
   }{2}\right)}(n\cdot\partial_{\bar 1})^Sx_{\bar{1}\bar{2}}^{-\frac{3
   d}{2}+\Delta +S}\;,
\end{align}
where in the second relation we replaced the factor of $(n\cdot x_{\bar 2\bar 1})$
by derivatives acting on $x_{\bar 1}$.

Next integrating by parts we can swap  $\partial_{\bar 1}$ into the derivative in $x_{\bar 3}$ and pull it out of the integral
\beq
I=c^4(x_{0\bar{3}} x_{0\bar{4}})^{d/2}( n\cdot\partial_{\bar 3})^S\int d^d x_{\bar 1} d^d x_{\bar 2} \left(\frac{1}{x_{\bar{1}\bar{3}}
   x_{\bar{2}\bar{4}}}\right)^{d/2} \frac{\Gamma \left(\frac{3
   d}{4}-\frac{S}{2}-\frac{\Delta }{2}\right) }{\Gamma \left(\frac{3 d}{4}+\frac{S}{2}-\frac{\Delta
   }{2}\right)}x_{\bar{1},\bar{2}}^{-\frac{3
   d}{2}+\Delta +S}\;.
\eeq
After that the integration can be performed by applying the star-triangle identity \eq{starttriangle2} twice, leading to
\beq
I = c^4 \pi ^d
\frac{  \Gamma
   \left(\frac{d}{4}-\frac{S}{2}-\frac{\Delta }{2}\right) \Gamma
   \left(-\frac{d}{4}+\frac{S}{2}+\frac{\Delta }{2}\right)
   }{\Gamma \left(\frac{3
   d}{4}+\frac{S}{2}-\frac{\Delta }{2}\right) \Gamma
   \left(\frac{d}{4}+\frac{S}{2}+\frac{\Delta }{2}\right)}
   (x_{0\bar{3}} x_{0\bar{4}})^{d/2}( n\cdot\partial_{\bar 3})^Sx_{\bar{3},\bar{4}}^{-\frac{d}{2}+\Delta +S}\,.
\eeq
Evaluating the derivatives and comparing with \eq{waveGeneral0} we arrive at 
\beq
E_{\ga}=c^4 \pi ^d \frac{\Gamma \left(\frac{d}{4}+\frac{S}{2}-\frac{\Delta
   }{2}\right) \Gamma \left(-\frac{d}{4}+\frac{S}{2}+\frac{\Delta
   }{2}\right)}{\Gamma \left(\frac{3 d}{4}+\frac{S}{2}-\frac{\Delta
   }{2}\right) \Gamma \left(\frac{d}{4}+\frac{S}{2}+\frac{\Delta
   }{2}\right)}\;.
\eeq
In the particular case $d=4$ we obtain \eq{E0}. The last relation 
agrees with the results of \cite{Kazakov:2018qbr} and generalizes the \(d=4\) result of \cite{Grabner:2017pgm}

\subsection{Eigenvalue $E_{\gb}$}\la{app:derE1}

The calculation of the eigenvalue of the graph-building operator for the correlation function $\Gb$ is similar to the previous case.
As before, we perform the calculation for general $d$. The graph-building operator 
$H_\gb$ looks as
\begin{align}
{}& H_{\gb}(x_1,x_2|x_3,x_4)=c^2\delta^{(d)}(x_4-x_1)\;
(x_{42}x_{23})^{-d/2}\;.
\end{align}
Its eigenstate is given by \eq{waveFGeneral} with $\Delta_1=d/2$ and $\Delta_2=d/4$ 
\begin{align}
\Phi_{\nu,S,x_0}(x_1,x_2)=
x_{01}^{-\Delta+S-d/4} x_{02}^{-\Delta+S+d/4} x_{12}^{\Delta-S -3d/4}
\left(
\frac{2 (nx_{02}) }{x_{02}^2}
-\frac{2 (nx_{01}) }{x_{01}^2}
\right)^S\;.  
\end{align}   
Like in the previous case, to find the eigenvalue we need to evaluate the integral
\begin{align}\la{waveGeneral01}
I=c^2\int d^d x_2\; (x_{42}x_{23})^{-d/2}\Phi_{\nu,S,x_0}(x_2,x_4)={E_{\gb}}
\Phi_{\nu,S,x_0}(x_4,x_3)\;.
\end{align}
Again the integrand simplifies after the inversion around $\vec x_0$
\beq
I=c^2 2^S x_{0\bar{3}}^{d/2} x_{0\bar{4}}^d\int d^d x_{\bar 2}\;
 x_{\bar{2}\bar{3}}^{-d/2}
x_{\bar{2}\bar{4}}^{-\frac{5
		d}{4}+\Delta -S}
\left(\vec n\cdot\vec x_{\bar{4}\bar{2}}\right)^S \;.
\eeq
Again we can absorb the last term under the integral into the derivatives in $\vec x_{\bar 4}$
and evaluate the integral using \eq{starttriangle2}
\beq
I= c^2 \pi ^{d/2} (-1)^S 
\frac{\Gamma \left(\frac{3
		d}{8}-\frac{S}{2}-\frac{\Delta }{2}\right) \Gamma
	\left(-\frac{d}{8}+\frac{S}{2}+\frac{\Delta }{2}\right)
}{\Gamma \left(\frac{5
		d}{8}+\frac{S}{2}-\frac{\Delta }{2}\right) \Gamma
	\left(\frac{d}{8}+\frac{S}{2}+\frac{\Delta }{2}\right)}
x_{0\bar{3}}^{d/2} x_{0\bar{4}}^d
	(\vec n\cdot \partial_{\bar 4})^Sx_{\bar{3}\bar{4}}^{-\frac{3 d}{4}+\Delta +S}
 \;.
\eeq
Finally, computing the derivatives and comparing the result with \eq{waveGeneral01} we arrive at
\beq\la{E1generald}
E_{\Delta,S}=c^2 \pi ^{d/2} (-1)^S \frac{\Gamma \left(\frac{3
		d}{8}+\frac{S}{2}-\frac{\Delta }{2}\right) \Gamma
	\left(-\frac{d}{8}+\frac{S}{2}+\frac{\Delta }{2}\right)}{\Gamma
	\left(\frac{5 d}{8}+\frac{S}{2}-\frac{\Delta }{2}\right) \Gamma
	\left(\frac{d}{8}+\frac{S}{2}+\frac{\Delta }{2}\right)}\;,
\eeq
which  reduces at $d=4$ to \eq{E1}.

\subsection{Eigenvalue $E_{\gc}$}\label{app:EB}

Introducing for convenience  $I=(4\pi)^4 E_{\gc}$ and defining dual coordinates, $p=x_{12}$, $k_1=x_{31}$ and $k_2=x_{41}$, we
can rewrite \re{EB-int} as a two-loop Feynman integral
\begin{align}\label{EB-int-dual}
I(\nu,S) = {1\over\pi^4}   \int {d^4 k_1 d^4 k_2 \, (n k_{12}) ^S \over k_1^2 (p+k_1)^2  k_2^2 (p+k_2)^2   (k_{12}^2)^{1-i\nu +S/2}} \,,
\end{align}
where $p^2=(pn)=1$ and $k_{12}=k_1-k_2$. Due to symmetry of the integrand under $k_1\leftrightarrow k_2$,  $I(\nu,S)$ vanishes for odd $S$.

For $S=0$ the integral \re{EB-int} is known as a massless two-loop self-energy Feynman integral. It can be 
expressed in a closed form in terms of ${}_3 F_2-$hypergeometric function
(see e.g. review \cite{Grozin:2012xi})
\begin{align}\label{Ia}
I(\nu,0)=
2 \Gamma (-i\nu) \Gamma (i\nu) \left(\pi  \cot (i \pi \nu)-\frac{\,
        _3F_2(1,2,1+i\nu;2+i\nu,2+i\nu;1)}{(1+i\nu) \Gamma (1-i\nu) \Gamma (2+i\nu)}\right).
\end{align}
This representation is not convenient however to elucidate analytic properties of $I(\nu,0)$, that why we derive below another 
equivalent representation.

We start with rewriting \re{EB-int-dual} with all Lorentz indices uncontracted
\begin{align}
I(\nu,S) p^{(\mu_1}\dots p^{\mu_S)}= {1\over\pi^4}   \int {d^4 k_1 d^4 k_2 \, k_{12}^{(\mu_1}\dots k_{12}^{\mu_S)} \over k_1^2 (p+k_1)^2  k_2^2 (p+k_2)^2   (k_{12}^2)^{1-i\nu +S/2}} \,,
\end{align}
where $ p^{(\mu_1}\dots p^{\mu_S)}$ denotes a symmetric traceless tensor and $S$ is even. Then, we project all Lorentz indices
on the vector $p$ and take into account the identity
\begin{align}\label{kp}
k_{12}^{(\mu_1}\dots k_{12}^{\mu_S)} p_{\mu_1} \dots p_{\mu_S} = (pk_{12})^S + c_1 
(pk_{12})^{S-2} p^2 k_{12}^2 + \ldots + c_{S/2} (p^2 k_{12}^2)^{S/2}\,.
\end{align}
The coefficients $c_k$ can be found from the requirement for the expression on the right-hand
side to vanish upon hitting it with $\partial^2/\partial p^\mu\partial p_\mu$. Replacing $k_{12}$ with $p$
in  \re{kp} we get 
\begin{align}\label{sum-c}
p^{(\mu_1}\dots p^{\mu_S)} p_{\mu_1} \dots p_{\mu_S} = (p^2)^S(1 + c_1 + \ldots + c_{S/2})\,.
\end{align}
In this way, we obtain another representation for $I(\nu,S)$ that differs from \re{EB-int-dual} in that $(nk_{12})^S$ is
replaced by the expression on the right-hand side of \re{kp} divided by the normalization factor that enters \re{sum-c}.
The advantage of this representation is that we can use the identity
\begin{align}
2(k_{12}p) = (k_1+p)^2 -k_1^2-(k_2+p)^2+k_2^2 
\end{align}
to express $I(\nu,S)$ in terms of scalar Feynman integrals.  Some of the integrals coincide with those contributing to $I(\nu,0)$ whereas the remaining integrals have one propagator less and can be easily 
evaluated. As a consequence, $I(\nu,S)$ is given by a linear combination of $I(\nu,0)$ and some rational functions
of $\nu$.

Going through the calculation of $I(\nu,S)$ for $S=2$ and $S=4$ we find
\begin{align}\notag
I(\nu,2)= -\frac13 I(\nu,0)+\frac{8}{3 \left(\nu ^2+1\right)^2}\,,
\\
I(\nu,4)=-\frac35 I(\nu,2) +\frac{16}{5 \left(\nu ^2+4\right)^2}\,.
\end{align}
In general, for $S=2\ell$, we have
\begin{align}
I(\nu,2\ell)= -{2\ell-1\over 2\ell+1} I(\nu,2\ell-2) +{8\ell\over (2\ell+1)(\nu^2+\ell^2)^2}\,.
\end{align}
The solution to this recurrence relation looks as
\begin{align}\label{large-ell}
I(\nu,2\ell)={(-1)^\ell\over 2\ell+1} \left[ I(\nu,0) +\sum_{n=1}^\ell {8(-1)^n n\over (n^2+\nu^2)^2} \right]\,.
\end{align}
Assuming that $I(\nu,2\ell)$ vanishes at large $\ell$ faster then $1/\ell$ (this property can be verified a posteriori)
we find from \re{large-ell} that $I(\nu,0)$ is given by
\begin{align} \label{EB-sum}I(\nu,0) {}&=\sum_{n=1}^\infty {8(-1)^{n-1} n\over (n^2+\nu^2)^2} 
\\=\frac{i\left(
        \psi^{(1)}\left(\frac{1+i\nu}{2}\right)
        -\psi ^{(1)}\left(\frac{i \nu }{2}\right)\right)}{2 \nu } + (\nu \to -\nu)\,.
\end{align}
 We verified that the two representations \re{Ia} and \re{EB-sum} are equivalent.

Substituting \re{EB-sum} into \re{large-ell} we arrive at 
\begin{align}\notag\label{EB1-sum}
I(\nu,2\ell) {}&= {1\over 2\ell+1}  \sum_{n=\ell+1}^\infty  (-1)^{n-\ell-1} {8n\over (n^2+\nu^2)^2} 
\\
{}&=\frac{i \left[
        \psi
        ^{(1)}\left(\frac{1}{2} (\ell+i \nu +1)\right)-\psi ^{(1)}\left(\frac{1}{2}
        (\ell+i \nu +2)\right)\right]}{2 (2 \ell +1)\nu} + (\nu \to -\nu)\,.
\end{align}

\section{Analytic properties of the correlation functions}\label{analGa}

We can apply \re{GGgen} and\re{Ggen} to understand analytic properties of $G(x_1,x_2|x_3,x_4)$ as a function of the coupling constant $\xi^2$. The integrand of \re{Ggen} is a meromorphic function of $\chi$ which is just a power of $\xi^2$ dependent on the type of the correlation function. 

The integral in \re{Ggen} is well-defined as soon as the physical poles \re{forspec} 
are away from the real axis. 
As soon as the physical
pole \re{forspec} approaches the real axis, the integral \re{Ggen} generates a branch cut in $\chi$.
The discontinuity across the cut can be found from \re{Ggen}
\begin{align}\label{disc1}
{\rm disc}_{\chi} {\cal G}(u,v)=
c^{-4}\sum_{S\ge 0}(-1)^S\int_{-\infty}^\infty\frac{d\nu}{c_2(\nu,S)} {E_{\Delta,S}^n} 
g_{\Delta,S}(u,v)\delta(1-\chi E_{\Delta,S})   \,.
\end{align}
In virtue of \re{Esym}, the integral localizes at two points $\Delta=2+2i\nu_\star$ and $4-\Delta=2-2i\nu_\star$ satisfying \re{forspec}.
 Taking into account \re{Cstruct} we obtain
\begin{align}\label{disc2}
{\rm disc}_{\chi}{\cal G}(x_1,x_2|x_3,x_4) = \sum_{S\ge 0} {C_{2+2i\nu_\star,S}g_{2+2i\nu_\star,S}(u,v)-C_{2-2i\nu_\star,S}g_{2-2i\nu_\star,S}(u,v)\over 2\pi i}\,.
\end{align} 
According to \re{E0} and \re{E2}, $E_\ga$ and $E_\gc$ are positive definite functions
of $\nu$ and, therefore, the
integral on the right-hand side of \re{disc1} vanishes for $\chi<0$ for zero- and two-magnon functions, ${\cal G}_\ga$ and ${\cal G}_\gc$, respectively. Moreover,
the function $E_\gc$ satisfies the relation $0< E_\gc\le {3 \zeta (3)}/({128 \pi ^4})$
on the real $\nu-$axis and, as a consequence,  ${\rm disc}_{\chi} {\cal G}_\gc(u,v)\neq 0$
for $\chi> 128 \pi ^4/{(3 \zeta (3))}$.

\section{Renormalization group flow of the coupling constants}

The \(\gamma\)-deformed \(\)\({\cal N}=4\)~SYM  remains a unitary theory as long as  the \(\gamma\)-deformation angles are real and all the couplings, 't'Hooft coupling and   the double-trace couplings, are real. 
On the other hand, this is not a CFT anymore since the real double-scaling couplings run with the scale even in the leading 't~Hooft limit.  Looking at the leading order beta-function for the  interaction term  
\(\alpha_{jj}^2\tr(\phi_j\phi_j)\tr(\phi^\dagger_j\phi_j^{\dagger })\) the one-loop beta-function 
 is given by \cite{Fokken:2014soa}
\begin{equation}
 \beta_{\alpha^2_{jj}}={}{g^4\over \pi^2}\sin^2\gamma_j^+\sin^2\gamma_j^-+{\alpha_{jj}^4\over  4\pi^2}\,,
\end{equation}
where \(\gamma^\pm_1=\mp\frac{1}{2}(\gamma_2\pm\gamma_3),\,\,\gamma^\pm_2=\mp\frac{1}{2}(\gamma_3\pm\gamma_1),\) and \( \gamma^\pm_3=\mp\frac{1}{2}(\gamma_1\pm\gamma_2)\) are the parameters of \(\gamma\)-twist,
we expect the critical double-trace couplings at $\beta_{\alpha^2_{jj}}(\alpha_{jj})=0$, where the theory is conformal,   to have two complex conjugate  fixed points  
\begin{equation}\label{fp}
\alpha_{jj\,\pm}^{2}=\pm 2i g^2 \sin\gamma_j^+\sin\gamma_j^-+O(g^4) .
\end{equation}
 Using Callan-Symanzik RG equation we arrive, at the  critical point, to the following complex  anomalous dimension
\begin{equation}\label{gammafull}
\gamma_{J=2}(g)=\mp { ig^2\over 2\pi^2} \sin\gamma_j^+\sin\gamma_j^-
+O(g^4).
\end{equation} At a finite \(g\), then two exact (in a given scheme) critical couplings are comnplex conjugate functions of \(g\)~:  \begin{equation}\label{fpexact}
\alpha_{jj\,\pm}^{2}=\pm 2i A(g)+B(g) . 
\end{equation} where \(A(g)\) and \(B(g)\) are real functions  of the couplping and \(\gamma_j\).

It is interesting to understand how the vicinity of these critical points to the real axis may influence the RG flow of \(\alpha_{jj}\) coupling (we remind that the 't~Hooft coupling \(g\) stays fixed with RG, at least in the large \(N_c\) limit). 
To get a better qualitative picture of what happens in \(\gamma\)-deformed \({\cal N}=4\)~SYM  we can study the RG of the similar coupling \(\alpha_1(\xi)\) in the bi-scalar model where we have more of the exact data, and only one coupling \(\alpha_1^2\) is running.
The beta function
for this coupling
was computed in \cite{Grabner:2017pgm} up
to a few orders of PT in \(\epsilon=4-D\) regularisation, in  MS scheme, and the result is quadratic in \(\alpha_1^2\), as was predicted in~\cite{Pomoni:2008de} for a generic large \(N\) CFT of this type \footnote{ This mechanism was first established in \cite{Dymarsky:2005uh} an the example of non-supersymmetric orbifold theories}, with non-running single-trace four-scalar interaction and  a running double trace coupling: 
\begin{align}\label{beta}
\beta_{1}  {}&=
a(\xi) + \alpha_1^2\, b(\xi) + \alpha_1^4\, c(\xi)\,=c(\alpha_1^2-\alpha_{1,+}^2)(\alpha_1^2-\alpha_{1,-}^2)
\end{align}
where \(\alpha_{1,\pm }^2 = -{1\over 2c} (b \pm \sqrt{b^2-4ac})\) and  the functions $a$, $b$, and $c$ are given by 
\begin{align}\notag\label{func}
{}& a= -\xi^4+\xi^8-\frac{4}{3}\xi^{12} + O(\xi^{16})\,,
\\\notag
{}& b=-4\xi^4+4\xi^8-\frac{88}{15}\xi^{12}+ O(\xi^{16})\,,
\\
{}& c=-4-4\xi^4+\frac{4}{3} \xi^8+ O(\xi^{12})\,.
\end{align}

\begin{figure}[t]
\begin{center}
\includegraphics[width = .3\textwidth]{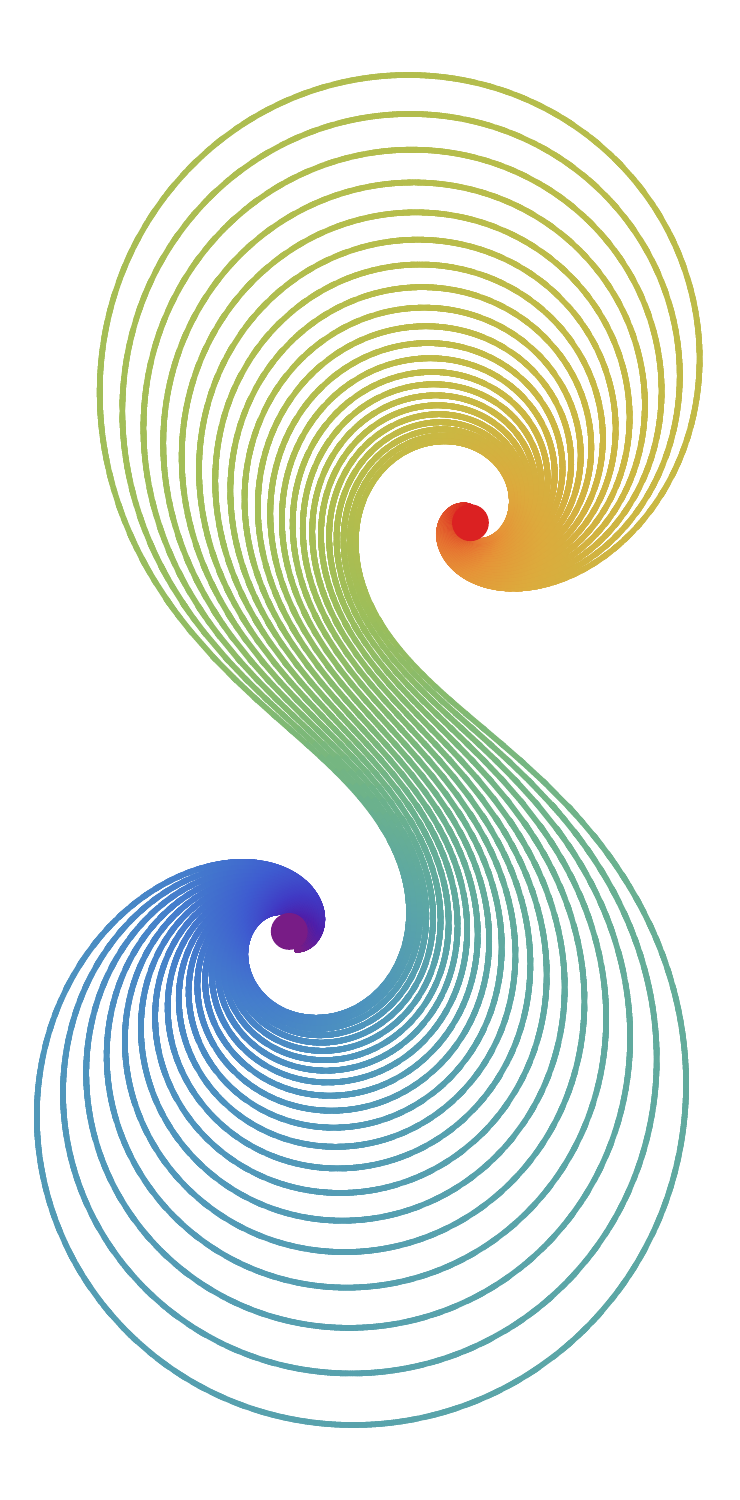}    
\end{center}
\caption{Flow of $\alpha^2(\mu)$ from  $\mu=0$ (red point) to $\mu=\infty$ (purpule point) for $\xi=1/3-i/14$ on a complex $\alpha^2-$plane. 
Different curves correspond to different initial conditions,
$\alpha^2(\mu=1)= -0.1+0.01\,k$ with $k=0,1,\dots,20$.}
\label{fig:RGflows}\end{figure}

Moreover, there exists the following exact relation between the
functions \eq{func} and the scaling dimensions \eq{Dexact} at the fixed point~ \cite{Pomoni:2008de,Grabner:2017pgm}:\footnote{This relation between the scaling dimension of the operator  \(\tr X^2\) at the critical point and the discriminant of quadratic equation on zeros of the beta-function follows directly from the eqs.(2.47-48) of the paper~\cite{Pomoni:2008de}. It is valid for a generic large \(N\) CFT of this type. }
\begin{align}\label{mag}
 b^2-4 ac = 4\gamma^2.
\end{align}  where \(\gamma=\Delta-2\) is the anomalous dimension of \(\tr (X^2)\) at the corresponding fixed point, so that
\begin{align}\label{alphaPM}
\qquad  \alpha_{1,\pm }^2 = -{1\over 2c} (b \pm \sqrt{b^2-4ac})\,=\bar \alpha_{1 }^2 \pm \frac{\gamma(\xi)}{c(\xi)}
\end{align}
 where we introduced two real functions of \(\xi\) \begin{align}\label{alphaRG}
\bar \alpha_{1 }^2 = -{b\over 2c}, \qquad   \gamma=  2i\sqrt{\sqrt{1+\xi^4}-1}\,.  
\end{align} 
We can write the RG equation in the following form
\begin{align}\label{alphaRGsq}
\frac{\p \alpha_1^2}{\p\log \mu}  {}&=
c(\xi) \left[\left(\alpha_1^2-\bar \alpha_1^2\, (\xi)\right)^2-\frac{\gamma^2(\xi)}{c^2(\xi)}\right]\,.
\end{align}
Let us introduce a new coupling \(\alpha^2=c\,\alpha_1^2\), as well as \(\bar\alpha^2=c\,\bar\alpha_1^2= -{b\over 2}\),  which can be considered as changing the RG scheme.  Then the RG equation can be written as \begin{align}\label{alphaRGren}
\frac{\p \alpha^2}{\p\log\mu}  =
\left(\alpha^2-\bar\alpha^2\, (\xi)\right)^2-\gamma^2(\xi)\,.
\end{align}

Solving this equation in this  specially chosen renormalization scheme, we find an RG flow: \begin{equation}\label{univFlow}
\alpha^2-\bar\alpha^2=-\gamma(\xi) \tanh\left[(\gamma(\xi)\log(\mu/\mu_0)\right]\,.
\end{equation}  
which is  defined only through the universal quantity - the anomalous dimension \(\gamma(\xi)\). This type of RG flow for the double-trace coupling also first appeared in the papers~\cite{Dymarsky:2005uh,Pomoni:2008de}. 

The physical quantities of the theory have a cut along the half-axis \(\xi>0\) and poles. For generic complex $\xi$'s there is no such problem. Assuming the coupling to have a little negative imaginary  \(\xi\to \xi -i\epsilon\) the RG flow \eq{univFlow} will avoid the poles in the r.h.s. and interpolate between two fixed point \(\alpha_{1,+}\) or \(\alpha_{1,-}\) (one of them is IR fixed point, another is the UV fixed point, depending on the sign of \(\epsilon\)).  We illustrate the possible RG flows in Fig.\ref{fig:RGflows}.

\bibliographystyle{JHEP}
\bibliography{4points_main}

\providecommand{\href}[2]{#2}\begingroup\raggedright\begin{thebibliography}{10}

\bibitem{Rychkov:2016iqz}
S.~Rychkov, \emph{{EPFL Lectures on Conformal Field Theory in D\&gt;= 3
  Dimensions}}, SpringerBriefs in Physics. 2016,
  \href{https://doi.org/10.1007/978-3-319-43626-5}{10.1007/978-3-319-43626-5},
  [\href{https://arxiv.org/abs/1601.05000}{{\ttfamily 1601.05000}}].

\bibitem{Shaposhnikov:2018xkv}
M.~Shaposhnikov and A.~Shkerin, \emph{{Conformal symmetry: towards the link
  between the Fermi and the Planck scales}},
  \href{https://arxiv.org/abs/1803.08907}{{\ttfamily 1803.08907}}.

\bibitem{Braun:2003rp}
V.~M. Braun, G.~P. Korchemsky and D.~Mueller, \emph{{The Uses of conformal
  symmetry in QCD}},
  \href{https://doi.org/10.1016/S0146-6410(03)90004-4}{\emph{Prog. Part. Nucl.
  Phys.} {\bfseries 51} (2003) 311}
  [\href{https://arxiv.org/abs/hep-ph/0306057}{{\ttfamily hep-ph/0306057}}].

\bibitem{DiFrancesco1997}
P.~Di~Francesco, P.~Mathieu and D.~Senechal, \emph{{Conformal Field Theory}},
  Graduate Texts in Contemporary Physics. Springer-Verlag, New York, 1997,
  \href{https://doi.org/10.1007/978-1-4612-2256-9}{10.1007/978-1-4612-2256-9}.

\bibitem{Maldacena:1997re}
J.~M. Maldacena, \emph{{The Large N limit of superconformal field theories and
  supergravity}}, \href{https://doi.org/10.1023/A:1026654312961,
  10.4310/ATMP.1998.v2.n2.a1}{\emph{Int. J. Theor. Phys.} {\bfseries 38} (1999)
  1113} [\href{https://arxiv.org/abs/hep-th/9711200}{{\ttfamily
  hep-th/9711200}}].

\bibitem{Leigh:1995ep}
R.~G. Leigh and M.~J. Strassler, \emph{{Exactly marginal operators and duality
  in four-dimensional N=1 supersymmetric gauge theory}},
  \href{https://doi.org/10.1016/0550-3213(95)00261-P}{\emph{Nucl. Phys.}
  {\bfseries B447} (1995) 95}
  [\href{https://arxiv.org/abs/hep-th/9503121}{{\ttfamily hep-th/9503121}}].

\bibitem{Lunin:2005jy}
O.~Lunin and J.~M. Maldacena, \emph{{Deforming field theories with U(1) x U(1)
  global symmetry and their gravity duals}},
  \href{https://doi.org/10.1088/1126-6708/2005/05/033}{\emph{JHEP} {\bfseries
  0505} (2005) 033} [\href{https://arxiv.org/abs/hep-th/0502086}{{\ttfamily
  hep-th/0502086}}].

\bibitem{Kazakov:2015efa}
V.~Kazakov, S.~Leurent and D.~Volin, \emph{{T-system on T-hook: Grassmannian
  Solution and Twisted Quantum Spectral Curve}},
  \href{https://arxiv.org/abs/1510.02100}{{\ttfamily 1510.02100}}.

\bibitem{Cordova:2015}
C.~Cordova, T.~T. Dumitrescu and K.~Intriligator, \emph{{Multipletsltiplets of
  Superconformal Symmetry in Diverse Dimensions}},
  \href{https://doi.org/10.1007/JHEP10(2016)080}{\emph{arXiv:1612.00809} (2016)
  165} [\href{https://arxiv.org/abs/1506.03807}{{\ttfamily 1506.03807}}].

\bibitem{Cordova:2016}
C.~Cordova, T.~T. Dumitrescu and K.~Intriligator, \emph{{Multiplets of
  Superconformal Symmetry in Diverse Dimensions}},
  \href{https://doi.org/10.1007/JHEP11(2016)135}{\emph{arXiv:1612.00809}
  {\bfseries 11} (2016) 135}
  [\href{https://arxiv.org/abs/1602.01217}{{\ttfamily 1602.01217}}].

\bibitem{Cordova:2016xhm}
C.~Cordova, T.~T. Dumitrescu and K.~Intriligator, \emph{{Deformations of
  Superconformal Theories}},
  \href{https://doi.org/10.1007/JHEP11(2016)135}{\emph{JHEP} {\bfseries 11}
  (2016) 135} [\href{https://arxiv.org/abs/1602.01217}{{\ttfamily
  1602.01217}}].

\bibitem{Beisert:2006ez}
N.~Beisert, B.~Eden and M.~Staudacher, \emph{{Transcendentality and Crossing}},
  \href{https://doi.org/10.1088/1742-5468/2007/01/P01021}{\emph{J. Stat. Mech.}
  {\bfseries 0701} (2007) P01021}
  [\href{https://arxiv.org/abs/hep-th/0610251}{{\ttfamily hep-th/0610251}}].

\bibitem{Gromov:2009tv}
N.~Gromov, V.~Kazakov and P.~Vieira, \emph{{Exact Spectrum of Anomalous
  Dimensions of Planar N=4 Supersymmetric Yang-Mills Theory}},
  \href{https://doi.org/10.1103/PhysRevLett.103.131601}{\emph{Phys. Rev. Lett.}
  {\bfseries 103} (2009) 131601}
  [\href{https://arxiv.org/abs/0901.3753}{{\ttfamily 0901.3753}}].

\bibitem{Beisert:2010jr}
N.~Beisert et~al., \emph{{Review of AdS/CFT Integrability: An Overview}},
  \href{https://doi.org/10.1007/s11005-011-0529-2}{\emph{Lett. Math. Phys.}
  {\bfseries 99} (2012) 3} [\href{https://arxiv.org/abs/1012.3982}{{\ttfamily
  1012.3982}}].

\bibitem{Cavaglia:2014exa}
C.~Andrea, D.~Fioravanti, N.~Gromov and R.~Tateo, \emph{{Quantum Spectral Curve
  of the $\mathcal N=$ 6 Supersymmetric Chern-Simons Theory}},
  \href{https://doi.org/10.1103/PhysRevLett.113.021601}{\emph{Phys. Rev. Lett.}
  {\bfseries 113} (2014) 021601}
  [\href{https://arxiv.org/abs/1403.1859}{{\ttfamily 1403.1859}}].

\bibitem{Gromov2014a}
N.~Gromov, V.~Kazakov, S.~Leurent and D.~Volin, \emph{{Quantum Spectral Curve
  for Planar $\mathcal{N} =$ Super-Yang-Mills Theory}},
  \href{https://doi.org/10.1103/PhysRevLett.112.011602}{\emph{Phys. Rev. Lett.}
  {\bfseries 112} (2014) 011602}.

\bibitem{Gromov:2014caa}
N.~Gromov, V.~Kazakov, S.~Leurent and D.~Volin, \emph{{Quantum spectral curve
  for arbitrary state/operator in AdS$_{5}$/CFT$_{4}$}},
  \href{https://doi.org/10.1007/JHEP09(2015)187}{\emph{JHEP} {\bfseries 09}
  (2015) 187} [\href{https://arxiv.org/abs/1405.4857}{{\ttfamily 1405.4857}}].

\bibitem{Gromov:2017blm}
N.~Gromov, \emph{{Introduction to the Spectrum of $N=4$ SYM and the Quantum
  Spectral Curve}},  \href{https://arxiv.org/abs/1708.03648}{{\ttfamily
  1708.03648}}.

\bibitem{Kazakov:2018ugh}
V.~Kazakov, \emph{{Quantum Spectral Curve of $\gamma$-twisted ${\cal N}=4$ SYM
  theory and fishnet CFT}},  \href{https://arxiv.org/abs/1802.02160}{{\ttfamily
  1802.02160}}.

\bibitem{Escobedo:2010xs}
J.~Escobedo, N.~Gromov, A.~Sever and P.~Vieira, \emph{{Tailoring Three-Point
  Functions and Integrability}},
  \href{https://doi.org/10.1007/JHEP09(2011)028}{\emph{JHEP} {\bfseries 09}
  (2011) 028} [\href{https://arxiv.org/abs/1012.2475}{{\ttfamily 1012.2475}}].

\bibitem{Basso:2015zoa}
B.~Basso, S.~Komatsu and P.~Vieira, \emph{{Structure Constants and Integrable
  Bootstrap in Planar N=4 SYM Theory}},
  \href{https://arxiv.org/abs/1505.06745}{{\ttfamily 1505.06745}}.

\bibitem{Cavaglia:2018lxi}
A.~Cavaglia, N.~Gromov and F.~Levkovich-Maslyuk, \emph{{Quantum Spectral Curve
  and Structure Constants in N=4 SYM: Cusps in the Ladder Limit}},
  \href{https://arxiv.org/abs/1802.04237}{{\ttfamily 1802.04237}}.

\bibitem{Giombi:2018qox}
S.~Giombi and S.~Komatsu, \emph{{Exact Correlators on the Wilson Loop in
  $\mathcal{N}=4$ SYM: Localization, Defect CFT, and Integrability}},
  \href{https://arxiv.org/abs/1802.05201}{{\ttfamily 1802.05201}}.

\bibitem{Fleury:2016ykk}
T.~Fleury and S.~Komatsu, \emph{{Hexagonalization of Correlation Functions}},
  \href{https://doi.org/10.1007/JHEP01(2017)130}{\emph{JHEP} {\bfseries 01}
  (2017) 130} [\href{https://arxiv.org/abs/1611.05577}{{\ttfamily
  1611.05577}}].

\bibitem{Eden:2016xvg}
B.~Eden and A.~Sfondrini, \emph{{Tessellating cushions: four-point functions in
  $\mathcal{N} $ = 4 SYM}},
  \href{https://doi.org/10.1007/JHEP10(2017)098}{\emph{JHEP} {\bfseries 10}
  (2017) 098} [\href{https://arxiv.org/abs/1611.05436}{{\ttfamily
  1611.05436}}].

\bibitem{Correa:2012hh}
D.~Correa, J.~Maldacena and A.~Sever, \emph{{The quark anti-quark potential and
  the cusp anomalous dimension from a TBA equation}},
  \href{https://doi.org/10.1007/JHEP08(2012)134}{\emph{JHEP} {\bfseries 08}
  (2012) 134} [\href{https://arxiv.org/abs/1203.1913}{{\ttfamily 1203.1913}}].

\bibitem{Gromov:2015dfa}
N.~Gromov and F.~Levkovich-Maslyuk, \emph{{Quantum Spectral Curve for a Cusped
  Wilson Line in N=4 SYM}},  \href{https://arxiv.org/abs/1510.02098}{{\ttfamily
  1510.02098}}.

\bibitem{Bargheer:2017nne}
T.~Bargheer, J.~Caetano, T.~Fleury, S.~Komatsu and P.~Vieira, \emph{{Handling
  Handles I: Nonplanar Integrability}},
  \href{https://arxiv.org/abs/1711.05326}{{\ttfamily 1711.05326}}.

\bibitem{Banks:1981nn}
T.~Banks and A.~Zaks, \emph{{On the Phase Structure of Vector-Like Gauge
  Theories with Massless Fermions}},
  \href{https://doi.org/10.1016/0550-3213(82)90035-9}{\emph{Nucl. Phys.}
  {\bfseries B196} (1982) 189}.

\bibitem{ElShowk:2012ht}
S.~El-Showk, M.~F. Paulos, D.~Poland, S.~Rychkov, D.~Simmons-Duffin and
  A.~Vichi, \emph{{Solving the 3D Ising Model with the Conformal Bootstrap}},
  \href{https://doi.org/10.1103/PhysRevD.86.025022}{\emph{Phys. Rev.}
  {\bfseries D86} (2012) 025022}
  [\href{https://arxiv.org/abs/1203.6064}{{\ttfamily 1203.6064}}].

\bibitem{Rattazzi:2008pe}
R.~Rattazzi, V.~S. Rychkov, E.~Tonni and A.~Vichi, \emph{{Bounding scalar
  operator dimensions in 4D CFT}},
  \href{https://doi.org/10.1088/1126-6708/2008/12/031}{\emph{JHEP} {\bfseries
  12} (2008) 031} [\href{https://arxiv.org/abs/0807.0004}{{\ttfamily
  0807.0004}}].

\bibitem{Frolov:2005dj}
S.~Frolov, \emph{{Lax pair for strings in Lunin-Maldacena background}},
  \href{https://doi.org/10.1088/1126-6708/2005/05/069}{\emph{JHEP} {\bfseries
  05} (2005) 069} [\href{https://arxiv.org/abs/hep-th/0503201}{{\ttfamily
  hep-th/0503201}}].

\bibitem{Beisert:2005if}
N.~Beisert and R.~Roiban, \emph{{Beauty and the twist: The Bethe ansatz for
  twisted N=4 SYM}},
  \href{https://doi.org/10.1088/1126-6708/2005/08/039}{\emph{JHEP} {\bfseries
  08} (2005) 039} [\href{https://arxiv.org/abs/hep-th/0505187}{{\ttfamily
  hep-th/0505187}}].

\bibitem{Sieg:2016vap}
C.~Sieg and M.~Wilhelm, \emph{{On a CFT limit of planar $\gamma_i$-deformed
  $\mathcal{N}=4$ SYM theory}},
  \href{https://doi.org/10.1016/j.physletb.2016.03.004}{\emph{Phys. Lett.}
  {\bfseries B756} (2016) 118}
  [\href{https://arxiv.org/abs/1602.05817}{{\ttfamily 1602.05817}}].

\bibitem{Grabner:2017pgm}
D.~Grabner, N.~Gromov, V.~Kazakov and G.~Korchemsky, \emph{{Strongly
  gamma-deformed N=4 SYM as an integrable CFT}},
  \href{https://arxiv.org/abs/1711.04786}{{\ttfamily 1711.04786}}.

\bibitem{Tseytlin:1999ii}
A.~A. Tseytlin and K.~Zarembo, \emph{{Effective potential in nonsupersymmetric
  SU(N) x SU(N) gauge theory and interactions of type 0 D3-branes}},
  \href{https://doi.org/10.1016/S0370-2693(99)00471-2}{\emph{Phys. Lett.}
  {\bfseries B457} (1999) 77}
  [\href{https://arxiv.org/abs/hep-th/9902095}{{\ttfamily hep-th/9902095}}].

\bibitem{Jin:2013baa}
Q.~Jin, \emph{{The Emergence of Supersymmetry in $\gamma_i$-deformed ${\cal
  N}=4$ super-Yang-Mills theory}},
  \href{https://arxiv.org/abs/1311.7391}{{\ttfamily 1311.7391}}.

\bibitem{Fokken:2013aea}
J.~Fokken, C.~Sieg and M.~Wilhelm, \emph{{Non-conformality of ${{\gamma
  }_{i}}$-deformed N = 4 SYM theory}},
  \href{https://doi.org/10.1088/1751-8113/47/45/455401}{\emph{J. Phys.}
  {\bfseries A47} (2014) 455401}
  [\href{https://arxiv.org/abs/1308.4420}{{\ttfamily 1308.4420}}].

\bibitem{Fokken:2014soa}
J.~Fokken, C.~Sieg and M.~Wilhelm, \emph{{A piece of cake: the ground-state
  energies in $\gamma_{i}$ -deformed $ \mathcal{N} $ = 4 SYM theory at leading
  wrapping order}}, \href{https://doi.org/10.1007/JHEP09(2014)078}{\emph{JHEP}
  {\bfseries 09} (2014) 78} [\href{https://arxiv.org/abs/1405.6712}{{\ttfamily
  1405.6712}}].

\bibitem{Gromov:2010dy}
N.~Gromov and F.~Levkovich-Maslyuk, \emph{{Y-system and $\beta$-deformed N=4
  Super-Yang-Mills}},
  \href{https://doi.org/10.1088/1751-8113/44/1/015402}{\emph{J. Phys.}
  {\bfseries A44} (2011) 015402}
  [\href{https://arxiv.org/abs/1006.5438}{{\ttfamily 1006.5438}}].

\bibitem{Ahn:2011xq}
C.~Ahn, Z.~Bajnok, D.~Bombardelli and R.~I. Nepomechie, \emph{{TBA, NLO Luscher
  correction, and double wrapping in twisted AdS/CFT}},
  \href{https://doi.org/10.1007/JHEP12(2011)059}{\emph{JHEP} {\bfseries 12}
  (2011) 059} [\href{https://arxiv.org/abs/1108.4914}{{\ttfamily 1108.4914}}].

\bibitem{Gurdogan:2015csr}
O.~Gurdogan and V.~Kazakov, \emph{{New integrable non-gauge 4D QFTs from
  strongly deformed planar N=4 SYM}},
  \href{https://arxiv.org/abs/1512.06704}{{\ttfamily 1512.06704}}.

\bibitem{Caetano:2016ydc}
J.~Caetano, O.~Gurdogan and V.~Kazakov, \emph{{Chiral limit of N = 4 SYM and
  ABJM and integrable Feynman graphs}},
  \href{https://arxiv.org/abs/1612.05895}{{\ttfamily 1612.05895}}.

\bibitem{Gromov:2017cja}
N.~Gromov, V.~Kazakov, G.~Korchemsky, S.~Negro and G.~Sizov,
  \emph{Integrability of conformal fishnet theory}, {\emph{arXiv:1706.04167
  [hep-th]} }.

\bibitem{Chicherin:2017cns}
D.~Chicherin, V.~Kazakov, F.~Loebbert, D.~Mueller and D.-l. Zhong,
  \emph{{Yangian Symmetry for Bi-Scalar Loop Amplitudes}},
  \href{https://arxiv.org/abs/1704.01967}{{\ttfamily 1704.01967}}.

\bibitem{Chicherin:2017frs}
D.~Chicherin, V.~Kazakov, F.~Loebbert, D.~Mueller and D.-l. Zhong,
  \emph{{Yangian Symmetry for Fishnet Feynman Graphs}},
  \href{https://doi.org/10.1103/PhysRevD.96.121901}{\emph{Phys. Rev.}
  {\bfseries D96} (2017) 121901}
  [\href{https://arxiv.org/abs/1708.00007}{{\ttfamily 1708.00007}}].

\bibitem{Kazakov:2018qbr}
V.~Kazakov and E.~Olivucci, \emph{{Bi-scalar integrable CFT at any dimension}},
   \href{https://arxiv.org/abs/1801.09844}{{\ttfamily 1801.09844}}.

\bibitem{Zamolodchikov:1980mb}
A.~B. Zamolodchikov, \emph{{'FISHNET' DIAGRAMS AS A COMPLETELY INTEGRABLE
  SYSTEM}}, \href{https://doi.org/10.1016/0370-2693(80)90547-X}{\emph{Phys.
  Lett.} {\bfseries 97B} (1980) 63}.

\bibitem{Gromov:2016rrp}
N.~Gromov and F.~Levkovich-Maslyuk, \emph{{Quark-anti-quark potential in $
  \mathcal{N} =$ 4 SYM}},
  \href{https://doi.org/10.1007/JHEP12(2016)122}{\emph{JHEP} {\bfseries 12}
  (2016) 122} [\href{https://arxiv.org/abs/1601.05679}{{\ttfamily
  1601.05679}}].

\bibitem{Derkachov:2001yn}
S.~E. Derkachov, G.~P. Korchemsky and A.~N. Manashov, \emph{{Noncompact
  Heisenberg spin magnets from high-energy QCD: 1. Baxter Q operator and
  separation of variables}},
  \href{https://doi.org/10.1016/S0550-3213(01)00457-6}{\emph{Nucl. Phys.}
  {\bfseries B617} (2001) 375} [\href{https://arxiv.org/abs/arXiv:0107193
  [hep-th]}{{\ttfamily arXiv:0107193 [hep-th]}}].

\bibitem{DeVega:2001pu}
H.~J. De~Vega and L.~N. Lipatov, \emph{{Interaction of reggeized gluons in the
  Baxter-Sklyanin representation}},
  \href{https://doi.org/10.1103/PhysRevD.64.114019}{\emph{Phys. Rev.}
  {\bfseries D64} (2001) 114019}
  [\href{https://arxiv.org/abs/hep-ph/0107225}{{\ttfamily hep-ph/0107225}}].

\bibitem{Basso:2017jwq}
B.~Basso and L.~J. Dixon, \emph{{Gluing Ladder Feynman Diagrams into
  Fishnets}}, \href{https://doi.org/10.1103/PhysRevLett.119.071601}{\emph{Phys.
  Rev. Lett.} {\bfseries 119} (2017) 071601}
  [\href{https://arxiv.org/abs/1705.03545}{{\ttfamily 1705.03545}}].

\bibitem{Tod:1977harm}
V.~K. Dobrev et~al., \emph{Harmonic analysis on the n-dimensional lorentz group
  and its application to conformal quantum field theory},
  \href{https://doi.org/10.1007/BFb0009678}{\emph{Lect.Notes Phys. 63}
  {\bfseries 12} (1977) 059}.

\bibitem{Dolan2011}
F.~A. Dolan and H.~Osborn, \emph{{Conformal Partial Waves: Further Mathematical
  Results}}, \href{https://doi.org/10.1016/j.physletb.2012.09.045}{\emph{Phys.
  Lett} {\bfseries 718} (2011) 169}
  [\href{https://arxiv.org/abs/arXiv:1108.6194 [hep-th]}{{\ttfamily
  arXiv:1108.6194 [hep-th]}}].

\bibitem{Dolan:2000ut}
F.~A. Dolan and H.~Osborn, \emph{{Conformal four point functions and the
  operator product expansion}},
  \href{https://doi.org/10.1016/S0550-3213(01)00013-X}{\emph{Nucl. Phys.}
  {\bfseries B599} (2001) 459} [\href{https://arxiv.org/abs/arXiv:0011040
  [hep-th]}{{\ttfamily arXiv:0011040 [hep-th]}}].

\bibitem{Korchemsky:2015cyx}
G.~P. Korchemsky, \emph{{On level crossing in conformal field theories}},
  \href{https://doi.org/10.1007/JHEP03(2016)212}{\emph{JHEP} {\bfseries 03}
  (2016) 212} [\href{https://arxiv.org/abs/1512.05362}{{\ttfamily
  1512.05362}}].

\bibitem{Escobedo2010}
J.~Escobedo, N.~Gromov, A.~Sever and P.~Vieira, \emph{Tailoring three-point
  functions and integrability},
  \href{https://arxiv.org/abs/1012.2475v2}{{\ttfamily 1012.2475v2}}.

\bibitem{Gromov2012}
N.~Gromov and P.~Vieira, \emph{Tailoring three-point functions and
  integrability iv. theta-morphism},
  \href{https://arxiv.org/abs/1205.5288v1}{{\ttfamily 1205.5288v1}}.

\bibitem{Caetano:TBP}
J.~Caetano, ``{To be published}.''.

\bibitem{Brown:2004ugm}
F.~C.~S. Brown, \emph{{Polylogarithmes multiples uniformes en une variable}},
  \href{https://doi.org/10.1016/j.crma.2004.02.001}{\emph{Compt. Rend. Math.}
  {\bfseries 338} (2004) 527}.

\bibitem{Dixon:2012yy}
L.~J. Dixon, C.~Duhr and J.~Pennington, \emph{{Single-valued harmonic
  polylogarithms and the multi-Regge limit}},
  \href{https://doi.org/10.1007/JHEP10(2012)074}{\emph{JHEP} {\bfseries 10}
  (2012) 074} [\href{https://arxiv.org/abs/1207.0186}{{\ttfamily 1207.0186}}].

\bibitem{Remiddi:1999ew}
E.~Remiddi and J.~A.~M. Vermaseren, \emph{{Harmonic polylogarithms}},
  \href{https://doi.org/10.1142/S0217751X00000367}{\emph{Int. J. Mod. Phys.}
  {\bfseries A15} (2000) 725}
  [\href{https://arxiv.org/abs/hep-ph/9905237}{{\ttfamily hep-ph/9905237}}].

\bibitem{Maitre:2005uu}
D.~Maitre, \emph{{HPL, a mathematica implementation of the harmonic
  polylogarithms}},
  \href{https://doi.org/10.1016/j.cpc.2005.10.008}{\emph{Comput. Phys. Commun.}
  {\bfseries 174} (2006) 222}
  [\href{https://arxiv.org/abs/hep-ph/0507152}{{\ttfamily hep-ph/0507152}}].

\bibitem{Gromov:2011jh}
N.~Gromov, A.~Sever and P.~Vieira, \emph{{Tailoring Three-Point Functions and
  Integrability III. Classical Tunneling}},
  \href{https://doi.org/10.1007/JHEP07(2012)044}{\emph{JHEP} {\bfseries 07}
  (2012) 044} [\href{https://arxiv.org/abs/1111.2349}{{\ttfamily 1111.2349}}].

\bibitem{Kazama:2012is}
Y.~Kazama and S.~Komatsu, \emph{{Wave functions and correlation functions for
  GKP strings from integrability}},
  \href{https://doi.org/10.1007/JHEP09(2012)022}{\emph{JHEP} {\bfseries 09}
  (2012) 022} [\href{https://arxiv.org/abs/1205.6060}{{\ttfamily 1205.6060}}].

\bibitem{Basso:2018agi}
B.~Basso and D.-l. Zhong, \emph{{Continuum limit of fishnet graphs and AdS
  sigma model}},  \href{https://arxiv.org/abs/1806.04105}{{\ttfamily
  1806.04105}}.

\bibitem{Hogervorst:2013sma}
M.~Hogervorst and S.~Rychkov, \emph{{Radial Coordinates for Conformal Blocks}},
  \href{https://doi.org/10.1103/PhysRevD.87.106004}{\emph{Phys. Rev.}
  {\bfseries D87} (2013) 106004}
  [\href{https://arxiv.org/abs/1303.1111}{{\ttfamily 1303.1111}}].

\bibitem{Medina-Rincon:2018wjs}
D.~Medina-Rincon, A.~A. Tseytlin and K.~Zarembo, \emph{{Precision matching of
  circular Wilson loops and strings in AdS$_{5}\times$ S$^{5}$}},
  \href{https://doi.org/10.1007/JHEP05(2018)199}{\emph{JHEP} {\bfseries 05}
  (2018) 199} [\href{https://arxiv.org/abs/1804.08925}{{\ttfamily
  1804.08925}}].

\bibitem{Kazakov2004a}
V.~A. Kazakov, A.~Marshakov, J.~A. Minahan and K.~Zarembo,
  \emph{Classical/quantum integrability in ads/cft},
  \href{https://arxiv.org/abs/hep-th/0402207v6}{{\ttfamily hep-th/0402207v6}}.

\bibitem{Beisert2005}
N.~Beisert, V.~A. Kazakov, K.~Sakai and K.~Zarembo, \emph{The algebraic curve
  of classical superstrings on ads_5xs^5},
  \href{https://arxiv.org/abs/hep-th/0502226v3}{{\ttfamily hep-th/0502226v3}}.

\bibitem{Kazakov2004}
V.~A. Kazakov and K.~Zarembo, \emph{Classical/quantum integrability in
  non-compact sector of ads/cft},
  \href{https://arxiv.org/abs/hep-th/0410105v3}{{\ttfamily hep-th/0410105v3}}.

\bibitem{Karateev:2017jgd}
D.~Karateev, P.~Kravchuk and D.~Simmons-Duffin, \emph{{Weight Shifting
  Operators and Conformal Blocks}},
  \href{https://doi.org/10.1007/JHEP02(2018)081}{\emph{JHEP} {\bfseries 02}
  (2018) 081} [\href{https://arxiv.org/abs/1706.07813}{{\ttfamily
  1706.07813}}].

\bibitem{Kim:2017sju}
M.~Kim, N.~Kiryu, S.~Komatsu and T.~Nishimura, \emph{{Structure Constants of
  Defect Changing Operators on the 1/2 BPS Wilson Loop}},
  \href{https://doi.org/10.1007/JHEP12(2017)055}{\emph{JHEP} {\bfseries 12}
  (2017) 055} [\href{https://arxiv.org/abs/1710.07325}{{\ttfamily
  1710.07325}}].

\bibitem{Costa:2011mg}
M.~S. Costa, J.~Penedones, D.~Poland and S.~Rychkov, \emph{{Spinning Conformal
  Correlators}}, \href{https://doi.org/10.1007/JHEP11(2011)071}{\emph{JHEP}
  {\bfseries 11} (2011) 071} [\href{https://arxiv.org/abs/1107.3554}{{\ttfamily
  1107.3554}}].

\bibitem{Giombi:2017dtl}
S.~Giombi, I.~R. Klebanov and G.~Tarnopolsky, \emph{{Bosonic tensor models at
  large $N$ and small $\epsilon$}},
  \href{https://doi.org/10.1103/PhysRevD.96.106014}{\emph{Phys. Rev.}
  {\bfseries D96} (2017) 106014}
  [\href{https://arxiv.org/abs/1707.03866}{{\ttfamily 1707.03866}}].

\bibitem{Gross:2017aos}
D.~J. Gross and V.~Rosenhaus, \emph{{All point correlation functions in SYK}},
  \href{https://doi.org/10.1007/JHEP12(2017)148}{\emph{JHEP} {\bfseries 12}
  (2017) 148} [\href{https://arxiv.org/abs/1710.08113}{{\ttfamily
  1710.08113}}].

\bibitem{Rosenhaus:2018dtp}
V.~Rosenhaus, \emph{{An introduction to the SYK model}},
  \href{https://arxiv.org/abs/1807.03334}{{\ttfamily 1807.03334}}.

\bibitem{WithDavidToAppear}
D.~Grabner, N.~Gromov, V.~Kazakov and G.~Korchemsky, \emph{{to appear}}, .

\bibitem{Lipa:1993pmr}
L.~N. Lipatov, \emph{{ High energy asymptotics of multi-color QCD and exactly
  sol vable lattice models}},
  \href{https://doi.org/10.1080/01422419708219641}{\emph{JETP Lett. 59 (1994)
  596-599;} {\bfseries 12} (1994) 059}
  [\href{https://arxiv.org/abs/arXiv:9311037v1 [hep-th]}{{\ttfamily
  arXiv:9311037v1 [hep-th]}}].

\bibitem{Faddeev:1994zg}
L.~D. Faddeev and G.~P. Korchemsky, \emph{{High-energy QCD as a completely
  integrable model}},
  \href{https://doi.org/10.1016/0370-2693(94)01363-H}{\emph{Phys. Lett.}
  {\bfseries B342} (1995) 311}
  [\href{https://arxiv.org/abs/hep-th/9404173}{{\ttfamily hep-th/9404173}}].

\bibitem{Balitsky:2013npa}
I.~Balitsky, V.~Kazakov and E.~Sobko, \emph{{Two-point correlator of twist-2
  light-ray operators in N=4 SYM in BFKL approximation}},
  \href{https://arxiv.org/abs/1310.3752}{{\ttfamily 1310.3752}}.

\bibitem{Balitsky:2015oux}
I.~Balitsky, V.~Kazakov and E.~Sobko, \emph{{Three-point correlator of twist-2
  light-ray operators in N=4 SYM in BFKL approximation}},
  \href{https://arxiv.org/abs/1511.03625}{{\ttfamily 1511.03625}}.

\bibitem{Balitsky:2015tca}
I.~Balitsky, V.~Kazakov and E.~Sobko, \emph{{Structure constant of twist-2
  light-ray operators in the Regge limit}},
  \href{https://doi.org/10.1103/PhysRevD.93.061701}{\emph{Phys. Rev.}
  {\bfseries D93} (2016) 061701}
  [\href{https://arxiv.org/abs/1506.02038}{{\ttfamily 1506.02038}}].

\bibitem{Liu:2018jhs}
J.~Liu, E.~Perlmutter, V.~Rosenhaus and D.~Simmons-Duffin,
  \emph{{$d$-dimensional SYK, AdS Loops, and $6j$ Symbols}},
  \href{https://arxiv.org/abs/1808.00612}{{\ttfamily 1808.00612}}.

\bibitem{Simmons-Duffin:2017nub}
D.~Simmons-Duffin, D.~Stanford and E.~Witten, \emph{{A spacetime derivation of
  the Lorentzian OPE inversion formula}},
  \href{https://arxiv.org/abs/1711.03816}{{\ttfamily 1711.03816}}.

\bibitem{Grozin:2012xi}
A.~G. Grozin, \emph{{Massless two-loop self-energy diagram: Historical
  review}}, \href{https://doi.org/10.1142/S0217751X12300189}{\emph{Int. J. Mod.
  Phys.} {\bfseries A27} (2012) 1230018}
  [\href{https://arxiv.org/abs/1206.2572}{{\ttfamily 1206.2572}}].

\bibitem{Pomoni:2008de}
E.~Pomoni and L.~Rastelli, \emph{{Large N Field Theory and AdS Tachyons}},
  \href{https://doi.org/10.1088/1126-6708/2009/04/020}{\emph{JHEP} {\bfseries
  04} (2009) 020} [\href{https://arxiv.org/abs/0805.2261}{{\ttfamily
  0805.2261}}].

\bibitem{Dymarsky:2005uh}
A.~Dymarsky, I.~R. Klebanov and R.~Roiban, \emph{{Perturbative search for fixed
  lines in large N gauge theories}},
  \href{https://doi.org/10.1088/1126-6708/2005/08/011}{\emph{JHEP} {\bfseries
  08} (2005) 011} [\href{https://arxiv.org/abs/hep-th/0505099}{{\ttfamily
  hep-th/0505099}}].

\end{thebibliography}\endgroup

\end{document}